\documentclass[11pt]{article}
\usepackage{fullpage}
\RequirePackage{amsthm,amsmath,amsfonts,amssymb,graphicx,subcaption,multirow,bm,placeins,xcolor}
\usepackage[hidelinks]{hyperref}
\usepackage[round]{natbib}
\usepackage{listings}
\usepackage{authblk}

\title{\vspace{-1cm}A Methodology for Exploring Deep Convolutional Features in Relation to Hand-Crafted Features \\ with an Application to Music Audio Modeling}
\author[1]{Anna K. Yanchenko}
\author[2]{Mohammadreza Soltani}
\author[2]{Robert J. Ravier}
\author[1]{Sayan Mukherjee}
\author[2]{Vahid Tarokh}

\affil[1]{Department of Statistical Science,
Duke University, Durham NC 27708-0251. U.S.A.}

\affil[2]{Department of Electrical and Computer Engineering,
Duke University, Durham NC 27708-0251. U.S.A.}
\date{\today}
\begin{document}
\maketitle

\begin{abstract}
Understanding the features learned by deep models is important from a model trust perspective, especially as deep systems are deployed in the real world. Most recent approaches for deep feature understanding or model explanation focus on highlighting input data features that are relevant for classification decisions. In this work, we instead take the perspective of relating deep features to well-studied, hand-crafted features that are meaningful for the application of interest.  We propose a methodology and set of systematic experiments for exploring deep features in this setting, where input feature importance approaches for deep feature understanding do not apply.  Our experiments focus on understanding which hand-crafted and deep features are useful for the classification task of interest, how robust these features are for related tasks and how similar the deep features are to the meaningful hand-crafted features.  Our proposed method is general to many application areas and we demonstrate its utility on orchestral music audio data.
\smallskip\smallskip

\end{abstract}

\section{Introduction}\label{sec:intro}
Understanding the features learned by deep models is important from a model trust perspective, especially as deep systems are deployed in the real world. The majority of existing explainable and interpretable methods \citep[e.g.][]{NEURIPS2020_71e9c662, NEURIPS2018_294a8ed2, Simonyan:2014, NEURIPS2019_adf7ee2d} focus on feature importance  approaches that select features of the input data that are relevant for the classification decisions of the model.  These type of techniques are especially applicable for \textit{natural images} (e.g. saliency based techniques that highlight important pixels in input images).  However, there are many settings where highlighting aspects of the input data that are important to the classification decision is not meaningful or interpretable for the application, for example, any setting where the input features are \textit{summary statistics}. 

Instead of feature importance-type approaches, we take the perspective of \textit{relating} deep features to \textit{well-studied, hand-crafted} features that are meaningful for the application of interest. In this work, we propose a method with systematic experiments to explore learned deep features in application areas where deep architectures achieve high accuracy on relevant classification tasks, yet existing feature importance-based explainability approaches don't apply.  We apply our methodology to orchestral music audio data, where mel-spectrograms are commonly used as input training data.  Methods like saliency maps don't apply in this setting, as highlighting aspects of a mel-spectrogram that correspond to a given composer is not easily interpretable.  Our proposed method and experiments are general to any setting where the focus is on discriminative tasks, there exist well-studied, hand-crafted features that are meaningful for the application and feature importance measures for the input data are not especially meaningful for the application. 

Our method leverages existing deep learning analysis approaches \citep{NEURIPS2020_71e9c662, pmlr-v97-kornblith19a} to explore deep convolutional features for these settings via three systematic experiments:
\begin{enumerate}
\item What is the baseline discriminative performance of the hand-crafted and deep features?  Are these features robust for different application-meaningful tasks?
\item How similar are the trained deep features to the application-meaningful, hand-crafted features?
\item How similar are the \textit{untrained} deep features to the hand-crafted features? Do untrained convolutional architectures have a useful inductive bias for the discriminative tasks of interest?
\end{enumerate}
Experiment \#1 baselines the (potential) gap in performance between the hand-crafted features and the deep features and answers the question: Can these deep features perform well across various target tasks, regardless of whether or not they are extracted from deep architectures trained on that task?  This experiment gives insight into how robust the deep features are for other down-stream or transfer learning settings.  Experiment \#2 measures the quantitative similarity between the deep and hand-crafted features, giving insight into the features learned by the deep architectures in reference to well-studied,  hand-crafted features.  This allows for interpretability in relation to features that are commonly used and/or meaningful for the application of interest.  Finally, Experiment \#3 explores the inductive bias of the convolutional architectures, again in relation to the hand-crafted features, complimenting deep prior approaches \citep[e.g][]{deep-image, dap}.  

Overall, these experiments allow for understanding deep features \textit{in the context} of the application-relevant hand-crafted features and the results of the experiments can then be leveraged to further improve the deep modeling. If the deep features are robust to the discriminative task, they could be used for a variety of downstream settings, without needing to retrain the deep models. If the results of all three experiments reveal that the hand-crafted features represent complementary information to the deep features, these two types of features could be combined to further improve accuracy on the discriminative task, and/or to steer the training of the deep models for increased similarity to the hand-crafted features for interpretable models, like \citet{NEURIPS2019_adf7ee2d}.

We apply our method to music audio classification and focus on \textit{orchestral} audio data, which is understudied in the literature.  We represent the input music audio as mel-spectrograms for various classification tasks and relate deep convolutional features to several hand-crafted features motivated by signal processing. 
This music audio setting is especially relevant for our methodology.  Unlike in natural image settings, it is challenging or impossible to visually inspect a spectrogram and determine the composer of the piece being performed, necessitating approaches other than saliency-based or other feature importance techniques to understand deep music audio features.  

\subsection{Contributions}

Our contributions are as follows:
\begin{itemize}
\item We propose a methodology and systematic set of experiments for exploring and understanding deep features in the context of well-studied, application-meaningful, hand-crafted features. Our experiments are designed for settings beyond natural image data where feature importance measures do not apply, for example, any setting where the input data consists of summary statistics.
\item We apply this method across various datasets, architectures, tasks, layers and initializations to explore the deep features learned by convolutional architectures on discriminative, \textit{orchestral} music audio tasks.  
\end{itemize}

In the context of these specific music tasks, our experiments reveal:
\begin{itemize}
\item Convolutional deep features are robust to discriminative tasks and achieve high classification accuracy and low predictive error on several distinct tasks, whether or not the deep features were extracted from architectures originally trained on that specific task. 
\item Deep features from earlier layers in the convolutional architecture are highly similar to the hand-crafted features that are motivated by classical signal processing.
\item Earlier layers have high similarity to the hand-crafted features even when extracted from \textit{untrained} deep models, suggesting that convolutions may have a useful inductive bias for these music audio tasks.
\end{itemize}

\subsection{Related Work}\label{sec:related_work}

The majority of existing approaches for deep model explainability or interpretability methods focus on input feature importance approaches, which are especially relevant for natural image data.  Some recent approaches are interpretable by design \citep{NEURIPS2019_adf7ee2d}, while numerous post-hoc explainability measures have been developed and compared for understanding deep architectures and visualizing classification decisions by convolutional models \citep{Simonyan:2014, NIPS2017_8a20a862, pmlr-v70-shrikumar17a, NEURIPS2019_77d2afcb, NEURIPS2020_c7bf0b7c, NEURIPS2019_fe4b8556, NEURIPS2018_294a8ed2, pmlr-v70-sundararajan17a}.  Many of these methods are designed specifically for image tasks (for example, utilizing image segmentation methods \citep{NEURIPS2019_77d2afcb}) and often visualize ``important'' parts of input images that are relevant for the classification decisions of the deep model using saliency map techniques, e.g. \citet{NEURIPS2018_294a8ed2} for a comparison of approaches.  This limits many of these approaches to only natural image data or to settings where the input data features are inherently interpretable. Our proposed method instead focuses on settings where input data feature importance is not inherently meaningful for the application due to the nature of the input data, for example, audio data represented as a mel-spectrogram. 

From a music perspective, prior work for analyzing deep audio features includes \citet{DBLP:journals/corr/abs-2007-11154}, which used pre-trained ImageNet weights for music genre classification and used integrated gradients to analyze mel-spectrograms, finding that convolutional deep architectures were able to detect sound events.  Additionally, \citet{dap} explored convolutions for denoising and sound separation for general audio signals and proposed a new type of convolution to better represent the structure in audio signals.  \citet{DBLP:conf/ismir/ChoiFSC17} showed that convolutional features are useful as a general purpose music representation across various transfer learning tasks for music tagging and genre classification, while \citet{8682475} also demonstrated the importance of audio-informed inputs for evaluating deep embeddings.  However, none of the above approaches use \textit{orchestral} audio data, which is an application-specific contribution of our work.  Our work additionally focuses on music audio data as a specific illustrative example of our general methodology.

\section{Setting Details}\label{sec:setting}

Our experimental approach is applicable to any setting where the focus is on discriminative tasks,  feature importance-based explainability methods are not applicable and there exist well-studied, hand-crafted features for the application of interest. For example, audio data represented as a spectrogram or data consisting of summary statistics both fall into this setting.    In this section, we elaborate on the above setting in general and then in the specific context of our music audio tasks; full details are given in the Appendix. 

\subsection{Data and Tasks}

Our experiments and methodology apply to understanding deep features learned for discriminative tasks.  We focus on convolutional deep features and assume 2D input data, where the representation of the input data is such that feature importance or saliency based methods for feature understanding do not apply.  
To demonstrate the utility of our method, we focus on music audio tasks, specifically, one widely used benchmark dataset \citep{engel_neural_2017} and two datasets comprised of real, \textit{orchestral} audio recordings.  Again, this setting provides an especially relevant demonstration of our method, as it is challenging or impossible to visually inspect a spectrogram and determine the composer of the piece being performed. The example discriminative tasks of interest focus on both low-level musical concepts (i.e. note pitch, instrument) and high-level, stylistic concepts. Full orchestral audio recordings for machine learning tasks are relatively under-explored in the literature (e.g. \citet{DBLP:conf/ismir/VermaT19} use symbolic representations), likely due to the complexity in terms of multiple instruments playing simultaneously. It is important to analyze orchestral audio recordings to evaluate the utility of features in settings that are much more complex than a single instrument \citep{engel_neural_2017} or music genre classification \citep{Liu:2021, oramas_multi-label_2017}.

The three datasets and tasks of interest are:

\noindent \textbf{(1) NSynth} \citep{engel_neural_2017}:  This dataset consists of examples of single instruments playing a single note pitch at a single velocity (volume).  We limit to only acoustic recordings and are mainly focused on classifying the \textbf{instrument family}, which is an 8-class classification task (the classes are brass, flute, guitar, keyboard, mallet, reed, string, vocal).  Related discriminative tasks discussed in Section~\ref{sec:experiment_1} include predicting the note pitch and note velocity performed in each audio sample. Our dataset consists of 50000 total examples and the focus of this dataset is the recovery of \textit{low-level} musical features (e.g., instrument, pitch, volume).

\noindent \textbf{(2) Composer}: This dataset consists of symphonic recordings by five composers (Beethoven, Brahms, Haydn, Sibelius, Tchaikovsky) performed by the Berlin Philharmonic under Herbert von Karajan. The main task is 5-class \textbf{composer} classification, with the goal of recovering differences in \textit{musical style} between these composers.  This is a much higher-level musical concept than what single instrument is playing a single pitch, as in the NSynth dataset.   This dataset consists of 19667 total examples.  

\noindent \textbf{(3) Beethoven}: This dataset consists of recordings of all nine Beethoven symphonies recorded by 10 different orchestras, first used in a different context in \citet{HMDS}.  The main task is 10-class classification of the \textbf{orchestra} performing each piece, focusing on the evaluation of \textit{performance style} following \citet{HMDS}. This dataset consists of 52999 total examples. 

All datasets start as 4 second audio clips that are transformed into mel-spectrograms with 128 mels on the dB scale.  Mel-spectrograms are widely used to represent audio data, especially with CNNs \citep{Dieleman:2014, 8682475, 7500246}.  Note: the presented results could be explored with alternative data representations (e.g. MFCC, CQT), but we focus on the above setting to illustrate the utility of our proposed methods. 

\subsection{Convolutional Architectures and Deep Features}

We focus on exploring deep features from convolutional architectures, as these are the most commonly used architectures for 2D data, whether natural images or not. To demonstrate the utility of our methodology, we consider a wide range of convolutions.  We explore 5 different deep convolutional architectures that are selected to be fast to train and reflective of commonly used convolutions for audio modeling.  We select smaller networks to (a) explore minimal architectures for extracting features, and (b) because of the moderate size of the datasets that we consider.  Especially in music audio modeling, it is not always possible to obtain more data; for example, we already consider all 6 Tchaikovsky symphonies in the Composer dataset, and meaningful data augmentation is non-trivial for orchestral music audio.  However, our methods apply to deeper architectures or other types of deep models beyond convolutional architectures, as long as deep features can be extracted.

Specifically, each architecture consists of 3 convolutional layers with batch normalization, and max pooling after the first 2 convolutional layers, similar to \citet{Luo:2018}.  The deep layers of interest for extracting features are then: \texttt{conv1} (10 channels), \texttt{pool1}, \texttt{conv2} (20 channels), \texttt{pool2}, and \texttt{conv3} (30 channels).  The last convolutional layer is followed by a fully connected layer with 120 units and then a fully connected layer to a softmax output with the appropriate dimension for each dataset's classification task.   The 5 architectures considered all have the same aforementioned structure and differ only in the type of convolution used.  The first architecture uses \textbf{Regular} convolutions and the second architecture uses \textbf{Deformable} convolutions~\citep{8237351} for the last convolutional layer, \texttt{conv3}. The \textbf{Dilated} architecture uses dilated convolutions at all three layers and is motivated by the success of dilated convolutions in the WaveNet architecture~\citep{oord_wavenet:_2016}. The last two architectures are musically motivated, following \citet{7500246}; the \textbf{1dF} architecture utilizes 1 dimensional dilated convolutions along the frequency dimension only, while the \textbf{1dT} architecture uses dilated convolutions along the time dimension only. The 1dF architecture specializes in learning frequency features, like pitch or timbre, while the 1dT architecture is meant for learning temporal features like rhythm and tempo   \citep{7500246}.  

For all deep architectures, we use 5 different initializations, as the features learned by deep architectures have been found to vary slightly based on initialization \citep{DBLP:journals/corr/LiYCLH15, NEURIPS2018_5fc34ed3}; we confirm this finding for our data in Section~ \ref{suppsec:experiment_4}.  The deep features correspond to the layer activations after each convolutional and pooling layer, and are extracted with the weights of the network frozen \citep{NEURIPS2020_71e9c662}.  We conduct experiments across all layers and focus on the deep features from the last convolutional layer (\texttt{conv3}). 

\subsection{Hand-Crafted Features}

Our experimental methodology explores deep features in relation to application-motivated, well-studied, hand-crafted features.  These features can be selected as appropriate for the application, but should provide some context for understanding the deep features.  We illustrate example hand-crafted features for our music audio tasks, motivated by signal processing. We explore numerous different hand-crafted features (Section~\ref{subsec:features}) and highlight a selection of top features in terms of classification accuracy here.  All features are calculated from the mel-spectrograms directly.  These features are: \textit{(1) Mean Power} of the mel-spectrogram over time (relates to the volume of the audio signal). \textit{(2) Time to -70 dB}: for each mel-frequency bin, at what time does that frequency reach -70 dB in power or less?  This is a measure of the decay of the audio signal for each frequency bin and relates to envelope features (i.e. the ADSR, attack, decay, sustain, release model) commonly used in music processing \citep{Muller:2015}. \textit{(3) Mean Wavelet (25)}: the mean coefficients across time from a Ricker wavelet with bandwidth 25 and centered at 0 applied to each frequency bin of the mel-spectrogram.  \textit{(4) Combined Wavelet}: the same procedure as feature (3), but instead of taking the mean of the coefficients across time, the standard deviation, variance, kurtosis, 25th quantile and 75th quantile are calculated and concatenated together.  These features capture additional information about the shape of the wavelet transform over time beyond just the mean coefficients.  \textit{(5) Top Combined}: all four of the above features are concatenated together.  The wavelet features considered here are fairly robust to the choice of bandwidth parameter, and taking the summary wavelet statistics over frequency instead of time gives higher classification accuracy (Section~\ref{suppsec:experiment_1}).

\section{Experiments and Methodology}\label{sec:methodology}
In this section, we motivate the analysis methods and experiments used in our proposed methodology.

\subsection{Decoding Analysis}
Following \citep{NEURIPS2020_71e9c662}, all features are analyzed in terms of their ability to achieve high classification accuracy on the main discriminative tasks for each dataset, using simple multi-class logistic regression.  All hand-crafted and deep features are normalized prior to training the logistic regression models.  The hand-crafted features are 1 dimensional vectors and fed directly into the logistic regression model, while the deep features are flattened over channel and dimension for the decoding.  That is, if the features after \texttt{conv3} are $C\times H \times W$ dimensional, the deep features fed to the logistic regression models are $CHW$ dimensional.  The decoding results are averaged over the classification accuracies for each of the deep feature architecture initializations; logistic regression models with 5 different initializations are trained for each of the hand-crafted features.

\subsection{Feature Similarity Analysis}
Ultimately, we wish to relate the deep features to the hand-crafted features to \textit{understand} what deep architectures are learning; we use linear centered kernel alignment (CKA) \citep{pmlr-v97-kornblith19a} to do so. Let $X\in\mathbb{R}^{n\times p_1}$ be one set of features, and $Y\in\mathbb{R}^{n\times p_2}$ be another set of features, where $n$ is the number of training examples and $p_1$ and $p_2$ are the dimensions of each feature.  Assume that the columns of $X$ and $Y$ are centered.  Then, the linear CKA similarity measure can be calculated as \citep{pmlr-v97-kornblith19a}:
\begin{equation}
\text{sim}(X, Y) = \dfrac{||Y^TX||_F^2}{||X^TX||_F||Y^TY||_F},
\end{equation}
where $||\cdot||_F$ is the Frobenius norm.  

The use of Linear CKA to compare hand-crafted features to deep features extends the original use of the method as proposed in \citet{pmlr-v97-kornblith19a}.  Linear CKA provides a quantitative measure of similarity between different sets of features, and can accommodate both settings where there are more features than training examples (i.e. $p_1, p_2 > n$) and settings where there is a different number of features for each representation (i.e. $p_1\neq p_2$).  Linear CKA is selected here as it is invariant to orthogonal transformation (which is important since this implies invariance to permutations, accommodating the symmetries of neural networks) and isotropic scaling, is not invariant to invertable linear transformations (which is important since this preserves scale information and does not degenerate when there are more features than training examples), passes basic sanity checks for deep feature similarity and performs very similarly to CKA alternatives with different kernels; for more details related to the above, see \citet{pmlr-v97-kornblith19a}. We explore additional similarity measures in Section~\ref{suppsec:experiment_2}. 

\subsection{Experiments and Motivations}
Our proposed methodology focuses on three experiments:
\begin{enumerate}
\item What is the baseline discriminative performance of the hand-crafted and deep features?  Are these features robust for different application-meaningful tasks?
\item How similar are the trained deep features to the application-meaningful, hand-crafted features?
\item How similar are the \textit{untrained} deep features to the hand-crafted features? Do untrained convolutional architectures have a useful inductive bias for the discriminative tasks of interest?
\end{enumerate}

Experiment \#1 baselines the gap in performance between the hand-crafted features and the deep features and explores if the deep features perform well across various target tasks, regardless of whether or not they are extracted from deep architectures trained on that task.  Experiment \#2 measures the quantitative similarity between the deep and hand-crafted features, giving insight into the features learned by the deep architectures in reference, to well-studied,  hand-crafted features.  This allows for understanding the deep features in relation to features that are commonly used and/or meaningful for the application of interest.  Finally, Experiment \#3 explores the inductive bias of the convolutional architectures, again in relation to the hand-crafted features, complimenting deep prior work \citep[e.g][]{deep-image, dap}. 


\section{Experimental Results for Music Audio Data}\label{sec:experiments}

In this section we apply our proposed methodology to our music audio data and highlight the insights that our method can provide for the application of interest.

\subsection{What Features Are Useful For the Discriminative Tasks? Are these Features Robust Across Tasks?}\label{sec:experiment_1}

The experiments in this section focus on what features are useful for the main discriminative tasks, in the sense of achieving high test accuracy. In particular, we try to understand if these features are useful across multiple tasks \citep{NEURIPS2020_71e9c662}. Table~\ref{tab:class-results} shows that across all three datasets, the hand-crafted features are able to achieve comparable accuracy to some of the deep features. Among the deep features from the last convolutional layer (\texttt{conv3}), the Deformable features achieve the highest classification accuracy across datasets, with the Regular convolutions second best. The Regular and Deformable architectures will be the main focus of the remaining results. The baseline single hidden layer, fully connected Feed Forward architecture in Table~\ref{tab:class-results} uses mel-spectrograms as features directly and is outperformed by both the hand-crafted and deep features, especially on the Composer and Beethoven datasets.  Furthermore, the deep convolutional features provide a significant dimensionality reduction compared to the mel-spectrograms themselves, for example 30 x 4 x 6 (Regular \texttt{conv3}) vs. 128 x 176 (mel-spectrograms) for the Composer and Beethoven datasets.

\begin{table*}[t]
\begin{center}
\begin{tabular}{lccc} \hline
& NSynth (8 Classes) & Composer (5 Classes) & Beethoven (10 Classes)  \\\hline
\textit{Random Guessing} & \textit{26.50\%} &  \textit{25.20\%} & \textit{10.4\%} \\\hline
Mean Power & $66.81\pm0.37$\% &  $27.50\pm0.04$\% &  $12.93\pm0.07$\% \\
Time to -70 dB & $56.95\pm0.03$\% &  $35.02\pm0.12$\% &  $23.85\pm0.10$\% \\
Mean Wavelet (25) & $60.69\pm0.08$\% & $62.65\pm0.05$\% & $\bm{74.14\pm0.04}$\textbf{\%} \\
Wavelet Combined & $74.44\pm0.18$\% &  $59.56\pm0.15$\% &  $66.32\pm0.05$\% \\
Top 4 Combined &  $\bm{84.05\pm0.07}$\textbf{\%} &  $\bm{64.04\pm0.09}$\textbf{\%} &  $73.33\pm0.04$\% \\\hline
Regular & $96.35\pm0.19$\% &  $72.41\pm1.21$\% &  $82.32\pm0.40$\% \\
Deformable & $\bm{97.80\pm0.08}$\textbf{\%} & $\bm{75.76\pm0.68}$\textbf{\%} & $\bm{84.11\pm0.70}$\textbf{\%} \\
Dilated & $95.02\pm0.42$\% &  $68.10\pm0.57$\% &  $75.98\pm0.58$\% \\
1dF & $95.60\pm0.12$\% &  $65.42\pm0.21$\% &  $72.84\pm1.08$\% \\
1dT & $94.28\pm0.69$\% &  $70.08\pm0.98$\% & $82.11\pm0.75$\% \\\hline
Feed Forward & $96.91\pm0.06$\% &  $64.86\pm2.07$\% &  $76.28\pm0.36$\% \\\hline
\end{tabular}
\end{center}
\caption{Test set accuracy for the top hand-crafted features and the deep features from the last convolutional layer (\texttt{conv3}) for each dataset.  The mean values across 5 initializations are reported with $1$ standard deviation. The hand-crafted features are able to achieve comparable accuracy to some of the deep features.  A Feed Forward, fully connected architecture with one hidden layer of 120 units is used as a baseline.}
\label{tab:class-results}
\end{table*}%

Next, we examine the robustness of deep features following \citet{NEURIPS2020_71e9c662}. Specifically, we compare how the deep features necessary for different targets are enhanced or suppressed when the discriminative task changes. The  tasks for the NSynth dataset include one classification task defined as multi-classification of instrument family and two regression tasks defined as the note pitch and note velocity (volume) prediction for each mel-spectrogram. We first train Regular and Deformable architectures separately for each of the above three tasks, and again extract the deep features from each layer for these architectures (the architectures for the regression tasks are identical to those described in Section 2.2 above, but with a linear last layer with one output and MSE loss for the regression tasks). Then, the goal is to explore how well all of these deep features perform at each discriminative task, whether or not the features are extracted from deep architectures that were originally trained on that task. For example, we want to see how the deep features that are trained for the note pitch regression perform on the note velocity task.

\begin{figure}
\centering
\begin{subfigure}[b]{\linewidth}
\includegraphics[width=\linewidth]{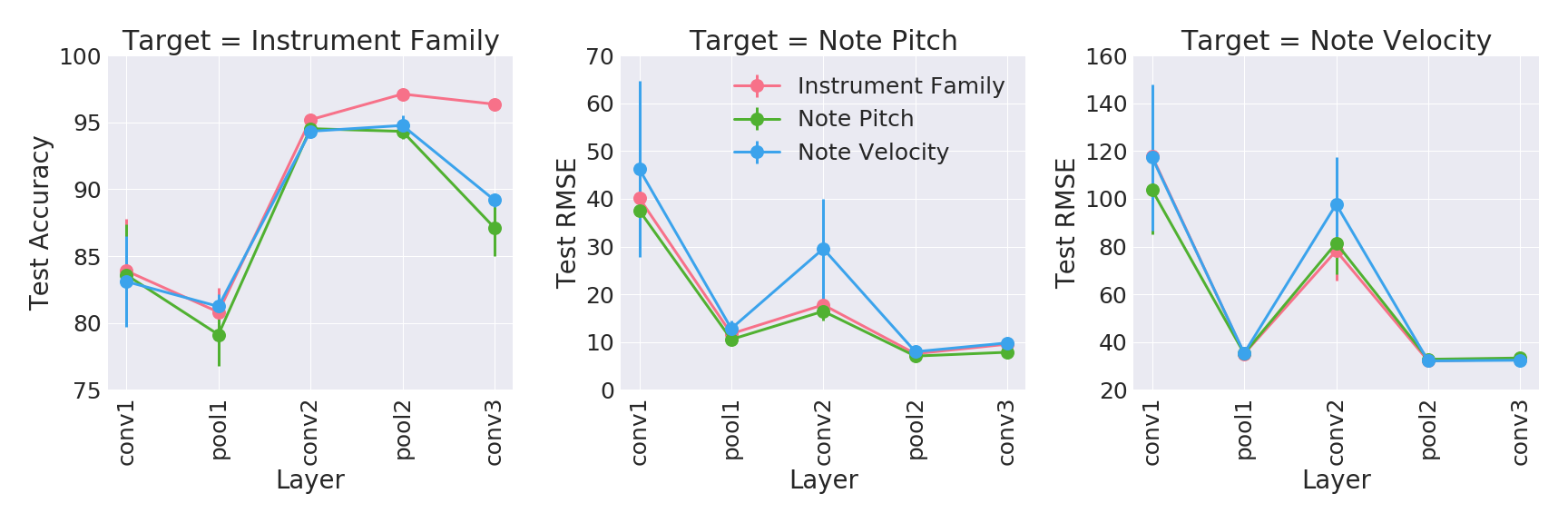}
\caption{Regular}\label{default}
\end{subfigure}

\begin{subfigure}[b]{\linewidth}
\includegraphics[width=\linewidth]{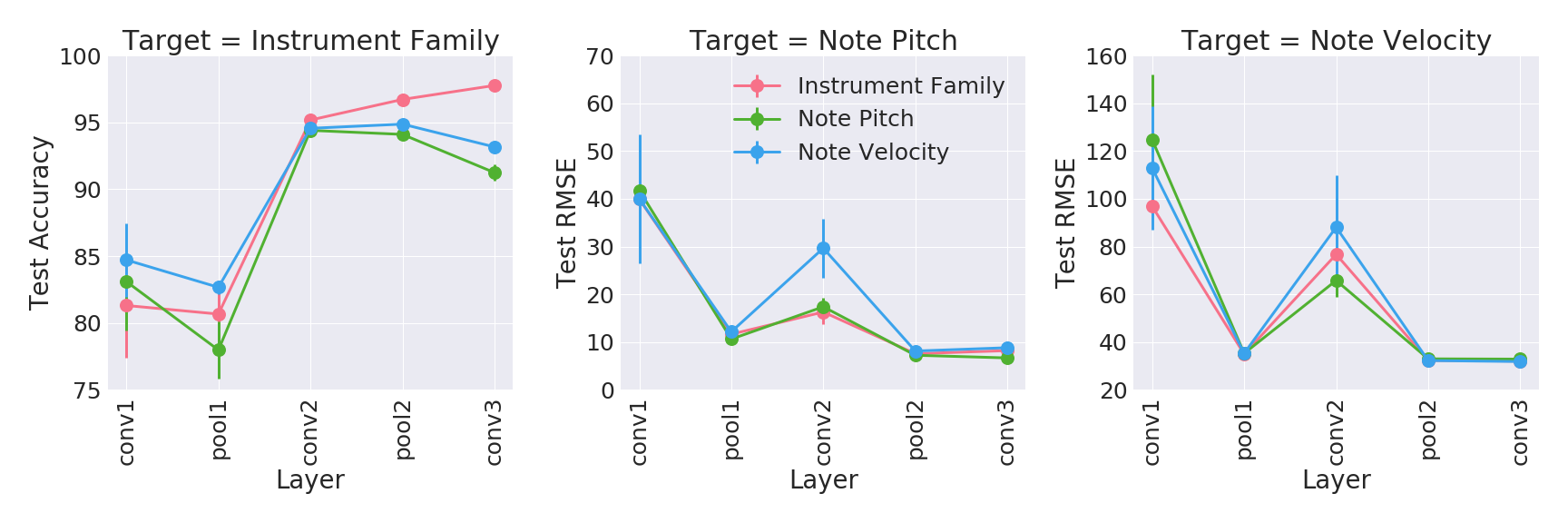}
\caption{Deformable}\label{default}
\end{subfigure}
\caption{Discriminative task results for deep features by layer for (a) Regular (\textbf{Top row}) and (b) Deformable (\textbf{Bottom row}) architectures trained on the NSynth data. Three identical CNN architectures are trained to perform each of the three target discriminative tasks: the pink curves are extracted from architectures trained to predict instrument family, the green curves are extracted from architectures trained to predict note pitch and the blue curves are extracted from architectures trained to predict note velocity. Then, these extracted features from all 3 models are used to classify instrument family (left column, 8-class test accuracy), note pitch (center column, RMSE) and note velocity (right column, RMSE).  For each target task or column, the three curves are overlapping or identical. \textbf{The deep features perform well across target tasks, whether or not they were extracted from deep architectures originally trained on that task}.  \texttt{conv1} and \texttt{pool1} features are averaged over channels, all other deep features are flattened over channels.  The mean accuracies/RMSE values are shown with one standard deviation as error bars; plots following \citet{NEURIPS2020_71e9c662}.}
\label{fig:enhanced-supressed}
\end{figure}

As illustrated in Figure~\ref{fig:enhanced-supressed}, we find that the deep features are not enhanced or suppressed based on the target of interest. This is in contrast to the results in \citet{NEURIPS2020_71e9c662} for natural images. That is, whether the deep features are trained on classifying the instrument family (pink curves), predicting the note pitch (green curves) or predicting the note velocity (blue curves), the test accuracy and root mean squared error (RMSE) are nearly identical across target tasks.  Even though our original deep architectures are trained to classify the instrument family, these extracted deep features are able to predict both the note pitch and the note velocity \textbf{as well as} the architectures that were trained directly on those tasks; this is true of both the Regular and Deformable architectures.  The robustness of the deep features to target task is also seen for the Composer and Beethoven datasets (Section~\ref{suppsec:experiment_1}); that is, deep features trained on one audio task are able to achieve accuracy comparable to identical deep architectures trained directly on another task. The hand-crafted features are also useful for multiple tasks; the Top Combined wavelet features achieve RMSE values of 11.57 and 33.53 for the note pitch and note velocity tasks, respectively, comparable to the deep features at later layers in Figure~\ref{fig:enhanced-supressed}.  These results suggest that both the deep and hand-crafted features could be used for several down-stream tasks, without retraining the deep models.

\subsection{Are the Learned Deep Features Similar to the Hand-Crafted Features?}\label{sec:experiment_2}

\begin{figure*}
\centering
\begin{subfigure}[b]{.325\linewidth}
\includegraphics[width=\linewidth]{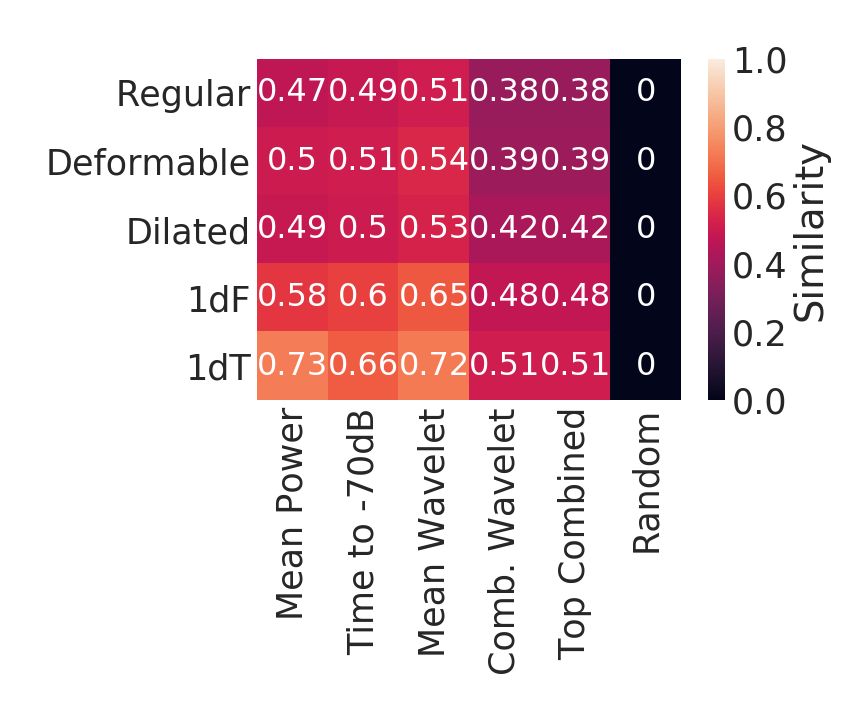}
\caption{NSynth}
\end{subfigure}
\begin{subfigure}[b]{.325\linewidth}
\includegraphics[width=\linewidth]{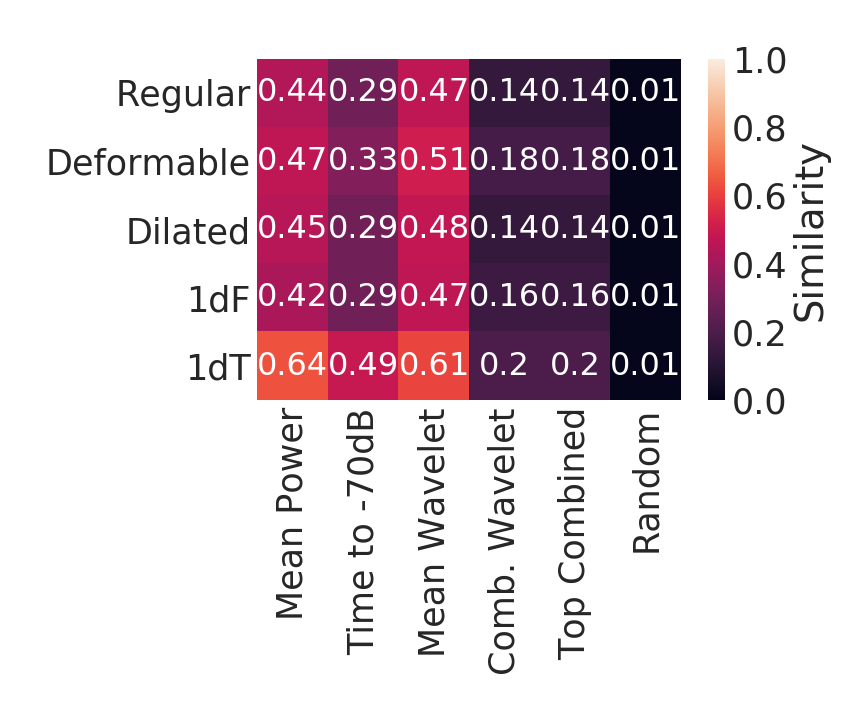}
\caption{Composer}
\end{subfigure}
\begin{subfigure}[b]{.325\linewidth}
\includegraphics[width=\linewidth]{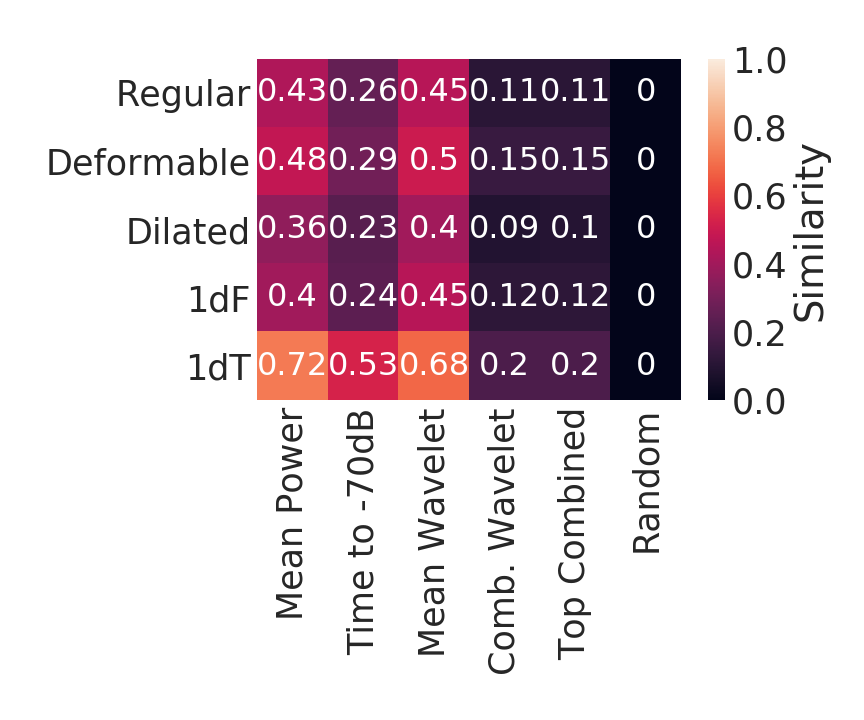}
\caption{Beethoven}
\end{subfigure}
\caption{Linear CKA similarity between the deep features from each architecture from the last convolutional layer (\texttt{conv3}) with the top hand-crafted features for the (a) NSynth, (b) Composer and (c) Beethoven datasets. We baseline with Random standard normal noise in the last column. The 1dT features are especially similar to the mean wavelet features by design. Plots are the similarity value averaged across the deep features from each initialization. }
\label{fig:sim-arch-main}
\end{figure*}

\begin{figure}
\centering
\begin{subfigure}[b]{.325\linewidth}
\includegraphics[width=\linewidth]{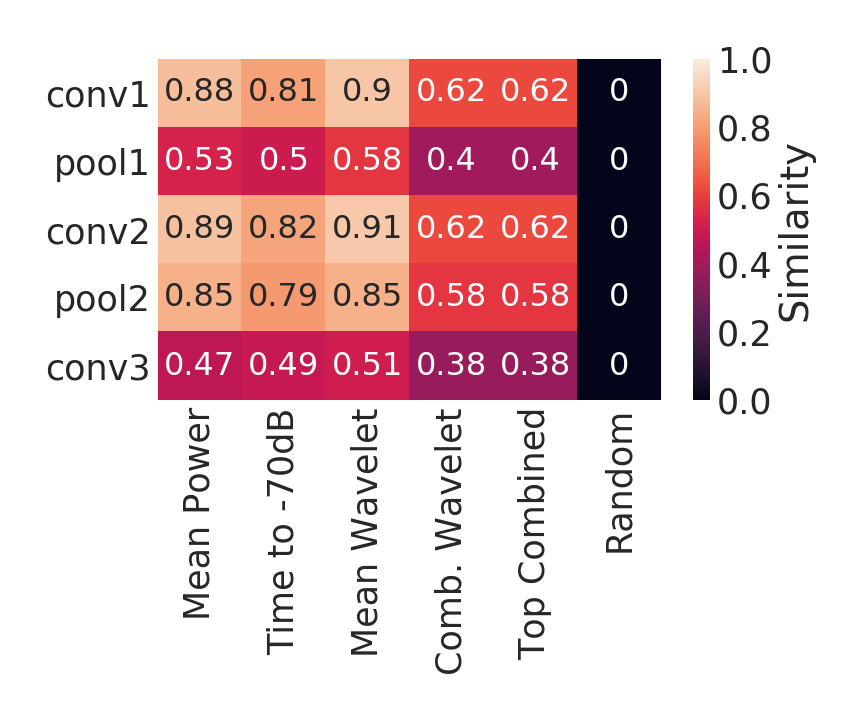}
\caption{Regular}
\end{subfigure}
\begin{subfigure}[b]{.325\linewidth}
\includegraphics[width=\linewidth]{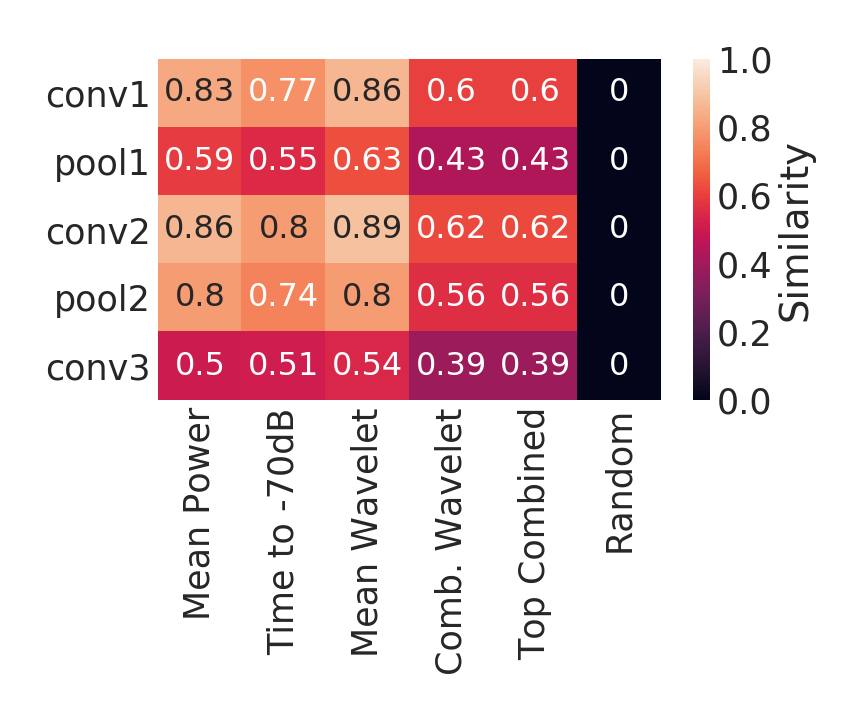}
\caption{Deformable}
\end{subfigure}
\caption{Linear CKA similarity between the deep features from each layer for the NSynth dataset for the (a) Regular and (b) Deformable architectures. The earlier layers exhibit very high similarity with the hand-crafted features. }
\label{fig:sim-NSynth-main}
\end{figure}

We next explore the similarity between the learned deep features and the hand-crafted features (which are useful for the main discriminative task, Table~\ref{tab:class-results}) using the linear CKA similarity measure \citep{pmlr-v97-kornblith19a}.  We first compare the similarity between the deep features from the last convolutional layer (\texttt{conv3}) across architectures to the top hand-crafted features in Figure~\ref{fig:sim-arch-main}.  Across all three datasets, there is relatively high similarity between all deep features and the mean wavelet feature, especially for the 1dT architecture.  This makes sense, since the 1dT architecture takes convolutions over time only and is thus very similar to the mean wavelet feature, which takes the mean wavelet coefficient for each frequency over time.  This suggests that hand-crafted wavelet features could be used in lieu of the musically motivated architectures, 1dF and 1dT \citep{7500246}. Furthermore, the Deformable deep features tend to be more similar to the hand-crafted features than the Regular deep features.  We confirm this similarity by concatenating the deep features to each hand-crafted feature for the Beethoven dataset.  The Regular deep features are most similar to the Mean Power and Mean Wavelet features, and when these features are concatenated together, the logistic regression accuracy on the main classification task does not improve beyond that of the deep Regular features alone ($82.32\pm0.40$\% for the Regular deep features, vs. $82.35\pm0.34$\% for the deep features concatenated with Mean Power).  However, the Regular features are less similar to the Wavelet Combined and Top 4 Combined features, and indeed, when combined with the Regular deep features, these hand-crafted features can further improve the classification accuracy ($83.44\pm0.42$\% and $84.23\pm0.31$\%, respectively).

\begin{figure}
\centering
\begin{subfigure}[b]{.325\linewidth}
\includegraphics[width=\linewidth]{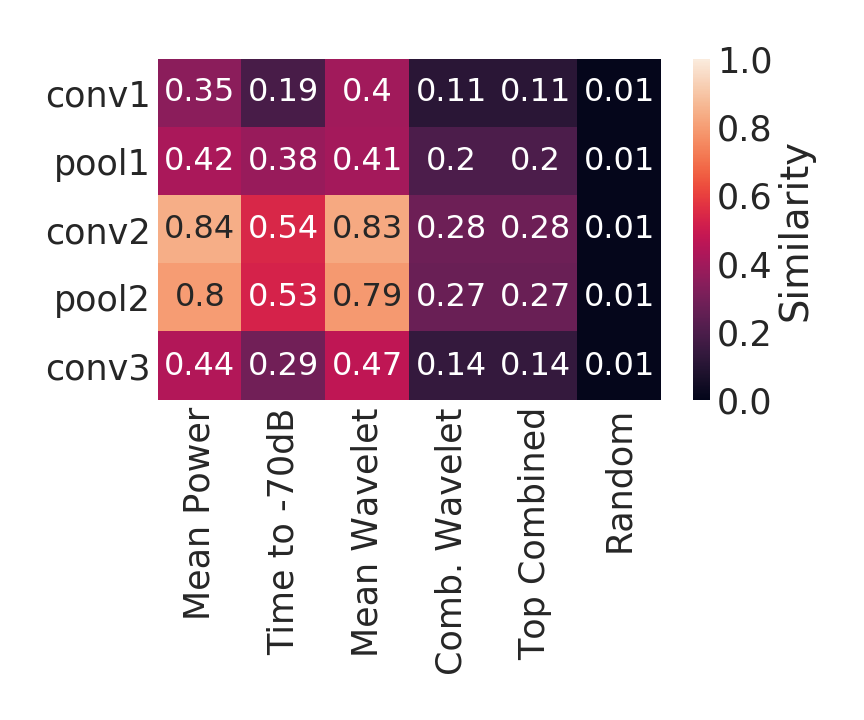}
\caption{Regular}
\end{subfigure}
\begin{subfigure}[b]{.325\linewidth}
\includegraphics[width=\linewidth]{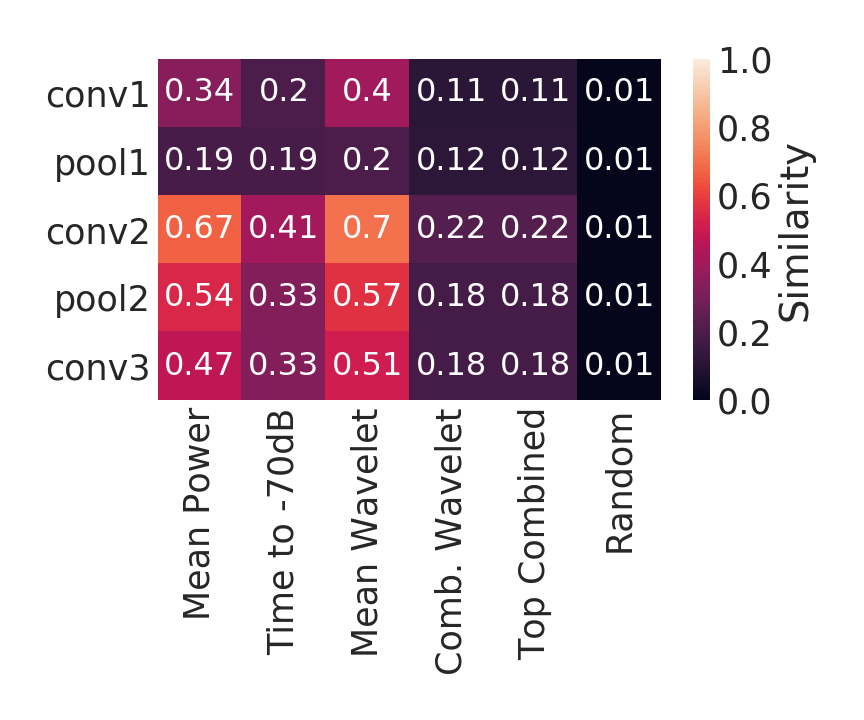}
\caption{Deformable}
\end{subfigure}
\caption{Linear CKA similarity between the deep features from each layer for the Composer dataset for the (a) Regular and (b) Deformable architectures. }
\label{fig:sim-Composer-main}
\end{figure}

We also compare the deep features at each layer to the top hand-crafted features for the Regular and Deformable convolutions on the NSynth (Figure~\ref{fig:sim-NSynth-main}) and Composer (Figure~\ref{fig:sim-Composer-main}) datasets. For the NSynth dataset, the deep features from the earlier layers exhibit a very high similarity with the hand-crafted features, while for the Composer dataset, the middle layers exhibit a very high similarity with the hand-crafted features.  
Overall, the last convolutional layer (\texttt{conv3}) for the Deformable architectures tends to be more similar to the hand-crafted features than the same Regular layer (\texttt{conv3} is the Deformable layer in this architecture, the earlier convolutions are regular).

\subsection{Do Untrained Convolutional Architectures Have a Useful Inductive Bias for Music Audio Discriminative Tasks?}\label{sec:experiment_3}

\begin{figure}
\centering
\begin{subfigure}[b]{0.49\linewidth}
\includegraphics[width=\linewidth]{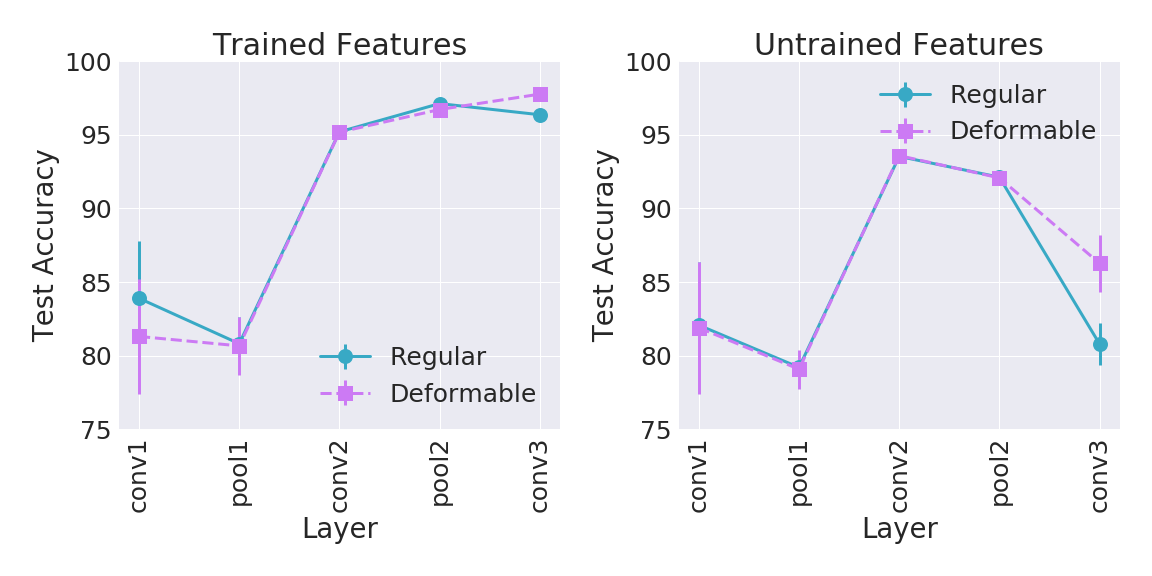}
\caption{NSynth}
\end{subfigure}
\begin{subfigure}[b]{0.49\linewidth}
\includegraphics[width=\linewidth]{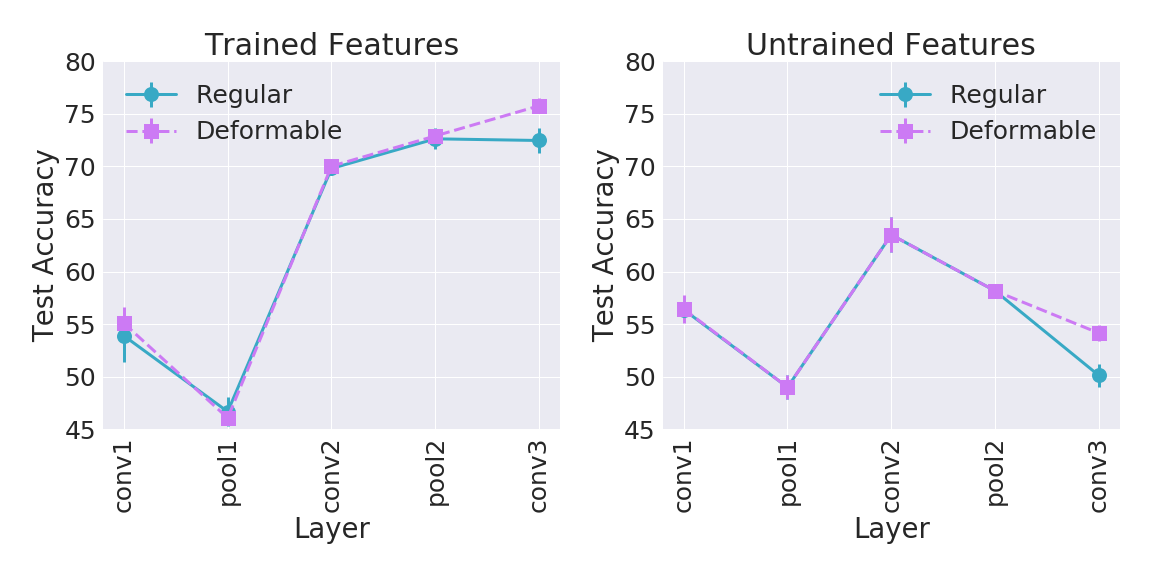}
\caption{Composer}\label{fig:composer-layers}
\end{subfigure}
\caption{Logistic regression classification accuracies by layer for the (a) NSynth and (b) Composer datasets using \textbf{untrained} deep features from the Regular and Deformable architectures (right side of each plot) compared to the trained deep features (left side of each plot). }
\label{fig:untrained-class}
\end{figure}

\begin{figure}[t]
\centering
\begin{subfigure}[b]{.325\linewidth}
\includegraphics[width=\linewidth]{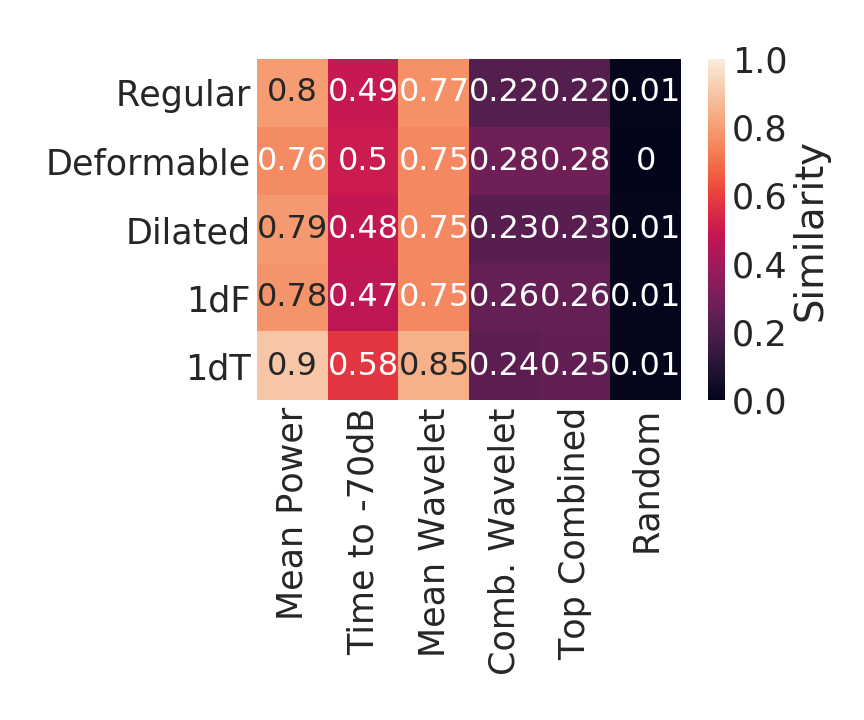}
\caption{Composer}
\end{subfigure}
\begin{subfigure}[b]{.325\linewidth}
\includegraphics[width=\linewidth]{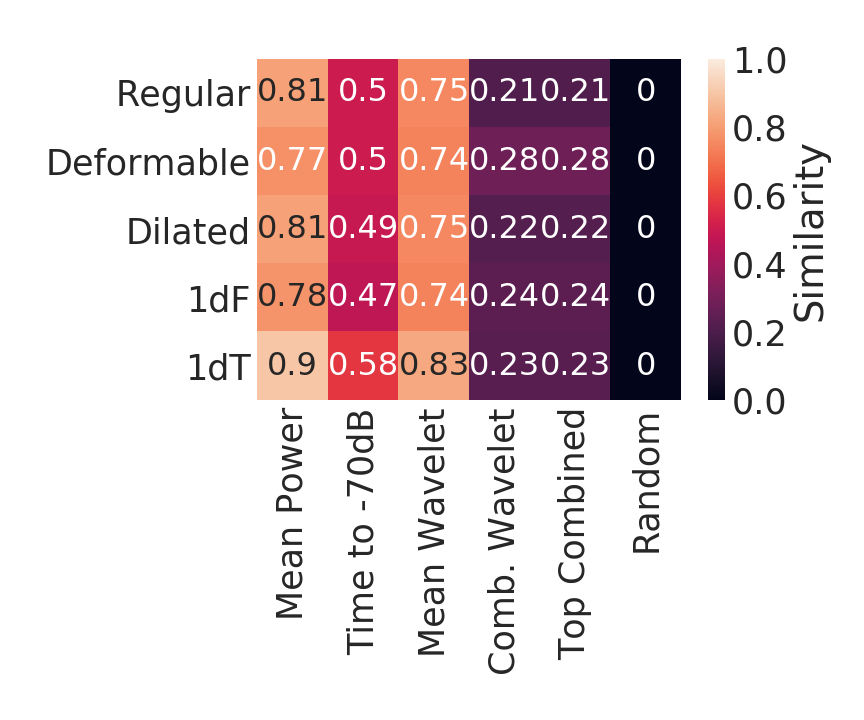}
\caption{Beethoven}
\end{subfigure}
\caption{Linear CKA similarity between the \textbf{untrained} deep features from each architecture from the last convolutional layer (\texttt{conv3}) with the top hand-crafted features for the (a) Composer and (b) Beethoven datasets.}
\label{fig:untrained-arch-sim-main}
\end{figure}

Finally, we explore how well features from the \textbf{untrained} deep architectures can perform the main classification tasks and how similar the untrained features are to the hand-crafted features.  These experiments explore the utility of the inductive bias of convolutions for music audio tasks, taking a different perspective than \citet{dap}.  First, in Figure~\ref{fig:untrained-class} we look at how well each layer for the Regular and Deformable untrained architectures performs for the main classification tasks. We find that all untrained layers are able to classify well above random guessing, and that only the later layers improve in accuracy after training.   This again confirms earlier results that the last convolutional layers are the most accurate and extract higher-level concepts than the earlier layers. In Figure~\ref{fig:untrained-arch-sim-main} we compare the untrained deep features from the last convolutional layer to the hand-crafted features for all architectures.  Across all datasets, the untrained deep features are \textbf{more} similar to the hand-crafted features than the trained deep features in  Figure~\ref{fig:sim-arch-main}.

\section{Conclusion}\label{sec:discussion}

We develop a methodology and set of systematic experiments to explore deep features in relation to hand-crafted features for data where feature importance methods do not apply.  Our experiments are motivated by understanding what deep features are useful for a given discriminative task, how robust these features are to different tasks, and how similar the trained and untrained deep features are to the hand-crafted features. The results of our experiments can be used to improve and inform further deep modeling for the application of interest. We apply our method to the specific setting of orchestral music audio data as an illustrative example of the application-specific insights that can be gained from our methods.  

Future work includes applying our methods to other application areas, for example time series data to compliment approaches like \citet{NEURIPS2020_47a3893c}, or data with privacy considerations.  From the specific application perspective, we could additionally explore other types of input data representation or related architectures \citep[e.g.][]{Anden:2014, Anden:2019}.

\bibliographystyle{plainnat} 
\bibliography{refs}

\clearpage
\appendix
\section{Summary of Additional Details and Experimental Results}

The results presented in this Appendix focus on specific details and results for the music audio application of our proposed methodology.  We include additional details about the data, architectures and results included in the main paper, as well as additional supporting experiments.  Specific details about the three datasets, the pre-processing procedure to calculate the mel-spectrograms and hand-crafted features, as well as details about all deep learning architectures, computing resources and code are discussed in Section~\ref{suppsec:methodology}.  Overall, we demonstrate that deep convolutional features perform well across various target tasks, whether or not they are extracted from deep architectures originally trained on that task. Additionally, deep features exhibit high similarity to hand-crafted wavelet features, whether the deep features are extracted from a trained or untrained model.  In Section~\ref{suppsec:experiment_1}, we include feature classification results for numerous additional hand-crafted features, supporting our selection of the top 4 features discussed in the main paper. The wavelet features, in particular, are found to be robust to the choice of the bandwidth parameter.  We also demonstrate that the deep features trained on one discriminative task are able to achieve high classification accuracy on other tasks that they were not trained on for the Composer and Beethoven datasets.  

In Section~\ref{suppsec:experiment_2}, we compare the similarity between the deep features for the last convolutional layer by architecture and the deep features from all layers for the Regular and Deformable architectures to all of the hand-crafted features.  We also explore additional similarity measures for comparison to the Linear CKA similarity of main focus.  Untrained deep features are compared by layer and by architecture to all of the hand-crafted features in Section~\ref{suppsec:experiment_3}.  We again find that the untrained features from the last convolutional layer across architectures and for all layers for the Regular and Deformable architectures are highly similar to the hand-crafted features, especially the wavelet features.  Finally, we include additional experiments in Section~\ref{suppsec:experiment_4} that explore the similarity of the deep features across architectures, layers and initializations.  We find that learned deep features are highly similar across initializations, but not identical, reinforcing the need for multiple initializations.   All code for the NSynth experiments is included on GitHub at \url{https://github.com/aky4wn/convolutions-for-music-audio}.

\section{Methodology}\label{suppsec:methodology}

\subsection{Data and Tasks}

\begin{figure}[htbp]
\begin{center}
\includegraphics[width=0.45\textwidth]{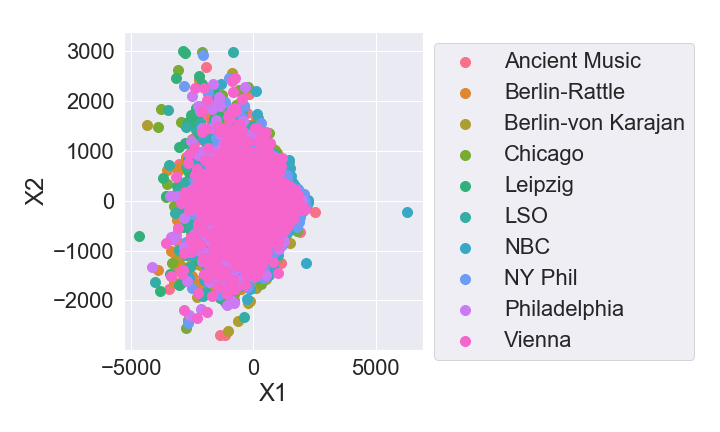}
\caption{PCA decomposition for the Beethoven training dataset, colored by the class.  The dataset is clearly not linearly separable when represented by the first two principal components of the flattened mel-spectrograms, so the orchestra classification task is non-trivial.}
\label{fig:Beethoven-EDA-PCA}
\end{center}
\end{figure}

All of the mel-spectrograms used as input data in this work are calculated from 4 second audio clips.  The sampling rate of the NSynth audio is 16 kHz and the sampling of the Composer and Beethoven datasets is 22050 Hz.  For all three datasets, the mel-spectrograms are calculated with 128 mels, a maximum frequency of 8000 Hz, a hop length of 502 and a Hanning window with FFT window length 2048 using the \texttt{librosa} package in Python \citep{librosa}.  This pre-processing results in mel-spectrograms of dimension 128 x 128 for the NSynth dataset, and of dimension 128 x 176 for the Composer and Beethoven datasets.  The mel-spectrogram is first calculated on the power scale, then converted to dB, with reference of the maximum power; the phase information is not used.  All datasets use a 70-30\% train-test split. PCA decompositions colored by class for the Beethoven dataset are given in Figure~\ref{fig:Beethoven-EDA-PCA} and are similar for the other datasets; the classes overlap indicating that the classification tasks for each dataset are non-trivial.

\subsubsection{NSynth}

\begin{figure}[h]
\begin{center}
\includegraphics[width=0.45\textwidth]{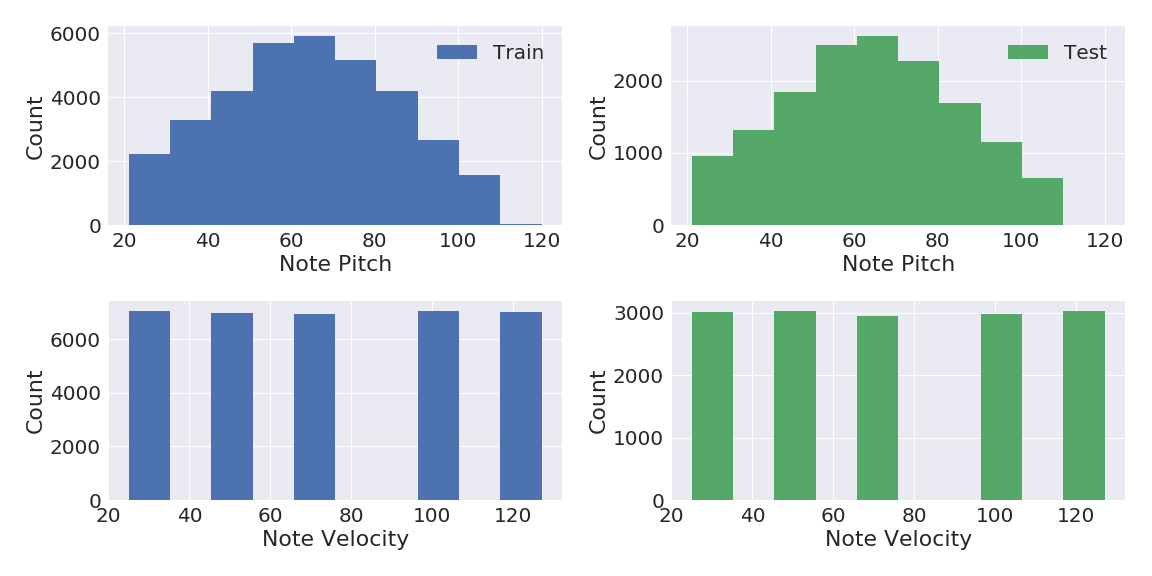}
\caption{Histograms of train (left) and test (right) distributions for the note pitch (top row) and note velocity = volume (bottom row) related tasks for the NSynth dataset.}
\label{fig:NSynth-hist}
\end{center}
\end{figure}

\begin{figure}[htbp]
\begin{center}
\includegraphics[width=0.6\textwidth]{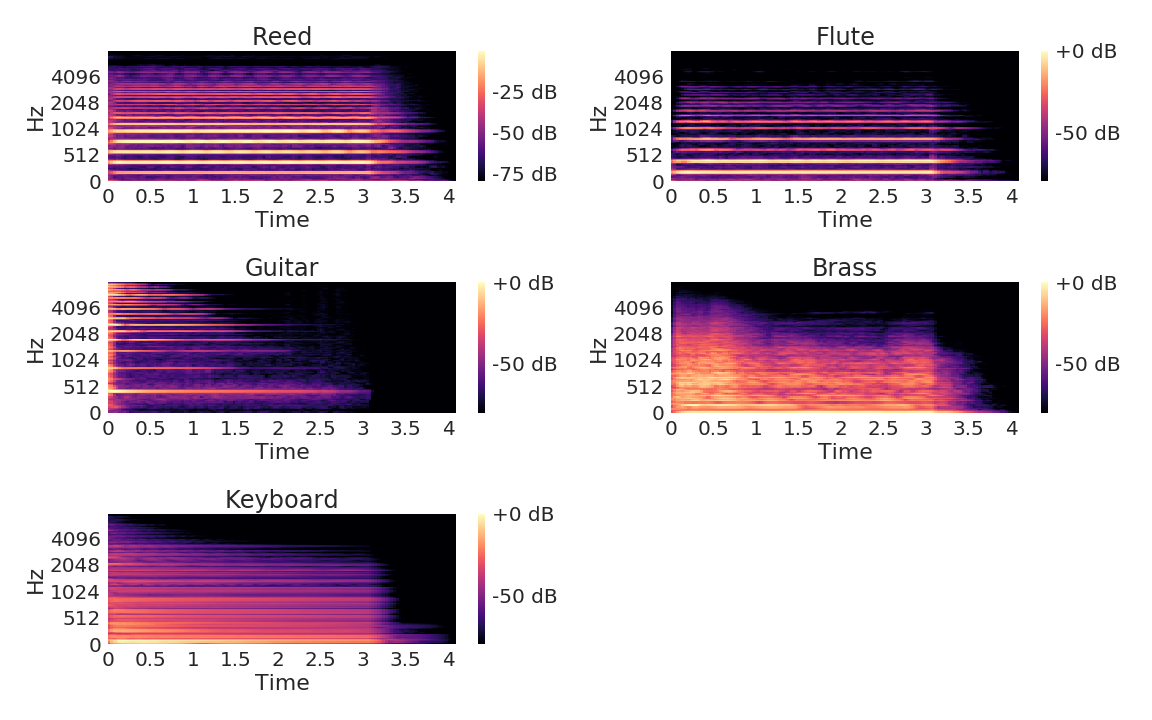}
\caption{Example mel-spectrograms for several instrument families for the NSynth dataset.  There are obvious visual differences in the mel-spectrograms based on the instrument family, especially between percussive (keyboard), string (guitar) and wind (reed, flute and brass) instruments.}
\label{fig:NSynth-mel}
\end{center}
\end{figure}

The NSynth dataset has 50000 total data examples.  The main task is an 8-class classification of instrument family; the class balance for the train and test set are given in Table~\ref{tab:NSynth-class-balance}.  The related tasks are Note Pitch and Note Velocity (volume) regression, with the distribution of these tasks given in Figure~\ref{fig:NSynth-hist}. The Note Pitch is an integer value that can be (in theory) between 0 and 127, inclusive, while the Note Velocity is also an integer velocity between 0 and 127 inclusive, though only 5 distinct values are observed among the acoustic instruments (Figure~\ref{fig:NSynth-hist}).  The data used is taken from the training portion of the NSynth dataset \citep{engel_neural_2017}, where 50000 randomly sampled acoustic examples are selected and the bass and organ instrument families are dropped since they do not have very many examples.  The acoustic examples are most similar to the real orchestral recordings and were selected for that reason.  The training data was subset to approximately match the amount of data for the Composer and Beethoven datasets. Example mel-spectrograms are shown in Figure~\ref{fig:NSynth-mel}.  The NSynth dataset is under a Creative Commons Attribution 4.0 International (CC BY 4.0) license (\url{https://magenta.tensorflow.org/datasets/nsynth#files}). 

\begin{table*}[t]
\begin{center}
\scalebox{0.8}{
\begin{tabular}{ccccccccc}\hline
& Brass & Flute & Guitar & Keyboard & Mallet & Reed & String &Vocal \\\hline
Train & 12.3\% & 6.3\% & 11.3\% & 8.0\% & 26.5\% & 13.0\% & 18.6\% & 4.0\% \\
Test & 12.7\% & 6.0\% & 11.4\% & 7.9\% & 26.6\% & 12.7\% & 19.3\% & 3.5\% \\\hline
\end{tabular}}
\end{center}
\caption{Train and test sets class balance for the NSynth dataset.}
\label{tab:NSynth-class-balance}
\end{table*}%

\begin{figure}
\centering
\begin{subfigure}[b]{0.49\linewidth}
\includegraphics[width=\linewidth]{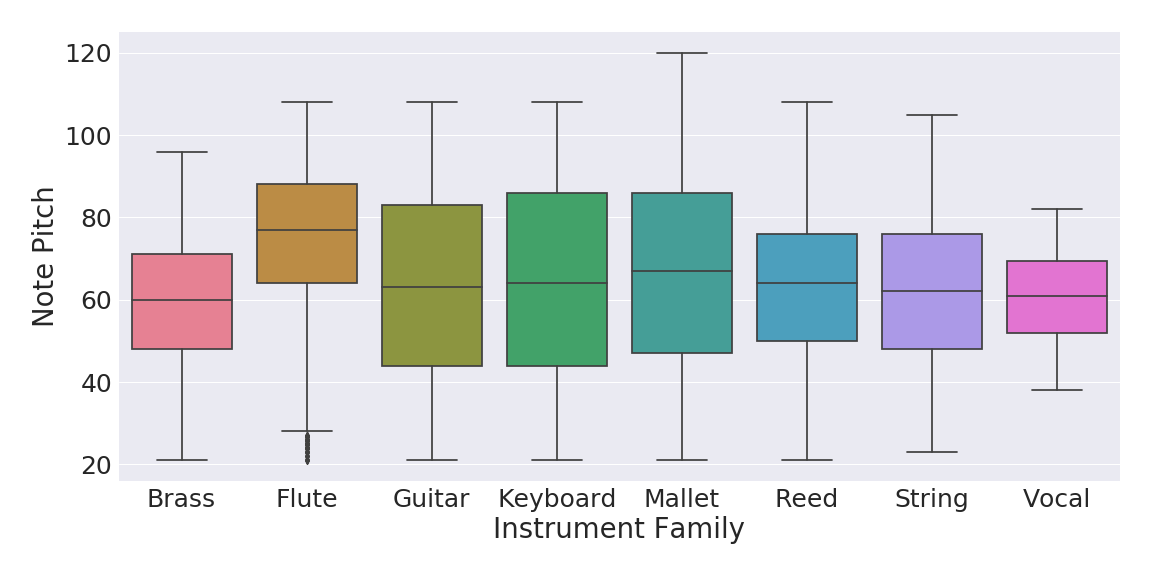}
\caption{Note Pitch.} \label{fig:NSynth-pitch-overlap}
\end{subfigure}
\begin{subfigure}[b]{0.49\linewidth}
\includegraphics[width=\linewidth]{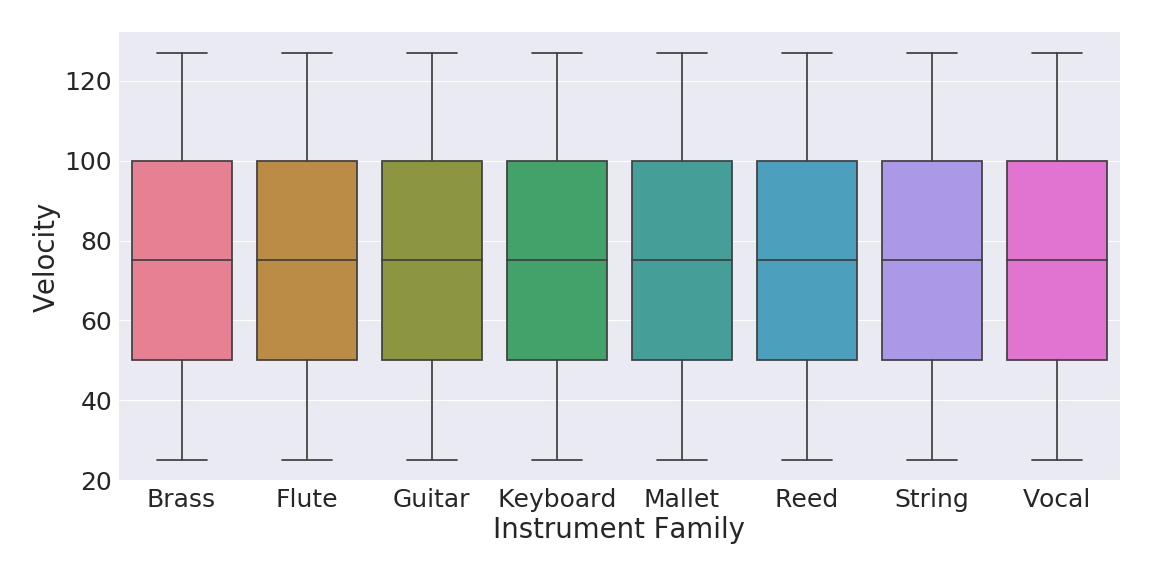}
\caption{Note Velocity.}\label{fig:NSynth-vel-overlap}
\end{subfigure}
\caption{Overlap between (a) note pitches and (b) note velocity by instrument family.  There are only a few note pitches that are performed by only one instrument family. All note velocities are observed by all instrument families.}
\label{fig:NSynth-overlap}
\end{figure}

There is significant overlap between the labels for the main classification task and the related tasks.  That is, almost all note pitches are observed by at least one other instrument family (Figure~\ref{fig:NSynth-pitch-overlap}) and all note velocities are observed with all instrument families (Figure~\ref{fig:NSynth-vel-overlap}).  This is important for the robustness of the deep features for Experiment \# 1 in Section 4.1 in the main paper; it is not the case that there is a lack of overlap between instrument family and note pitch or instrument family and note velocity that could be exploited to achieve these results.  Indeed, performing simple multi-class logistic regression and using note-pitch to predict instrument family leads to a test accuracy of 26.6\% which is equivalent to random guessing; thus, the deep features are truly robust to multiple tasks and are not using some artifact in the NSynth dataset.

\subsubsection{Composer}

\begin{figure}[htbp]
\begin{center}
\includegraphics[width=0.6\textwidth]{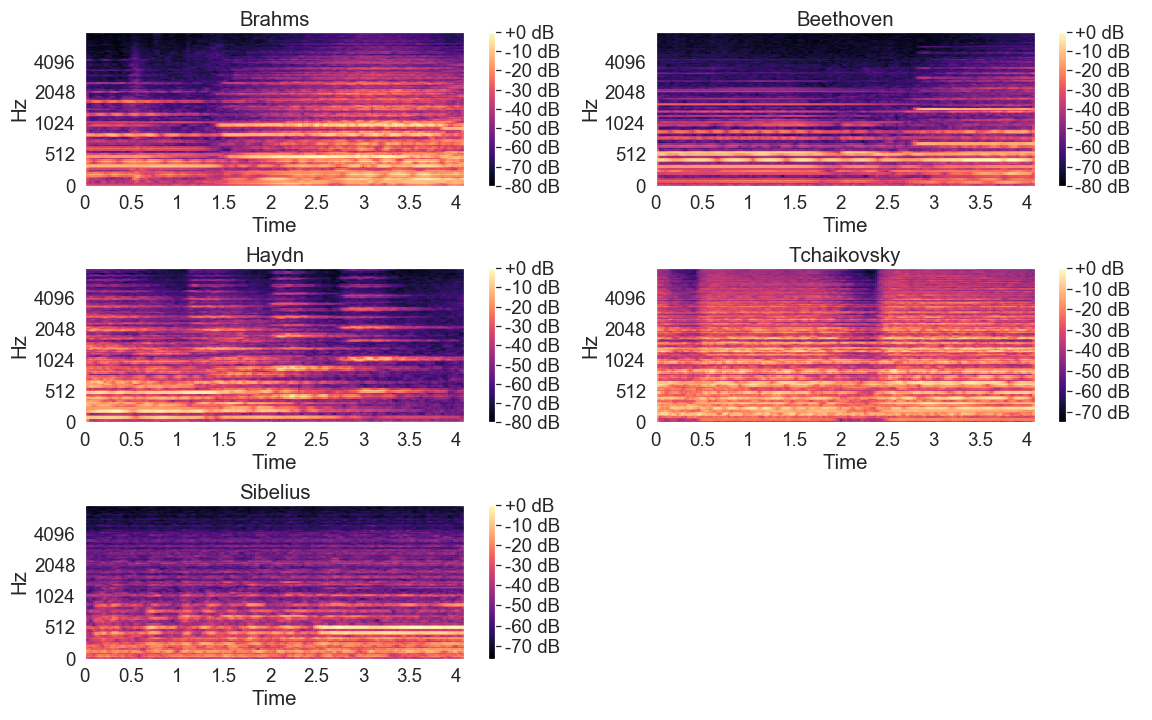}
\caption{Example mel-spectrograms for each composer for the Composer dataset.  There are not obvious visual differences in the mel-spectrograms based on the composer, indicating that visual explanation techniques are not likely to be meaningful for this type of data.}
\label{fig:Composer-mel}
\end{center}
\end{figure}

The Composer dataset is made up of 19667 total examples.  All of these orchestral pieces are from classical composers, recorded by the Berlin Philharmonic under Herbert von Karajan to eliminate performance style differences by different orchestras, which can vary considerably \citep{HMDS} and will be considered with the Beethoven dataset. Before calculating the mel-spectrograms, each audio recording of a piece is divided into 4 second non-overlapping chunks, with any chunks less than 4 seconds at the end of pieces dropped.  The class balance for the main task of 5-class composer classification is given in Table~\ref{tab:Composer-class-balance}.  The related task for the Composer dataset is classifying whether each example came from the beginning, middle or end of its respective piece, where the classes are determined by evenly splitting the total length of the audio recording of each piece in thirds and labeling each spectrogram accordingly.  The class balance for this task is given in Table~\ref{tab:Composer-part-balance} and again evaluates musical style differences across the length of each piece.  Again, the primary and related tasks have significant overlap; all pieces by all composers have a beginning, middle and end, so the robustness of the deep features in Table~\ref{tab:deep-transfer} are not due to an artifact in the class labels or data. That is, part of piece could not be used to predict composer with classification accuracy above random guessing.  Example composer spectrograms are given in Figure~\ref{fig:Composer-mel}.  A list of pieces used in the Composer dataset is below, and primarily consists of symphonic works by each composer.

\begin{table*}[t]
\begin{center}\scalebox{0.8}{
\begin{tabular}{cccccc}\hline
& Beethoven & Brahms & Haydn & Sibelius & Tchaikovsky \\\hline
Train & 22.8\% & 12.0\% & 25.2\% & 18.0\% & 22.1\% \\
Test & 23.1\% &  11.4\% & 25.5\% & 18.1\% & 21.9\% \\\hline
\end{tabular}}
\end{center}
\caption{Train and test sets class balance for the Composer dataset.}
\label{tab:Composer-class-balance}
\end{table*}%

\begin{table}[htp]
\begin{center}\scalebox{0.8}{
\begin{tabular}{cccc}\hline
& Beginning & Middle & End \\\hline
Train & 33.3\% & 33.3\% & 33.4\% \\
Test & 33.6\% & 33.6\% & 32.8\% \\\hline
\end{tabular}}
\end{center}
\caption{Train and test sets class balance for the related task of part of piece classification for the Composer dataset.}
\label{tab:Composer-part-balance}
\end{table}%

The Beethoven pieces in the Composer dataset are: Fidelio Overture Op. 72, Coriolan Overture Op. 62, Leonore No. 3 Overture Op. 72b, Egmont Overture, Symphony No 1 Op. 21 (4 movements), Symphony No 2 Op. 36 (4 movements), Symphony No 3 Op. 55 (4 movements), Symphony No 4 Op. 60 (4 movements), Symphony No 5 Op. 67 (4 movements), Symphony No 6 Op. 68 (5 movements), Symphony No 7 Op 92 (4 movements) and Symphony No 8 Op 93 (4 movements).

The Brahms pieces in the Composer dataset are:
Symphony No 1 Op. 68 (4 movements),
Symphony No 2 Op. 73 (4 movements),
Symphony No 3 Op. 90 (4 movements) and
Symphony No 4 Op. 98 (4 movements).

The Haydn pieces in the Composer dataset are:
Symphony No 93 H. 1/94 (4 movements),
Symphony No 94 H. 1/94 (4 movements),
Symphony No 95 H. 1/95 (4 movements),
Symphony No 96 H. 1/96 (4 movements),
Symphony No 97 H. 1/97 (4 movements),
Symphony No 98 H. 1/98 (4 movements),
Symphony No 99 H. 1/99 (4 movements),
Symphony No 100 H. 1/100 (4 movements),
Symphony No 101 H. 1/101 (4 movements),
Symphony No 102 H. 1/102 (4 movements),
Symphony No 103 H. 1/103 (4 movements) and
Symphony No 104 H. 1/104 (4 movements).

The Sibelius pieces in the Composer dataset are:
Karelia Suite Op. 11 (3 movements),
En Saga Op. 9,
Finlandia Op. 26,
Swan of Tuonela Op. 22-3,
Tapiola Op. 112,
Valse Triste Op. 44-1,
Symphony No 1 Op. 39 (4 movements),
Symphony No 2 Op. 43 (4 movements),
Symphony No 4 Op. 63 (4 movements) and
Symphony No 5 Op. 82 (3 movements).

The Tchaikovsky pieces in the Composer dataset are:
Capriccio Italien Op. 45,
Symphony No 1 Op. 13 (4 movements),
Symphony No 2 Op. 17 (4 movements),
Symphony No 3 Op. 29 (4 movements),
Symphony No 4 Op. 36 (4 movements),
Symphony No 5 Op. 64 (4 movements) and
Symphony No 6 Op. 74 (4 movements).

\subsubsection{Beethoven}

The Beethoven dataset is made up of 52999 total examples and uses the same audio recordings as in \citet{HMDS}.  The main classification task is the 10-class classification of orchestras performing each of the 9 Beethoven symphonies to focus on performance style of different orchestras.  The 10 orchestras considered are the Academy of Ancient Music under Hogwood, the Berlin Philharmonic under Rattle, the Berlin Philharmonic under von Karajan, the Chicago Symphony Orchestra under Solti, the Leipzig Gewandhaus Orchestra under Masur, the London Symphony Orchestra (LSO) under Haitink, the NBC Symphony Orchestra under Toscanini, the New York Philharmonic under Bernstein, the Philadelphia Orchestra under Muti and the Vienna Philharmonic under Rattle.  The class balance is given in Table~\ref{tab:Beethoven-class-balance}.  The recordings of each piece are of slightly different lengths (due to differences in tempo, for example), which is why the class balance is not exactly 10\%. The two related tasks are classifying the symphony number (9-classes, musical style task) and whether the orchestra is American or European (2-class, 41.4\% - 58.6\% test balance, performance style task), as European orchestras can have common performance styles that differ from American orchestras.  The class balance for the piece classification is given in Table~\ref{tab:Beethoven-number-balance}. Again, the primary and related tasks have significant overlap; all orchestras perform all 9 symphonies, so the robustness of the deep features in Table~\ref{tab:deep-transfer} are not due to an artifact in the class labels or data. That is, symphony number could not be used to predict orchestra with classification accuracy above random guessing.  Each orchestra can only belong to one continent, so there is less overlap here, but orchestra continent is a much broader categorization than specific orchestra, and again, orchestra continent could not be used to predict the specific orchestra with classification accuracy above random guessing.

\begin{table*}[t]
\begin{center}
\scalebox{0.8}{
\begin{tabular}{cp{1.2cm}p{1.2cm}p{2cm}ccccccc}\hline
& Ancient Music & Berlin-Rattle & Berlin-von Karajan & Chicago & Leipzig & LSO & NBC &NY Phil & Philadelphia & Vienna \\\hline
Train & 9.8\% & 9.7\% & 10.1\% & 10.4\% & 10.1\% & 9.7\% & 9.6\% & 10.3\% & 10.2\% & 10.1\% \\
Test & 9.5\% & 9.7\% & 10.0\% & 11.4\% & 9.8\% & 9.8\% & 9.6\% & 10.2\% & 10.2\% & 9.8\% \\\hline
\end{tabular}}
\end{center}
\caption{Train and test sets class balance for the Beethoven dataset.}
\label{tab:Beethoven-class-balance}
\end{table*}%

\begin{table*}[t]
\begin{center}\scalebox{0.8}{
\begin{tabular}{cccccccccc}\hline
& No 1 & No 2 & No 3 & No 4 & No 5 & No 6 & No 7 & No 8 & No 9\\\hline
Train & 7.3\% & 9.4\% & 14.3\% & 9.5\% & 9.1\% & 12.1\% & 11.5\% & 7.2\% & 19.5\% \\
Test & 7.3\% & 9.4\% & 14.2\% & 9.6\% & 9.7\% & 11.9\% & 11.2\% & 7.1\% & 19.6\% \\\hline
\end{tabular}}
\end{center}
\caption{Train and test sets class balance for the related task of symphony number classification for the Beethoven dataset.}
\label{tab:Beethoven-number-balance}
\end{table*}%

\subsection{Convolutional Architectures}

The exact convolutional architectures used for all three datasets are given below, displayed as the PyTorch model specification (\url{https://pytorch.org/}).  The architectures are slightly different between the NSynth and the Composer and Beethoven datasets due to the difference in input mel-spectrogram dimensions.  The Deformable convolution implementation used is from \url{https://github.com/4uiiurz1/pytorch-deform-conv-v2}.  The 1d Frequency and 1d Time architectures are motivated by \citet{7500246}. Again, these architectures are intended to be deep enough to learn meaningful features and perform the main task well, but not too large to overfit the moderate size datasets used here and in music audio modeling in general.

\subsubsection{NSynth Architectures}

\paragraph{Regular Convolution}
\begin{lstlisting}[
    basicstyle=\tiny, %or \small or \footnotesize etc.
]
CNN(
  (conv1): Conv2d(1, 10, kernel_size=(5, 5), stride=(1, 1))
  (conv1_bn1): BatchNorm2d(10, eps=1e-05, momentum=0.1, affine=True, track_running_stats=True)
  (pool1): MaxPool2d(kernel_size=2, stride=2, padding=0, dilation=1, ceil_mode=False)
  (conv2): Conv2d(10, 20, kernel_size=(5, 5), stride=(3, 3), padding=(3, 3))
  (conv2_bn2): BatchNorm2d(20, eps=1e-05, momentum=0.1, affine=True, track_running_stats=True)
  (pool2): MaxPool2d(kernel_size=2, stride=2, padding=0, dilation=1, ceil_mode=False)
  (conv3): Conv2d(20, 30, kernel_size=(5, 5), stride=(2, 2))
  (conv3_bn3): BatchNorm2d(30, eps=1e-05, momentum=0.1, affine=True, track_running_stats=True)
  (fc1): Linear(in_features=480, out_features=120, bias=True)
  (fc2): Linear(in_features=120, out_features=8, bias=True)
)
\end{lstlisting}

\paragraph{Deformable Convolution}
\begin{lstlisting}[
    basicstyle=\tiny, %or \small or \footnotesize etc.
]
CNN_deform(
  (conv1): Conv2d(1, 10, kernel_size=(5, 5), stride=(1, 1))
  (conv1_bn1): BatchNorm2d(10, eps=1e-05, momentum=0.1, affine=True, track_running_stats=True)
  (pool1): MaxPool2d(kernel_size=2, stride=2, padding=0, dilation=1, ceil_mode=False)
  (conv2): Conv2d(10, 20, kernel_size=(5, 5), stride=(3, 3), padding=(3, 3))
  (conv2_bn2): BatchNorm2d(20, eps=1e-05, momentum=0.1, affine=True, track_running_stats=True)
  (pool2): MaxPool2d(kernel_size=2, stride=2, padding=0, dilation=1, ceil_mode=False)
  (conv3): DeformConv2d(
    (zero_padding): ZeroPad2d(padding=(0, 0, 0, 0), value=0.0)
    (conv): Conv2d(20, 30, kernel_size=(5, 5), stride=(5, 5), bias=False)
    (p_conv): Conv2d(20, 50, kernel_size=(3, 3), stride=(2, 2), padding=(1, 1))
  )
  (conv3_bn3): BatchNorm2d(30, eps=1e-05, momentum=0.1, affine=True, track_running_stats=True)
  (fc1): Linear(in_features=1080, out_features=120, bias=True)
  (fc2): Linear(in_features=120, out_features=8, bias=True)
)
\end{lstlisting}

\paragraph{Dilated Convolution}
\begin{lstlisting}[
    basicstyle=\tiny, %or \small or \footnotesize etc.
]
CNN_dilated(
  (conv1): Conv2d(1, 10, kernel_size=(5, 5), stride=(1, 1), dilation=(3, 3))
  (conv1_bn1): BatchNorm2d(10, eps=1e-05, momentum=0.1, affine=True, track_running_stats=True)
  (pool1): MaxPool2d(kernel_size=2, stride=2, padding=0, dilation=1, ceil_mode=False)
  (conv2): Conv2d(10, 20, kernel_size=(5, 5), stride=(2, 2), padding=(3, 3), dilation=(2, 2))
  (conv2_bn2): BatchNorm2d(20, eps=1e-05, momentum=0.1, affine=True, track_running_stats=True)
  (pool2): MaxPool2d(kernel_size=2, stride=2, padding=0, dilation=1, ceil_mode=False)
  (conv3): Conv2d(20, 30, kernel_size=(5, 5), stride=(2, 2), padding=(1, 1))
  (conv3_bn3): BatchNorm2d(30, eps=1e-05, momentum=0.1, affine=True, track_running_stats=True)
  (fc1): Linear(in_features=1080, out_features=120, bias=True)
  (fc2): Linear(in_features=120, out_features=8, bias=True)
)
\end{lstlisting}

\paragraph{1d Frequency}
\begin{lstlisting}[
    basicstyle=\tiny, %or \small or \footnotesize etc.
]
CNN_1dF(
  (conv1): Conv2d(1, 10, kernel_size=(5, 1), stride=(1, 1), dilation=(3, 1))
  (conv1_bn1): BatchNorm2d(10, eps=1e-05, momentum=0.1, affine=True, track_running_stats=True)
  (pool1): MaxPool2d(kernel_size=(2, 1), stride=(2, 1), padding=0, dilation=1, ceil_mode=False)
  (conv2): Conv2d(10, 20, kernel_size=(5, 1), stride=(2, 1), padding=(3, 0), dilation=(2, 1))
  (conv2_bn2): BatchNorm2d(20, eps=1e-05, momentum=0.1, affine=True, track_running_stats=True)
  (pool2): MaxPool2d(kernel_size=(2, 1), stride=(2, 1), padding=0, dilation=1, ceil_mode=False)
  (conv3): Conv2d(20, 30, kernel_size=(5, 1), stride=(2, 1), padding=(1, 0))
  (conv3_bn3): BatchNorm2d(30, eps=1e-05, momentum=0.1, affine=True, track_running_stats=True)
  (fc1): Linear(in_features=23040, out_features=120, bias=True)
  (fc2): Linear(in_features=120, out_features=8, bias=True)
)
\end{lstlisting}

\paragraph{1d Time}
\begin{lstlisting}[
    basicstyle=\tiny, %or \small or \footnotesize etc.
]
CNN_1dT(
  (conv1): Conv2d(1, 10, kernel_size=(1, 5), stride=(1, 1), dilation=(1, 3))
  (conv1_bn1): BatchNorm2d(10, eps=1e-05, momentum=0.1, affine=True, track_running_stats=True)
  (pool1): MaxPool2d(kernel_size=(1, 2), stride=(1, 2), padding=0, dilation=1, ceil_mode=False)
  (conv2): Conv2d(10, 20, kernel_size=(1, 5), stride=(1, 2), padding=(0, 3), dilation=(1, 2))
  (conv2_bn2): BatchNorm2d(20, eps=1e-05, momentum=0.1, affine=True, track_running_stats=True)
  (pool2): MaxPool2d(kernel_size=(1, 2), stride=(1, 2), padding=0, dilation=1, ceil_mode=False)
  (conv3): Conv2d(20, 30, kernel_size=(1, 5), stride=(1, 2), padding=(0, 1))
  (conv3_bn3): BatchNorm2d(30, eps=1e-05, momentum=0.1, affine=True, track_running_stats=True)
  (fc1): Linear(in_features=23040, out_features=120, bias=True)
  (fc2): Linear(in_features=120, out_features=8, bias=True)
)
\end{lstlisting}

\subsubsection{Composer and Beethoven Architectures}
Note: for the Beethoven dataset experiments, \texttt{out\_features = 10}  for all architectures.
\paragraph{Regular Convolution}
\begin{lstlisting}[
    basicstyle=\tiny, %or \small or \footnotesize etc.
]
CNN(
  (conv1): Conv2d(1, 10, kernel_size=(5, 5), stride=(1, 1))
  (conv1_bn1): BatchNorm2d(10, eps=1e-05, momentum=0.1, affine=True, track_running_stats=True)
  (pool1): MaxPool2d(kernel_size=2, stride=2, padding=0, dilation=1, ceil_mode=False)
  (conv2): Conv2d(10, 20, kernel_size=(5, 5), stride=(3, 3), padding=(3, 3))
  (conv2_bn2): BatchNorm2d(20, eps=1e-05, momentum=0.1, affine=True, track_running_stats=True)
  (pool2): MaxPool2d(kernel_size=2, stride=2, padding=0, dilation=1, ceil_mode=False)
  (conv3): Conv2d(20, 30, kernel_size=(5, 5), stride=(2, 2))
  (conv3_bn3): BatchNorm2d(30, eps=1e-05, momentum=0.1, affine=True, track_running_stats=True)
  (fc1): Linear(in_features=720, out_features=120, bias=True)
  (fc2): Linear(in_features=120, out_features=5, bias=True)
)
\end{lstlisting}

\paragraph{Deformable Convolution}
\begin{lstlisting}[
    basicstyle=\tiny, %or \small or \footnotesize etc.
]
CNN_deform(
  (conv1): Conv2d(1, 10, kernel_size=(5, 5), stride=(1, 1))
  (conv1_bn1): BatchNorm2d(10, eps=1e-05, momentum=0.1, affine=True, track_running_stats=True)
  (pool1): MaxPool2d(kernel_size=2, stride=2, padding=0, dilation=1, ceil_mode=False)
  (conv2): Conv2d(10, 20, kernel_size=(5, 5), stride=(3, 3), padding=(3, 3))
  (conv2_bn2): BatchNorm2d(20, eps=1e-05, momentum=0.1, affine=True, track_running_stats=True)
  (pool2): MaxPool2d(kernel_size=2, stride=2, padding=0, dilation=1, ceil_mode=False)
  (conv3): DeformConv2d(
    (zero_padding): ZeroPad2d(padding=(0, 0, 0, 0), value=0.0)
    (conv): Conv2d(20, 30, kernel_size=(5, 5), stride=(5, 5), bias=False)
    (p_conv): Conv2d(20, 50, kernel_size=(3, 3), stride=(2, 2), padding=(1, 1))
  )
  (conv3_bn3): BatchNorm2d(30, eps=1e-05, momentum=0.1, affine=True, track_running_stats=True)
  (fc1): Linear(in_features=1440, out_features=120, bias=True)
  (fc2): Linear(in_features=120, out_features=5, bias=True)
)
\end{lstlisting}

\paragraph{Dilated Convolution}
\begin{lstlisting}[
    basicstyle=\tiny, %or \small or \footnotesize etc.
]
CNN_dilated(
  (conv1): Conv2d(1, 10, kernel_size=(5, 5), stride=(1, 1), dilation=(3, 3))
  (conv1_bn1): BatchNorm2d(10, eps=1e-05, momentum=0.1, affine=True, track_running_stats=True)
  (pool1): MaxPool2d(kernel_size=2, stride=2, padding=0, dilation=1, ceil_mode=False)
  (conv2): Conv2d(10, 20, kernel_size=(5, 5), stride=(2, 2), padding=(3, 3), dilation=(2, 2))
  (conv2_bn2): BatchNorm2d(20, eps=1e-05, momentum=0.1, affine=True, track_running_stats=True)
  (pool2): MaxPool2d(kernel_size=2, stride=2, padding=0, dilation=1, ceil_mode=False)
  (conv3): Conv2d(20, 30, kernel_size=(5, 5), stride=(2, 2), padding=(1, 1))
  (conv3_bn3): BatchNorm2d(30, eps=1e-05, momentum=0.1, affine=True, track_running_stats=True)
  (fc1): Linear(in_features=1620, out_features=120, bias=True)
  (fc2): Linear(in_features=120, out_features=5, bias=True)
)
\end{lstlisting}

\paragraph{1d Frequency}
\begin{lstlisting}[
    basicstyle=\tiny, %or \small or \footnotesize etc.
]
CNN_1dF(
  (conv1): Conv2d(1, 10, kernel_size=(5, 1), stride=(1, 1), dilation=(3, 1))
  (conv1_bn1): BatchNorm2d(10, eps=1e-05, momentum=0.1, affine=True, track_running_stats=True)
  (pool1): MaxPool2d(kernel_size=(2, 1), stride=(2, 1), padding=0, dilation=1, ceil_mode=False)
  (conv2): Conv2d(10, 20, kernel_size=(5, 1), stride=(2, 1), padding=(3, 0), dilation=(2, 1))
  (conv2_bn2): BatchNorm2d(20, eps=1e-05, momentum=0.1, affine=True, track_running_stats=True)
  (pool2): MaxPool2d(kernel_size=(2, 1), stride=(2, 1), padding=0, dilation=1, ceil_mode=False)
  (conv3): Conv2d(20, 30, kernel_size=(5, 1), stride=(2, 1), padding=(1, 0))
  (conv3_bn3): BatchNorm2d(30, eps=1e-05, momentum=0.1, affine=True, track_running_stats=True)
  (fc1): Linear(in_features=31680, out_features=120, bias=True)
  (fc2): Linear(in_features=120, out_features=5, bias=True)
)
\end{lstlisting}

\paragraph{1d Time}
\begin{lstlisting}[
    basicstyle=\tiny, %or \small or \footnotesize etc.
]
CNN_1dT(
  (conv1): Conv2d(1, 10, kernel_size=(1, 5), stride=(1, 1), dilation=(1, 3))
  (conv1_bn1): BatchNorm2d(10, eps=1e-05, momentum=0.1, affine=True, track_running_stats=True)
  (pool1): MaxPool2d(kernel_size=(1, 2), stride=(1, 2), padding=0, dilation=1, ceil_mode=False)
  (conv2): Conv2d(10, 20, kernel_size=(1, 5), stride=(1, 2), padding=(0, 3), dilation=(1, 2))
  (conv2_bn2): BatchNorm2d(20, eps=1e-05, momentum=0.1, affine=True, track_running_stats=True)
  (pool2): MaxPool2d(kernel_size=(1, 2), stride=(1, 2), padding=0, dilation=1, ceil_mode=False)
  (conv3): Conv2d(20, 30, kernel_size=(1, 5), stride=(1, 2), padding=(0, 1))
  (conv3_bn3): BatchNorm2d(30, eps=1e-05, momentum=0.1, affine=True, track_running_stats=True)
  (fc1): Linear(in_features=34560, out_features=120, bias=True)
  (fc2): Linear(in_features=120, out_features=5, bias=True)
)
\end{lstlisting}

All architectures are trained with cross entropy loss, the SGD optimizer with learning rate 0.005 and momentum 0.9, for 150 epochs with a batch size of 64.  The overall accuracies of each network architecture, as well as the number of parameters and the dimensions of the features of the last convolutional layer (\texttt{conv3}) are given for the NSynth (Table~\ref{tab:NSynth-overall-acc}), Composer (Table~\ref{tab:Composer-overall-acc}) and Beethoven (Table~\ref{tab:Beethoven-overall-acc}) datasets below.  For a baseline, a feed forward network with one hidden layer of 120 units with batch normalization and a softmax output with the appropriate number of classes is also trained (where the mel-spectrogram is flattened for the input).  All networks for all datasets are trained identically for 5 different initializations.

\begin{table*}[t]
\begin{center}\scalebox{0.8}{
\begin{tabular}{lcccc}\hline
Architecture & Train Acc. & Test Acc. & \# of Parameters & Feature Dim. \texttt{conv3} \\\hline
Regular & $98.91\pm0.02$\% &  $97.78\pm0.05$\% & 79118 & (30, 4, 4) \\
Deformable & $98.96\pm0.02$\% &  $98.20\pm0.10$\% & 160138 & (30, 6, 6) \\
Dilated & $96.11\pm2.75$\% & $94.79\pm2.46$\% & 151118 &  (30, 6, 6) \\
1dF & $98.78\pm0.07$\% &  $97.87\pm0.12$\% & 2770118 & (30, 6, 128) \\
1dT & $98.83\pm0.15$\% &  $96.99\pm0.28$\% & 2770118 & (30, 128, 6) \\\hline
Feed Forward & $98.89\pm0.02$\% & $96.91\pm0.06$\% & 1967408 & \\\hline
\end{tabular}}
\end{center}
\caption{Train and test accuracy for all 5 architectures trained on the NSynth dataset (8 classes).  Mean accuracy and standard deviation over 5 initializations is reported.  The number of parameters and feature dimensions of the last convolutional layer (\texttt{conv3}) are also reported.}
\label{tab:NSynth-overall-acc}
\end{table*}%

\begin{table*}[t]
\begin{center}\scalebox{0.8}{
\begin{tabular}{lcccc}\hline
Architecture & Train Acc. & Test Acc. & \# of Parameters & Feature Dim. \texttt{conv3} \\\hline
Regular & $96.31\pm0.15$\% & $75.78\pm1.13$\% & 107555 &  (30, 4, 6) \\
Deformable & $97.34\pm0.14$\% &  $79.79\pm0.88$\% & 202975 &  (30, 6, 8) \\
Dilated & $95.72\pm0.13$\% & $70.64\pm0.64$\% & 215555 &  (30, 6, 9) \\
1dF & $96.62\pm0.18$\% &  $70.24\pm0.72$\% & 3806555 & (30, 6, 176) \\
1dT & $98.24\pm0.13$\% &  $74.30\pm1.16$\% & 4152155 & (30, 128, 9) \\\hline
Feed Forward & $97.15\pm0.09$\% &  $64.86\pm2.07$\% & 2704325 & \\\hline
\end{tabular}}
\end{center}
\caption{Train and test accuracy for all 5 architectures trained on the Composer dataset (5 classes).  Mean accuracy and standard deviation over 5 initializations is reported.  The number of parameters and feature dimensions of the last convolutional layer (\texttt{conv3}) are also reported.}
\label{tab:Composer-overall-acc}
\end{table*}%

\begin{table*}[t]
\begin{center}\scalebox{0.8}{
\begin{tabular}{lcccc}\hline
Architecture & Train Acc. & Test Acc. & \# of Parameters & Feature Dim. \texttt{conv3} \\\hline
Regular & $95.49\pm3.62$\% & $85.85\pm2.51$\% & 108160 & (30, 4, 6) \\
Deformable & $96.39\pm3.99$\% &  $87.50\pm3.72$\% & 203580 & (30, 6, 8) \\
Dilated & $86.76\pm0.13$\% &  $75.92\pm0.37$\% & 216160 & (30, 6, 9) \\
1dF & $97.04\pm0.19$\% & $79.34\pm0.30$\% & 3807160 & (30, 6, 176) \\
1dT & $99.03\pm0.07$\% & $89.08\pm0.44$\% & 4152760 & (30, 128, 9) \\\hline
Feed Forward & $96.94\pm0.09$\% & $76.28\pm0.36$\% & 2704930 & \\\hline
\end{tabular}}
\end{center}
\caption{Train and test accuracy for all 5 architectures trained on the Beethoven dataset (10 classes).  Mean accuracy and standard deviation over 5 initializations is reported.  The number of parameters and feature dimensions of the last convolutional layer (\texttt{conv3}) are also reported.}
\label{tab:Beethoven-overall-acc}
\end{table*}%

\subsubsection{Number of Channels Comparison}
For the NSynth dataset and the Regular convolutional architecture shown above, we explore the impact of the number of channels on the overall classification accuracy, with all other details fixed.  The architectures with more channels result in a higher classification accuracy (Table~\ref{tab:NSynth-channel-acc}).

\begin{table*}[t]
\begin{center}\scalebox{0.8}{
\begin{tabular}{lccc}\hline
\# of Channels & Train Acc. & Test Acc. & \# of Parameters  \\\hline
(5, 10, 10) & $89.41\pm1.45$\% &  $88.84\pm1.20$\% & 24238 \\
(10, 10, 10) & $95.46\pm2.32$\%  & $94.28\pm1.95$\% & 25628 \\
\textbf{(10, 20, 30)} & $\bm{98.93\pm0.03}$\textbf{\%} & $\bm{97.82\pm0.04}$\textbf{\%} & \textbf{79118} \\\hline
\end{tabular}}
\end{center}
\caption{Train and test accuracy for all the Regular convolution architecture trained on the NSynth dataset (8 classes).  Mean accuracy and standard deviation over 3 initializations is reported for a different number of channels at each convolutional layer.  The number of parameters is also reported. The selected (10, 20, 30) channels gives the highest accuracy and has a reasonable number of parameters.}
\label{tab:NSynth-channel-acc}
\end{table*}%

\subsection{Features}\label{subsec:features}

\paragraph{Deep Features}
The dimensions of the deep features for the last convolutional layers are given in Table~\ref{tab:NSynth-overall-acc}, Table~\ref{tab:Composer-overall-acc} and Table~\ref{tab:Beethoven-overall-acc}.

\paragraph{Hand-Crafted Features}

A variety of hand-crafted features are calculated and compared.  The top hand-crafted features are discussed in the main paper, with the remaining features discussed here.  Classification results for all features are given in Section~\ref{suppsec:experiment_1} below.  Several of the hand-crafted features are calculated using the \texttt{librosa} package \citep{librosa} in Python. These features are the root mean square for each spectrogram frame (RMS), spectral centroid, spectral bandwidth, spectral flatness and spectral rolloff (threshold is 85\%).  All of these features are calculated on the raw audio, with the same parameters as used to calculate the mel-spectrograms (i.e. hop length is 502, etc.).  See \url{https://librosa.org/doc/main/feature.html} for details about each feature.  

The remaining features are calculated directly from the unnormalized mel-spectrograms.  These features include the median power over time, the mean power over time and the time to -80 dB (approximately 0 power), -75 dB and -70 dB.  These last three features capture the decay of the power in the audio signal and are calculated for each frequency bin as the first time at which that frequency goes to each of the respective dB thresholds.  

Several wavelet features are also considered.  All wavelet features are calculated with the 1D Ricker mother wavelet centered at 0.  Various bandwidths are considered.  The wavelet features are either calculated over frequency (included in the main paper) or over time.  That is, for each frequency (time) the wavelet transform is calculated over time (frequency) and a summary statistic of the wavelet coefficients is calculated.  The summary statistics considered are mean, median, standard deviation, variance, kurtosis, 25th quantile and 75th quantile. 

\begin{figure*}[htbp]
\begin{center}
\includegraphics[width=0.8\textwidth]{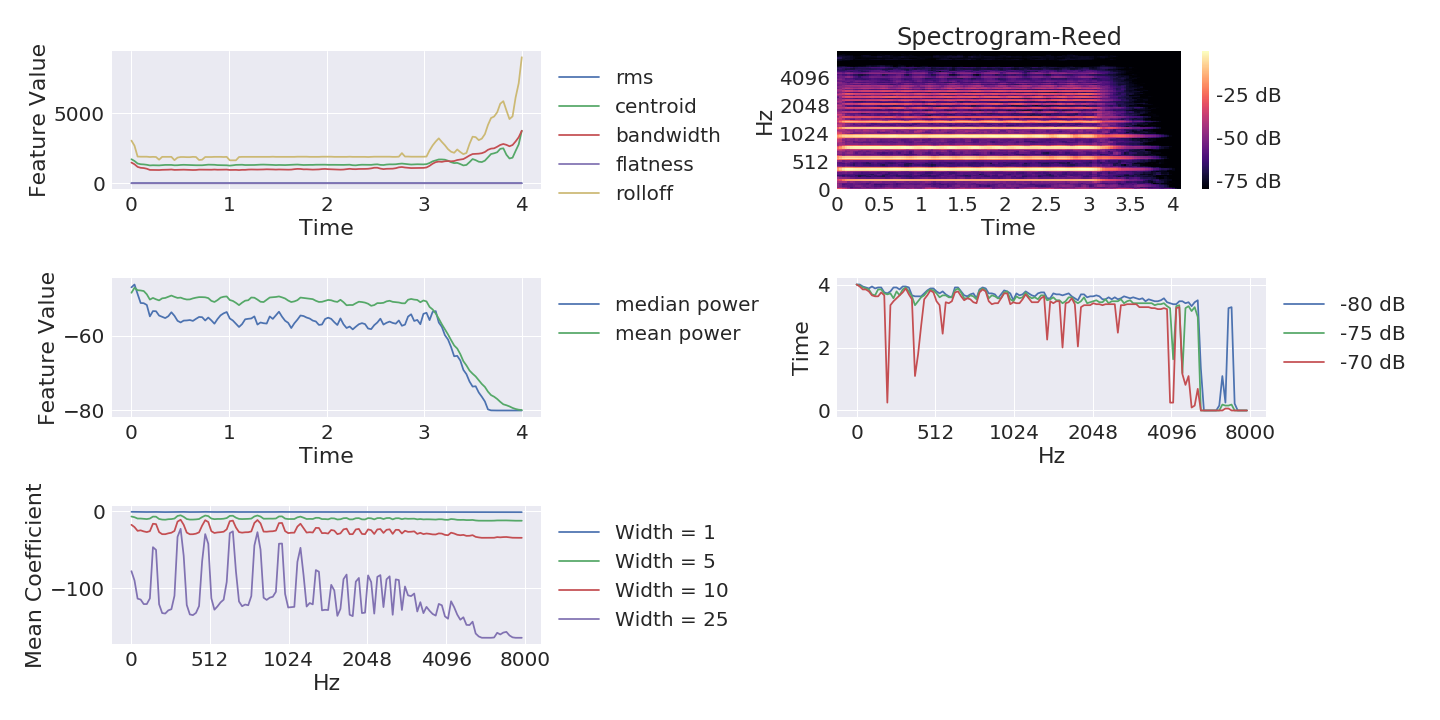}
\caption{Mel-spectrogram and various hand-crafted features for an example Reed instrument family for the NSynth dataset.  The \texttt{librosa} features are plotted in the top left, the median and mean power over time are plotted in the middle left, the time to specific dB features are plotted in the middle right and the mean wavelet coefficients (by frequency) for 4 different bandwidths are plotted in the bottom left.}
\label{fig:NSynth-EDA-features}
\end{center}
\end{figure*}

\begin{figure*}[htbp]
\begin{center}
\includegraphics[width=0.8\textwidth]{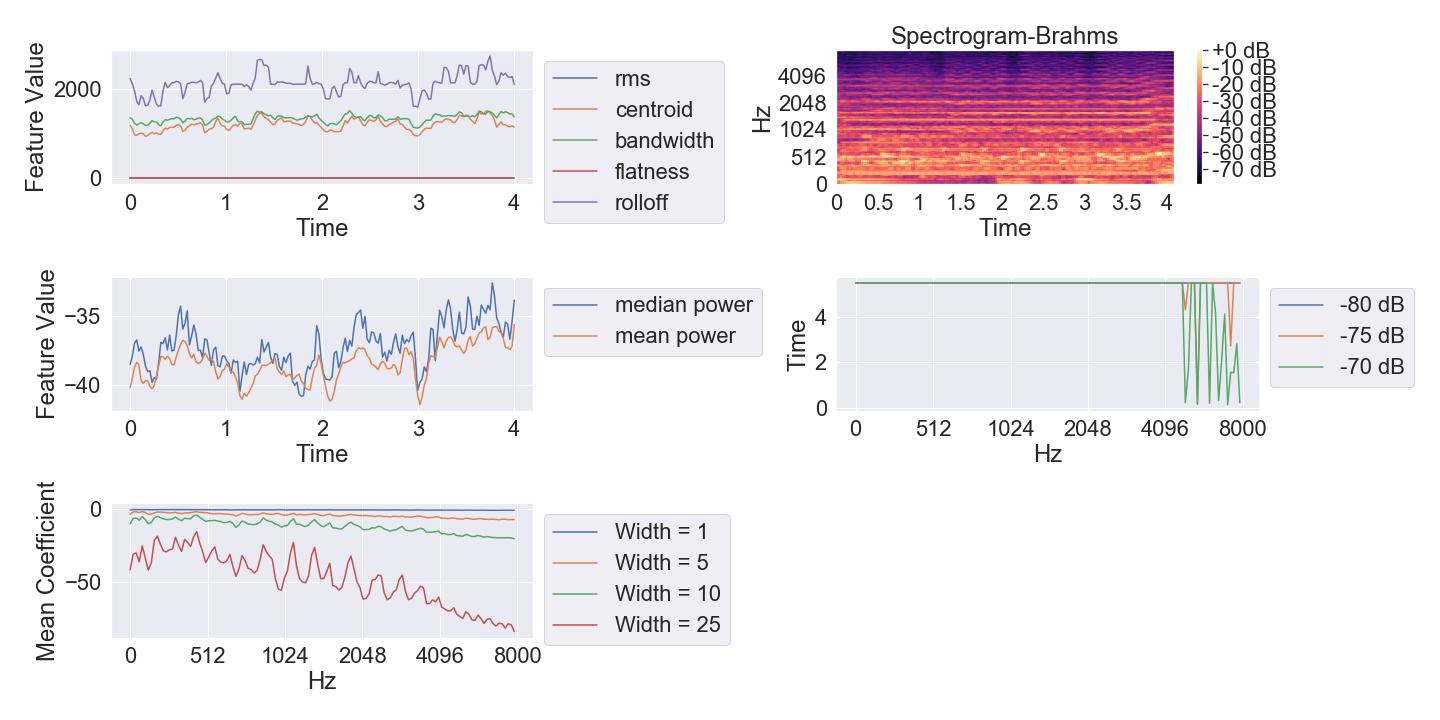}
\caption{Mel-spectrogram and various hand-crafted features for an example Brahms training mel-spectrogram for the Composer dataset.  The \texttt{librosa} features are plotted in the top left, the median and mean power over time are plotted in the middle left, the time to specific dB features are plotted in the middle right and the mean wavelet coefficients (by frequency) for 4 different bandwidths are plotted in the bottom left.}
\label{fig:Orchestral-EDA-features}
\end{center}
\end{figure*}

\begin{figure}[htbp]
\begin{center}
\includegraphics[width=0.48\textwidth]{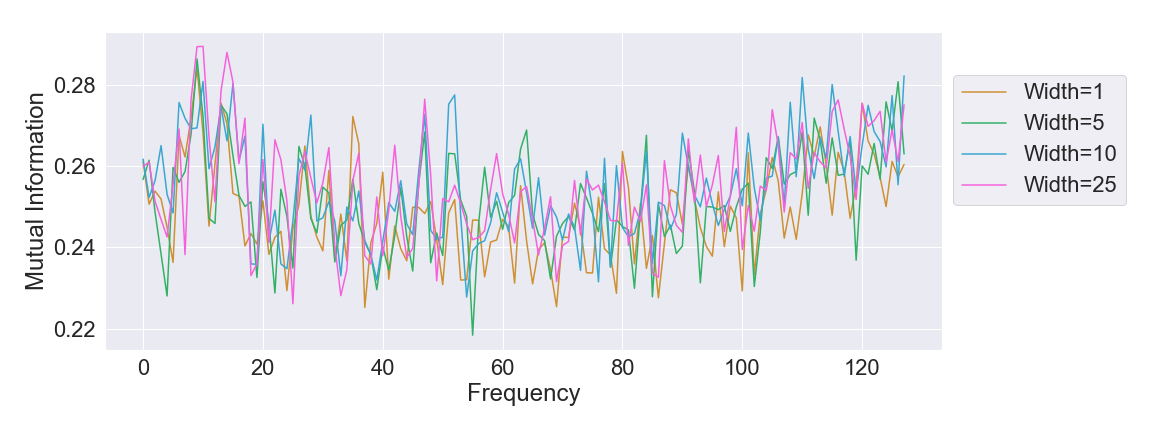}
\caption{Mutual information between the Composer dataset hand-crafted features of mean wavelet coefficient over frequency for various bandwidths and the output class (5 composer classes).  The x-axis is labeled by the mel frequency bin (0-128).}
\label{fig:Orchestral-MI-3}
\end{center}
\end{figure}

Plots of a subset of these features are plotted for a single input mel-spectrogram for the NSynth dataset (Figure~\ref{fig:NSynth-EDA-features}) and Composer dataset (Figure~\ref{fig:Orchestral-EDA-features}).  The mutual information between a subset of the wavelet features and the output class for the Composer dataset is plotted as well (Figure~\ref{fig:Orchestral-MI-3}). 

\subsection{Analysis Methods}

\paragraph{Feature Classification} All features, both deep and hand-crafted, are normalized to have mean 0 and standard deviation 1 prior to the feature logistic regression.  The multi-class logistic regression is implemented in PyTorch, and the models for all features are trained using cross entropy loss, with the SGD optimizer with a learning rate of 0.01 and momentum 0.9 for 100 epochs with a batch size of 64.  The deep features are flattened across channels as input to the multi-class logistic regression.  The same initialization of the logistic regression classifier is used for each of the five different initializations of the deep features for each architecture, while five different initializations of the logistic regression classifier are used for each hand-crafted feature.   

\paragraph{Feature Similarity} In addition to the linear CKA feature similarity, several additional similarity measures defined in \citet{pmlr-v97-kornblith19a} are also considered in  Section~\ref{suppsec:experiment_2} below.  The linear regression similarity measure is:
\begin{equation}\label{eq:lr-sim}
R^2_{LR} = \text{sim}(X, Y) = \dfrac{||Q^T_YX||^2_F}{||X||_F^2},
\end{equation}

\noindent where $Y = Q_YR_Y$ is the QR decomposition of $Y$ and $||\cdot||_F$ is the Frobenius norm.  Two Canonical Correlation Analysis (CCA) similarities are:
\begin{equation}\label{eq:cca-sim}
\begin{split}
R^2_{CCA} &=  \dfrac{||Q^T_YQ_X||^2_F}{p_1}, \\
\bar{\rho}_{CCA} &= \dfrac{||Q^T_YQ_X||_*}{p_1},
\end{split}
\end{equation}
where $X = Q_XR_X$ is the QR decomposition of $X$, $p_1$ is the number of columns of $X$ and $||\cdot||_*$ is the Nuclear norm, following the notation of \citet{pmlr-v97-kornblith19a}.  Singular vector CCA (SVCCA) is also calculated, as follows:
\begin{equation}\label{eq:svcca-sim}
\begin{split}
R^2_{SVCCA} &=  \dfrac{||U^T_YU_X||^2_F}{\min(d_X, d_Y)},\\
\bar{\rho}_{SVCCA} &= \dfrac{||U^T_YU_X||_*}{\min(d_X, d_Y)},
\end{split}
\end{equation}
where $U_Y$ and $U_X$ are the left-singular vectors of $Y$ and $X$, respectively, sorted in decreasing order and the first $d_Y$ and $d_X$ singular vectors are selected to explain 99\% of the variance, following \citet{pmlr-v97-kornblith19a}. That is, $U_Y$ is of dimension $n\times d_Y$ and $U_X$ is of dimension $n\times d_X$, where the largest $d_Y$ singular vectors explain 99\% of the variation in $Y$ and the largest $d_X$ singular vectors explain 99\% of the variation in $X$. 

\subsection{Computing Details}

All pre-processing of the data and analysis of results is performed on CPUs.  All deep models and the logistic regression models for decoding are performed on 1 GeForce RTX 2080 Ti GPU.  The computation of the mel-spectrograms and all the hand-crafted features took approximately 1 hour for the Composer dataset (smallest) and 2 hours for the Beethoven dataset (largest).  Extracting the deep features takes approximately 1 hour for all 5 trained and untrained initializations for each dataset and each architecture.  The linear CKA calculations for comparing the last layer of the deep features for each architecture to the top hand-crafted features takes approximately 1.5 hours for the Composer dataset (smallest) and 2.5 hours for the Beethoven dataset (largest) using the linear algebra functionality in PyTorch.  Other linear CKA calculations that use smaller dimensional features are faster.  Exact training times for the deep architectures are given in Table~\ref{tab:train-deep-time}.  Training times for the classification of the deep features, as well as training times for the hand-crafted features averaged across all features for all datasets are given in Table~\ref{tab:train-lr-time}; training times for the related tasks are included in the deep feature times and are similar for the hand-crafted features.  To classify all layers for the NSynth dataset main and related tasks took 30065 seconds for the Regular architecture and 22278 seconds for the Deformable architecture, and to classify all layers over all initializations for the main task for the Composer dataset took 12711 seconds for the Regular architecture and 10680 seconds for the Deformable architecture.  The CPUs and GPU used are on an internal cluster.

\begin{table*}[t]
\begin{center}\scalebox{0.8}{
\begin{tabular}{lccc} \hline
& NSynth (8 Classes) & Composer (5 Classes) & Beethoven (10 Classes) \\\hline
Regular & 10943 & 9529 & 22379 \\
Deformable & 18516 & 14268 & 14315 \\
Dilated & 9825 & 9454    & 25315 \\
1dF & 9376 &  5061 & 13862 \\
1dT & 9010 & 4782 & 19206 \\\hline
Feed Forward & 6148 & 2106 & 7720 \\\hline
\end{tabular}}
\end{center}
\caption{Exact training times in seconds for 5 initializations of the 5 deep architectures for each dataset (times are rounded to the nearest second). }
\label{tab:train-deep-time}
\end{table*}%

\begin{table*}[t]
\begin{center}\scalebox{0.8}{
\begin{tabular}{lccc} \hline
& NSynth (8 Classes) & Composer (5 Classes) & Beethoven (10 Classes) \\\hline
Hand-Crafted & 2420& 935 & 2722 \\\hline
Regular & 2330 & 2013 & 8440 \\
Deformable & 2493 & 2249 & 11558 \\
Dilated & 2490 & 2285 & 12004 \\
1dF & 12163 & 11749 & 18570 \\
1dT & 17743 & 12600 & 39184 \\\hline
\end{tabular}}
\end{center}
\caption{Training time in seconds for the logistic regression models to classify the deep and hand-crafted features.  The hand-crafted features results are averaged across the training times for all of the top features. The training times are for all five initializations for each feature. Times are rounded to the nearest second and the deep feature training times include the main task and all related tasks.  The times for training the related tasks for the hand-crafted features are similar to those reported here (the linear regression times for the hand-crafted features for the NSynth related tasks are very fast).}
\label{tab:train-lr-time}
\end{table*}%

\subsection{Code}

Several existing code resources are used in this work.  Much of the data pre-processing uses code from the \texttt{librosa} package in Python \citep{librosa}, \url{https://librosa.org/doc/latest/index.html}, package version 0.8.0. The wavelet features are calculated using the \texttt{scipy.signal} package \url{https://docs.scipy.org/doc/scipy/reference/generated/scipy.signal.cwt.html}, package version 1.1.0.  Loading the data into PyTorch uses code from \url{https://discuss.pytorch.org/t/input-numpy-ndarray-instead-of-images-in-a-cnn/18797/3}, and PyTorch version 1.4.0 is used throughout.  The Deformable convolution implementation is from \url{https://github.com/4uiiurz1/pytorch-deform-conv-v2}, under an MIT License.  

Extracting the deep features after each layer uses code from \url{https://discuss.pytorch.org/t/how-can-l-load-my-best-model-as-a-feature-extractor-evaluator/17254/6}.  The logistic regression model in PyTorch uses code following \url{https://pytorch.org/tutorials/beginner/nlp/deep_learning_tutorial.html}. Finally, information about the NSynth dataset can be found at \url{https://magenta.tensorflow.org/datasets/nsynth#files}.  All code for the NSynth experiments is included on GitHub at \url{https://github.com/aky4wn/convolutions-for-music-audio}.  The Composer and Beethoven experiments are analogous.

\section{What Features are Useful for the Discriminative Tasks?}\label{suppsec:experiment_1}

Classification accuracies for all of the considered hand-crafted features on the main classification tasks, as well as the related tasks for all datasets are discussed below.  Training procedures for the logistic regressions for all of the hand-crafted features are identical to those for the top features considered in the main paper.  Confusion matrices for the deep architectures are also included.

\subsection{Hand-Crafted Features}

Classification accuracies using the hand-crafted features described in Section~\ref{subsec:features} for the main classification tasks for all datasets are given in Table~\ref{tab:hand-crafted}. 

\begin{table*}[t]
\begin{center}\scalebox{0.8}{
\begin{tabular}{lccc} \hline
& NSynth (8 Classes) & Composer (5 Classes) & Beethoven (10 Classes) \\\hline
\textit{Random Guessing} & \textit{26.50\%} &  \textit{25.20\%} & \textit{10.4\%} \\\hline
RMS & $53.43\pm1.08$\% & $26.98\pm0.22$\% & $12.96\pm0.07$\% \\
Spectral Centroid & $47.95\pm1.01$\% & $32.08\pm0.07$\% &  $17.75\pm0.10$\% \\
Spectral Bandwidth & $55.68\pm0.10$\% &  $30.51\pm0.11$\% &  $17.94\pm0.11$\% \\
Spectral Flatness & $42.02\pm0.04$\% &  $31.01\pm0.09$\% & $16.10\pm0.04$\% \\
Spectral Rolloff & $48.11\pm0.10$\% &  $31.76\pm0.07$\% &  $17.54\pm0.20$\% \\
Median Power & $58.90\pm0.72$\% &  $28.23\pm0.08$\% &  $13.13\pm0.13$\% \\
\textbf{Mean Power} &  $\bm{66.81\pm0.37}$\textbf{\%} &  $\bm{27.50\pm0.04}$\textbf{\%} &  $\bm{12.93\pm0.07}$\textbf{\%} \\
Time to -80 dB &  $56.07\pm0.05$\% &  $33.70\pm0.06$\% &  $18.65\pm0.04$\% \\
Time to -75 dB & $57.18\pm0.17$\% & $34.93\pm0.10$\% &  $22.29\pm0.09$\% \\
\textbf{Time to -70 dB} & $\bm{56.95\pm0.03}$\textbf{\%} & $\bm{35.02\pm0.12}$\textbf{\%} & $\bm{23.85\pm0.10}$\textbf{\%} \\
Mean Wavelet (1) &  $54.79\pm1.11$\% &  $51.80\pm0.03$\% &  $56.38\pm0.03$\% \\
Mean Wavelet (5) & $61.76\pm0.10$\% &  $58.20\pm0.05$\% &  $66.94\pm0.05$\% \\
Mean Wavelet (10) & $62.76\pm0.08$\% &  $59.91\pm0.07$\% &  $70.37\pm0.03$\% \\
\textbf{Mean Wavelet (25)} & $\bm{60.69\pm0.08}$\textbf{\%} &  $\bm{62.65\pm0.05}$\textbf{\%} &  $\bm{74.14\pm0.04}$\textbf{\%} \\\hline
\end{tabular}}
\end{center}
\caption{Test accuracies using a logistic regression classifier for the hand-crafted features on the main classification tasks for each dataset.  Mean values over five initializations of the logistic regression classifier and 1 standard error are reported.  Top features included in the main paper are in bold. The wavelet features are the mean value of the wavelet coefficients over time for each frequency and the bandwidth is the number in parentheses. }
\label{tab:hand-crafted}
\end{table*}%

\subsection{Wavelet Features}

\begin{figure}
\centering
\begin{subfigure}[b]{0.49\linewidth}
\includegraphics[width=\linewidth]{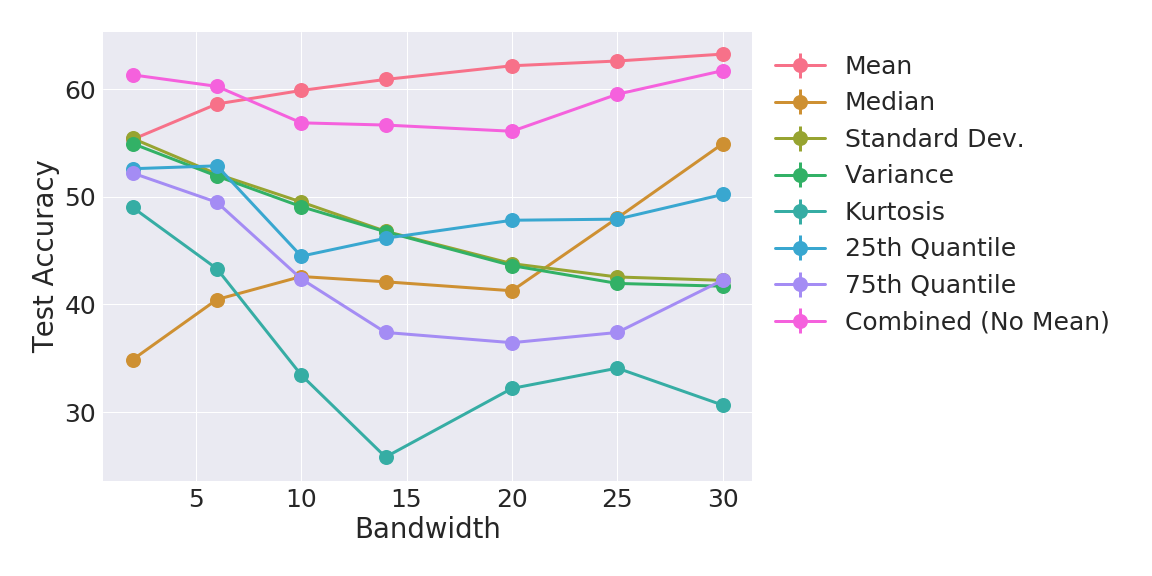}
\caption{Average over Frequency.} \label{fig:wavelet-frequency}
\end{subfigure}
\begin{subfigure}[b]{0.49\linewidth}
\includegraphics[width=\linewidth]{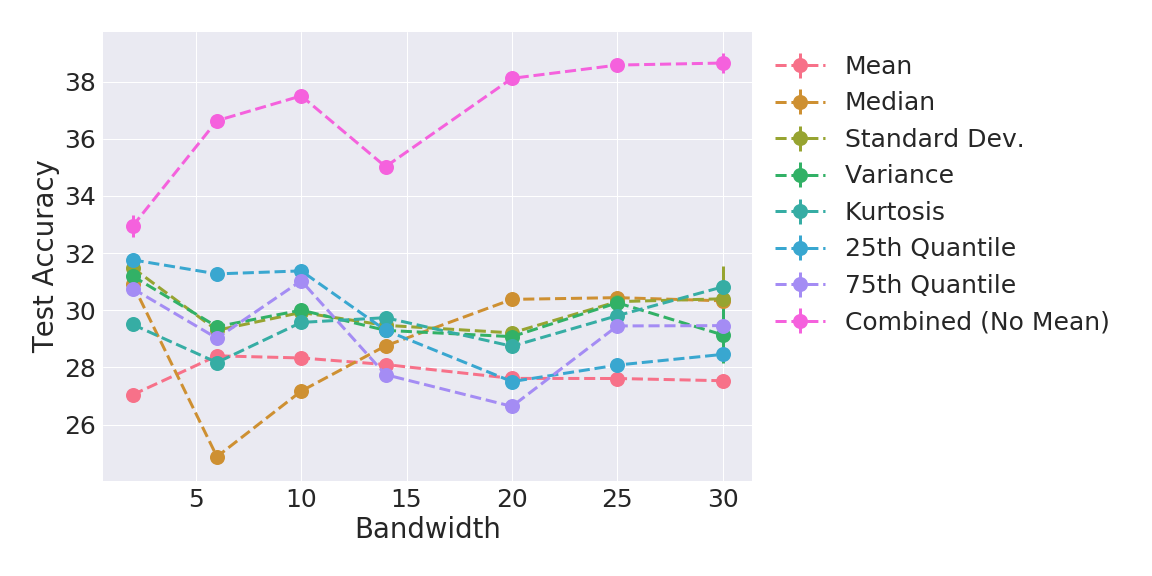}
\caption{Average over Time.}\label{fig:wavelet-time}
\end{subfigure}
\caption{Test accuracy for the main Composer 5-class classification task for wavelet features calculated for each (a) frequency, then a summary statistic is taken over time, and (b) time, then a summary statistic is taken over frequencies, for several different bandwidths.  The different colors are each of the different summary statistics.  The main features used, mean coefficients and combined, are relatively robust in accuracy for different bandwidths.}
\label{fig:wavelets-compare}
\end{figure}

Wavelet features are explored for various bandwidths and various summary statistics for the 1D Ricker mother wavelet centered at 0. Wavelets are either calculated for each frequency and then a summary statistics is taken over time (results for main paper), or for each time and a summary statistic is taken over frequency.  The test set accuracies for each dataset for the frequency wavelet features used in the main paper (bandwidth is 25) is given in Table~\ref{tab:wavelet}.  The mean wavelet coefficients and the combined other summary statistics give the highest test set accuracies across datasets. The frequency wavelet coefficients result in higher classification accuracies across bandwidths as compared to the time-based wavelet features (Figure~\ref{fig:wavelet-frequency} vs. Figure~\ref{fig:wavelet-time}).  Additionally, the top features used in the main paper, the mean wavelet coefficients and the combined wavelet summary features, are relatively consistent in test set accuracy across bandwidths.  Finally, for an additional comparison, we calculate 2D discrete wavelet transforms using the \texttt{db10} and \texttt{bior1.3} wavelet families from Version 0.5.2 of the PyWavelets package (\url{https://pywavelets.readthedocs.io/en/latest/}).  The horizontal and vertical components of each type of wavelet are separately flattened and fed into the Logistic Regression classifier.  The classification results for the Composer dataset are given in Table~\ref{tab:2d-wave}; the 2D wavelets do not improve over the 1D wavelet features considered, though again, the horizontal coefficients outperform the vertical coefficients, as seen for the 1D wavelets with the frequency summary statistics outperforming the time summary wavelet statistics.

\begin{table*}[t]
\begin{center}\scalebox{0.8}{
\begin{tabular}{lccc} \hline
& NSynth (8 Classes) & Composer (5 Classes) & Beethoven (10 Classes) \\\hline
\textit{Random Guessing} & \textit{26.50\%} &  \textit{25.20\%} & \textit{10.4\%} \\\hline
\textbf{Mean} & $\bm{60.69\pm0.08}$\textbf{\%} &  $\bm{62.65\pm0.05}$\textbf{\%} &  $\bm{74.14\pm0.04}$\textbf{\%} \\
Median & $59.04\pm0.06$\% &  $48.05\pm0.05$\% &  $46.44\pm0.04$\% \\
Standard Deviation & $55.08\pm0.06$\% &  $42.56\pm0.10$\% &  $33.19\pm0.09$\% \\
Variance & $54.11\pm0.14$\% &  $41.96\pm0.14$\% & $33.63\pm0.05$\% \\
Kurtosis & $49.81\pm0.12$\% &  $34.08\pm0.25$\% & $25.69\pm0.10$\% \\
25th Quantile & $59.34\pm0.06$\% &  $47.94\pm0.10$\% &  $48.52\pm0.06$\% \\
75th Quantile & $58.98\pm0.08$\% &  $37.40\pm0.10$\% &  $31.40\pm0.05$\% \\
\textbf{Combined (No Mean)} & $\bm{74.44\pm0.18}$\textbf{\%} & $\bm{59.56\pm0.15}$\textbf{\%} & $\bm{66.32\pm0.05}$\textbf{\%}
\\\hline
\end{tabular}}
\end{center}
\caption{Test accuracies using a logistic regression classifier for the hand-crafted wavelet features on the main classification tasks for each dataset.  Mean values over five initializations of the logistic regression classifier and 1 standard error are reported.  Top features included in the main paper are in bold. The wavelet features are the summary statistic calculated for each frequency over time for a bandwidth of 25. The median, standard deviation, variance, kurtosis and 25th and 75th quantiles are combined to form the last row.}
\label{tab:wavelet}
\end{table*}%

\begin{table}[htp]
\begin{center}\scalebox{0.8}{
\begin{tabular}{lcc}\hline
& Horizontal & Vertical \\\hline
\texttt{bior1.3} & $44.90\pm 0.35$ & $25.68\pm 0.20$ \\
\texttt{db10} & $43.69\pm 0.70$ & $25.06\pm 0.51$ \\\hline
\end{tabular}}
\end{center}
\caption{Test accuracy for 2D wavelets using either the horizontal or vertical coefficients on the Composer dataset.  These 2D coefficients do not improve over the summary statistics of the 1D wavelets.}
\label{tab:2d-wave}
\end{table}%

\subsection{Related Tasks}
The deep and hand-crafted features are also used for several related tasks for each dataset.  For the NSynth dataset, the related tasks are predicting the Note Pitch and the Note Velocity.  Both of these are linear regression tasks and the root mean squared error (RMSE) is reported for the hand-crafted features in Table~\ref{tab:NSynth-feature-transfer} and Table~\ref{tab:NSynth-wavelet-transfer}.  The linear regression is performed once for each input feature and does not have error bars. The related task for the Composer dataset is classifying which part of the piece (beginning, middle or end) each mel-spectrogram comes from. Finally, the related tasks for the Beethoven dataset are classifying the number of the symphony of each mel-spectrogram (10 classes) and whether each example is by an American or European orchestra. Results for the Composer and Beethoven datasets are shown in Table~\ref{tab:Composer-Beethoven-feature-transfer} and Table~\ref{tab:Composer-Beethoven-wavelet-transfer}.  All classification results are again using the same logistic regression procedure as for the main classification tasks.  Top features included in the main paper are in bold throughout and the wavelet features especially perform well in the related tasks, and can even outperform some of the deep features (Table~\ref{tab:deep-transfer}). As seen in the NSynth results, the deep features trained for one task perform well when used for another task, that is, the deep features in Table~\ref{tab:deep-transfer} achieve similar accuracies to Regular CNNs trained on each related task directly (displayed in the last row of Table~\ref{tab:deep-transfer}). 

\begin{table*}
\begin{center}
\scalebox{0.8}{
\begin{tabular}{p{2.8cm}ccccc}\hline
 & \multicolumn {2}{c}{NSynth} & Composer & \multicolumn {2}{c}{Beethoven} \\\hline
 & Note Pitch  & Note Velocity & Part of Piece  & \mbox{Symphony \#}  & Orchestra \\ 
 & (RMSE) & (RMSE) & \mbox{(3-class)} & \mbox{(9-class)} & \mbox{Continent} \mbox{(2-class)} \\\hline
\textit{Baseline / Random Guessing} & \textit{21.00} &  \textit{36.03} & \textit{33.5\%} &  \textit{19.5\%} & \textit{59.5\%}\\\hline
Mean Power & 18.97 & 34.68 & $37.64\pm0.09$\% &  $20.57\pm0.07$\% & $58.53\pm0.03$\% \\
Time to -70 dB & 16.01 &  35.34 &  $35.85\pm0.14$\% & $24.74\pm0.13$\% &  $63.09\pm0.01$\% \\
Mean Wavelet (25) & 14.87 & 35.37 &  $\bm{43.48\pm0.09}$\textbf{\%} &  $53.07\pm0.08$\% &  $79.79\pm0.01$\% \\
Wavelet Combined & 12.57 & 34.52 & $40.59\pm0.24$\% &  $50.47\pm0.11$\% &  $76.87\pm0.02$\% \\
Top 4 Combined & \textbf{11.57} &  \textbf{33.53} &  $\bm{41.34\pm0.35}$\textbf{\%} &  $\bm{53.61\pm0.24}$\textbf{\%} & $\bm{80.39\pm0.19}$\textbf{\%} \\\hline
Regular & $9.52\pm0.11$ &  $32.74\pm0.10$ & $40.33\pm0.32$\% &  $47.41\pm0.30$\% &  $85.11\pm0.44$\% \\
Deformable & $\bm{8.17\pm0.31}$ &  $\bm{31.93\pm0.07}$ &  $40.43\pm0.46$\% &  $\bm{49.29\pm0.65}$\textbf{\%} & 
$\bm{86.02\pm0.74}$\textbf{\%} \\
Dilated & $9.25\pm0.17$ & $33.16\pm0.15$ &  $40.13\pm0.43$\% & $51.51\pm0.52$\% &  $82.65\pm0.28$\% \\
1dF & $62.18\pm5.71$ &  $247.07\pm31.06$ &  $40.73\pm0.35$\% &  $49.89\pm0.56$\% &  $80.34\pm0.57$\% \\
1dT & $37.06\pm2.80$ &  $125.23\pm5.29$ &  $40.54\pm0.53$\% &  $\bm{58.45\pm0.73}$\textbf{\%} &  $\bm{86.62\pm1.17}$\textbf{\%} \\\hline
Regular (Trained) & $7.84\pm0.11$ &  $32.33\pm0.10$ &  $38.63\pm0.44$\% &  $62.04\pm1.37$\% &  $91.41\pm0.33$\% \\\hline
\end{tabular}}
\end{center}
\caption{Test RMSE for the related tasks of note pitch prediction and note velocity (volume) prediction for the NSynth dataset and test accuracies using a logistic regression classifier for the hand-crafted features on the related classification tasks for the Composer (part of piece, 3-class) and Beethoven (symphony number, 9-class and continent of orchestra, 2-class) datasets.  The baseline RMSE is predicting the mean value on the training data for all test examples. The hand-crafted features are the top features included in the main paper.  The last row is a regular CNN trained directly on each related task.  (Lower values of RMSE and higher classification accuracies are better).}
\label{tab:deep-transfer}
\end{table*}%

\begin{table}
\begin{center}\scalebox{0.8}{
\begin{tabular}{lcc}\hline
& Note Pitch & Note Velocity \\\hline
\textit{Baseline (Mean)} & \textit{21.000} & \textit{36.025} \\\hline
RMS & 19.886 & 35.858 \\
Spectral Centroid & 14.647 & 35.122 \\
Spectral Bandwidth & 19.079 & 34.519 \\
Spectral Flatness & 20.069 & 35.507 \\
Spectral Rolloff & 16.658 & 34.960 \\
Median Power & 19.636 & 35.020 \\
\textbf{Mean Power} & \textbf{18.972} & \textbf{34.676} \\
Time to -80 dB & 16.853 & 35.423 \\
Time to -75 dB & 16.427 & 35.320 \\
\textbf{Time to -70 dB} & \textbf{16.007} & \textbf{35.340} \\
Mean Wavelet (1)  & 14.528 & 34.826 \\
Mean Wavelet (5) & 13.139 & 34.889 \\
Mean Wavelet (10) & 13.599 &  35.097 \\
\textbf{Mean Wavelet (25)} & \textbf{14.866} & \textbf{35.365} \\\hline
\end{tabular}}
\end{center}
\caption{Test RMSE for the related tasks of note pitch prediction and note velocity (volume) prediction for the NSynth dataset.  The baseline RMSE is predicting the mean value on the training data for all test examples. Top features included in the main paper are in bold. The wavelet features are the mean value of the wavelet coefficients over time for each frequency and the bandwidth is the number in parentheses.  (Lower values are better).}
\label{tab:NSynth-feature-transfer}
\end{table}%

\begin{table}
\begin{center}\scalebox{0.8}{
\begin{tabular}{lcc}\hline
& Note Pitch & Note Velocity \\\hline
\textit{Baseline (Mean)} & \textit{21.000} & \textit{36.025} \\\hline
\textbf{Mean} & \textbf{14.866} & \textbf{35.365} \\
Median & 15.118 & 35.381 \\
Standard Deviation & 16.477 & 35.603 \\
Variance & 16.561 & 35.610 \\
Kurtosis & 19.224 & 35.507 \\
25th Quantile & 16.792 & 35.373 \\
75th Quantile & 15.190 & 35.528 \\
\textbf{Combined (No Mean)} & \textbf{12.565} & \textbf{34.520} \\\hline
\end{tabular}}
\end{center}
\caption{Test RMSE for the related tasks of note pitch prediction and note velocity (volume) prediction for the NSynth dataset.  The baseline RMSE is predicting the mean value on the training data for all test examples. Top features included in the main paper are in bold. The wavelet features are the summary statistic calculated for each frequency over time for a bandwidth of 25. The median, standard deviation, variance, kurtosis and 25th and 75th quantiles are combined to form the last row.}
\label{tab:NSynth-wavelet-transfer}
\end{table}%

\begin{table*}[t]
\begin{center}\scalebox{0.8}{
\begin{tabular}{lccc} \hline
& Composer & \multicolumn {2}{c}{Beethoven} \\\hline
& Part of Piece (3-class) & Symphony \# (9-class) & Orchestra Continent (2-class) \\\hline
\textit{Random Guessing} & \textit{33.5\%} &  \textit{19.5\%} & \textit{59.5\%} \\\hline
RMS & $37.94\pm0.11$\% &  $19.75\pm0.15$\% &  $59.07\pm0.03$\% \\
Spectral Centroid & $35.61\pm0.13$\% &  $19.75\pm0.09$\% & $63.03\pm0.02$\% \\
Spectral Bandwidth & $35.72\pm0.09$\% &  $19.22\pm0.14$\% &  $60.99\pm0.03$\% \\
Spectral Flatness & $37.75\pm0.01$\% &  $19.29\pm0.06$\% & $61.83\pm0.02$\% \\
Spectral Rolloff & $36.04\pm0.12$\% &  $19.43\pm0.17$\% &  $63.39\pm0.01$\% \\
Median Power & $36.94\pm0.084$\% &  $20.53\pm0.07$\% &  $58.50\pm0.02$\% \\
\textbf{Mean Power} & $\bm{37.64\pm0.09}$\textbf{\%} & $\bm{20.57\pm0.07}$\textbf{\%} & $\bm{58.53\pm0.03}$\textbf{\%} \\
Time to -80 dB & $35.14\pm0.11$\% &  $21.11\pm0.07$\% &  $58.96\pm0.04$\% \\
Time to -75 dB & $35.61\pm0.32$\% &  $22.34\pm0.06$\% &  $62.23\pm0.03$\% \\
\textbf{Time to -70 dB} & $\bm{35.85\pm0.14}$\textbf{\%} & $\bm{24.74\pm0.13}$\textbf{\%} &  $\bm{63.09\pm0.01}$\textbf{\%} \\
Mean Wavelet (1) & $41.94\pm0.06$\% &  $43.91\pm0.05$\% &  $73.71\pm0.01$\% \\
Mean Wavelet (5) & $43.25\pm0.08$\% &  $48.47\pm0.07$\% &  $76.95\pm0.01$\% \\
Mean Wavelet (10) & $43.05\pm0.08$\% &  $50.62\pm0.08$\% &  $78.23\pm0.01$\% \\
\textbf{Mean Wavelet (25)} & $\bm{43.48\pm0.09}$\textbf{\%} & $\bm{53.07\pm0.08}$\textbf{\%} &  $\bm{79.79\pm0.01}$\textbf{\%} \\\hline

\end{tabular}}
\end{center}
\caption{Test accuracies using a logistic regression classifier for the hand-crafted features on the related classification tasks for the Composer (part of piece, 3-class) and Beethoven (symphony number, 9-class and continent of orchestra, 2-class) datasets.  Mean values over five initializations of the logistic regression classifier and 1 standard error are reported.  Top features included in the main paper are in bold. The wavelet features are the mean value of the wavelet coefficients over time for each frequency and the bandwidth is the number in parentheses.}
\label{tab:Composer-Beethoven-feature-transfer}
\end{table*}%

\begin{table*}[t]
\begin{center}\scalebox{0.8}{
\begin{tabular}{lccc} \hline
& Composer & \multicolumn {2}{c}{Beethoven} \\\hline
& Part of Piece (3-class) & Symphony \# (9-class) & Orchestra Continent (2-class) \\\hline
\textit{Random Guessing} & \textit{33.5\%} &  \textit{19.5\%} & \textit{59.5\%} \\\hline
\textbf{Mean} & $\bm{43.48\pm0.09}$\textbf{\%} & $\bm{53.07\pm0.08}$\textbf{\%} &  $\bm{79.79\pm0.01}$\textbf{\%} \\
Median & $38.77\pm0.03$\% &  $39.99\pm0.09$\% &  $69.24\pm0.00$\% \\
Standard Deviation & $37.85\pm0.05$\% &  $30.85\pm0.07$\% &  $65.28\pm0.00$\% \\
Variance & $37.52\pm0.01$\% &  $30.24\pm0.10$\% &  $65.42\pm0.00$\% \\
Kurtosis & $36.19\pm0.12$\% &  $25.32\pm0.07$\% &  $63.61\pm0.00$\% \\
25th Quantile & $39.46\pm0.03$\% &  $38.43\pm0.06$\% &  $70.16\pm0.00$\% \\
75th Quantile & $36.73\pm0.02$\% &  $30.58\pm0.08$\% &  $64.23\pm0.00$\% \\
\textbf{Combined (No Mean)} & $\bm{40.59\pm0.24}$\textbf{\%} &  $\bm{50.47\pm0.11}$\textbf{\%} &  $\bm{76.87\pm0.02}$\textbf{\%} \\\hline
\end{tabular}}
\end{center}
\caption{Test accuracies using a logistic regression classifier for the hand-crafted features on the related classification tasks for the Composer (part of piece, 3-class) and Beethoven (symphony number, 9-class and continent of orchestra, 2-class) datasets.  Mean values over five initializations of the logistic regression classifier and 1 standard error are reported.  Top features included in the main paper are in bold. The wavelet features are the summary statistic calculated for each frequency over time for a bandwidth of 25. The median, standard deviation, variance, kurtosis and 25th and 75th quantiles are combined to form the last row.}
\label{tab:Composer-Beethoven-wavelet-transfer}
\end{table*}%

\subsection{Confusion Matrices for Deep Features}

\begin{figure}[htbp]
\begin{center}
\includegraphics[width=0.495\textwidth]{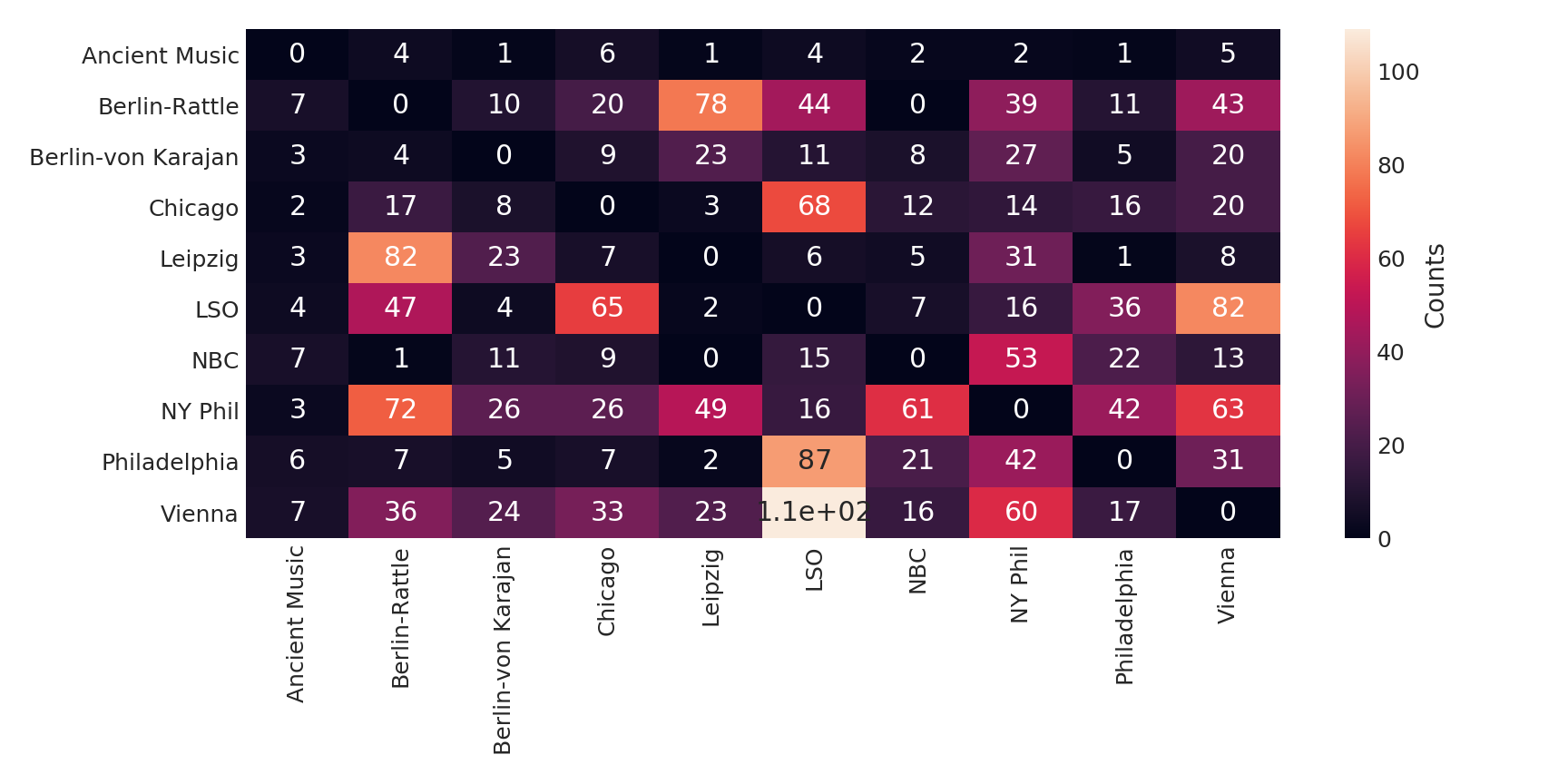}
\caption{Confusion matrices on the test set for the Beethoven dataset for the Regular deep architecture.  The counts on the diagonal are set to 0 for display purposes.}
\label{fig:Beethoven-Regular-confusion}
\end{center}
\end{figure}

\begin{figure}[htbp]
\begin{center}
\includegraphics[width=0.495\textwidth]{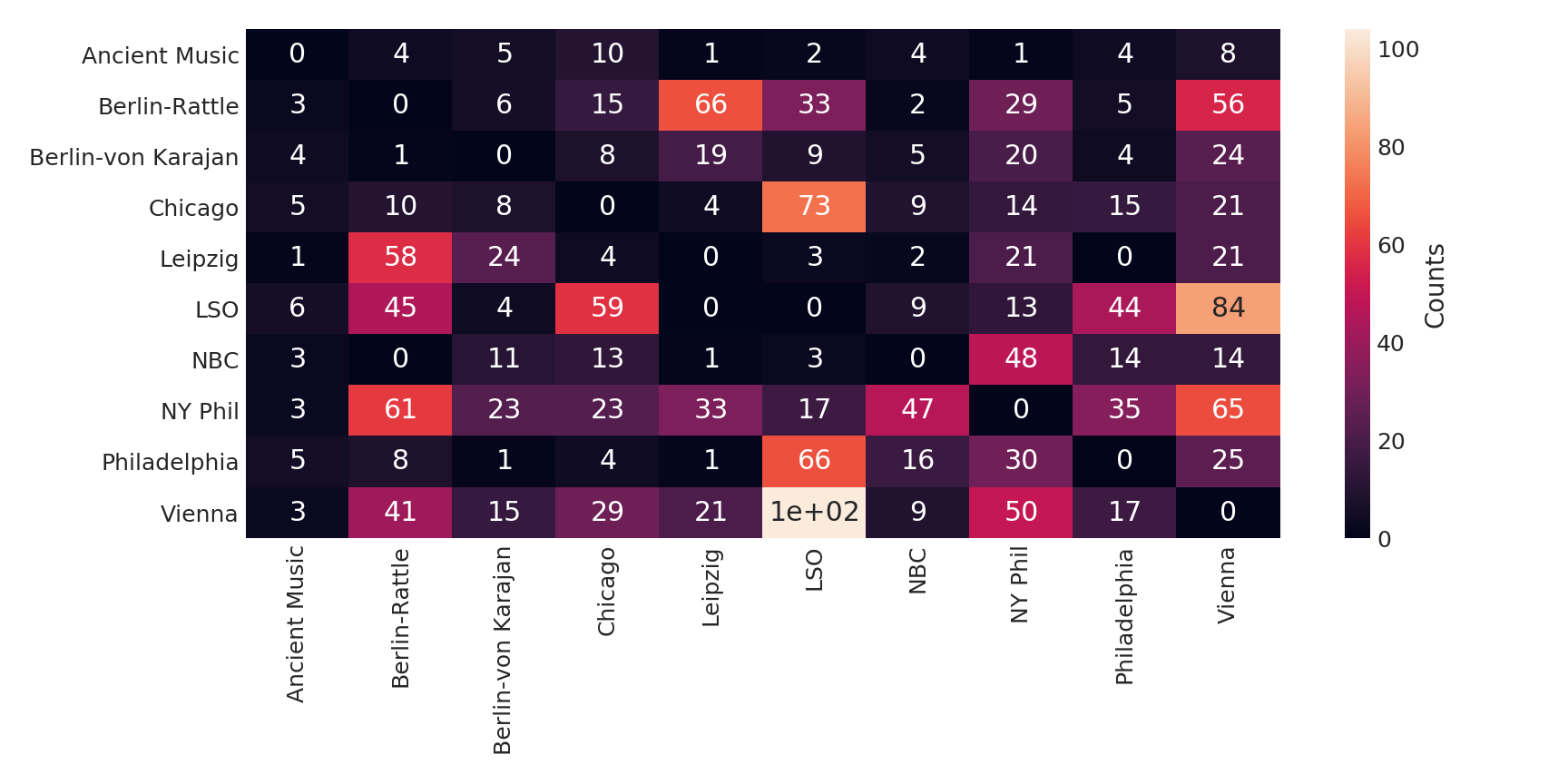}
\caption{Confusion matrices on the test set for the Beethoven dataset for the Deforamble deep architecture.  The counts on the diagonal are set to 0 for display purposes.}
\label{fig:Beethoven-Deformable-confusion}
\end{center}
\end{figure}

\begin{figure*}[t]
\centering
\begin{subfigure}[b]{.32\linewidth}
\includegraphics[width=\linewidth]{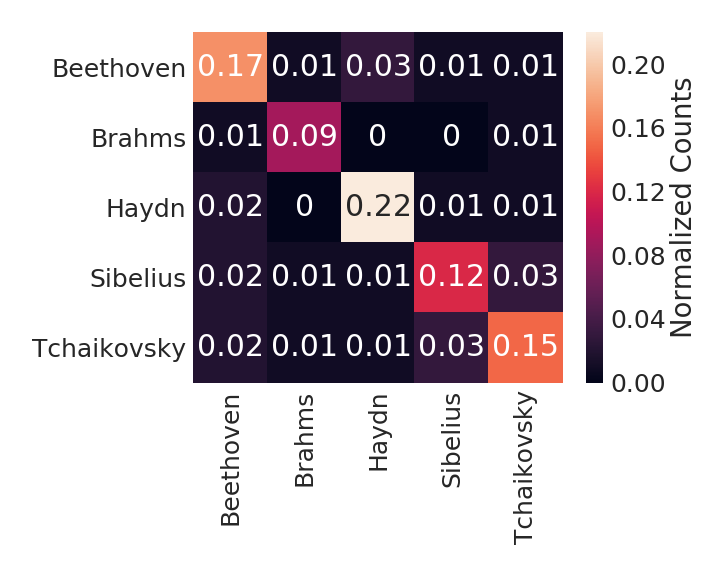}
\caption{Regular}
\end{subfigure}
\begin{subfigure}[b]{.32\linewidth}
\includegraphics[width=\linewidth]{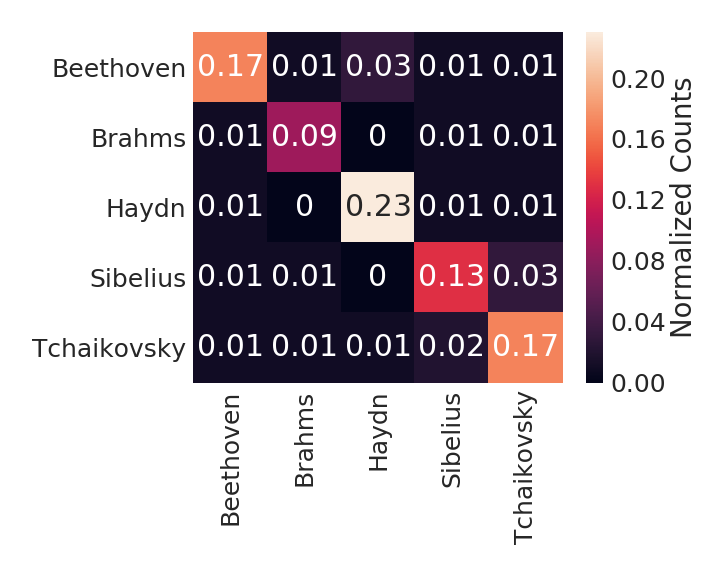}
\caption{Deformable}
\end{subfigure}
\begin{subfigure}[b]{.32\linewidth}
\includegraphics[width=\linewidth]{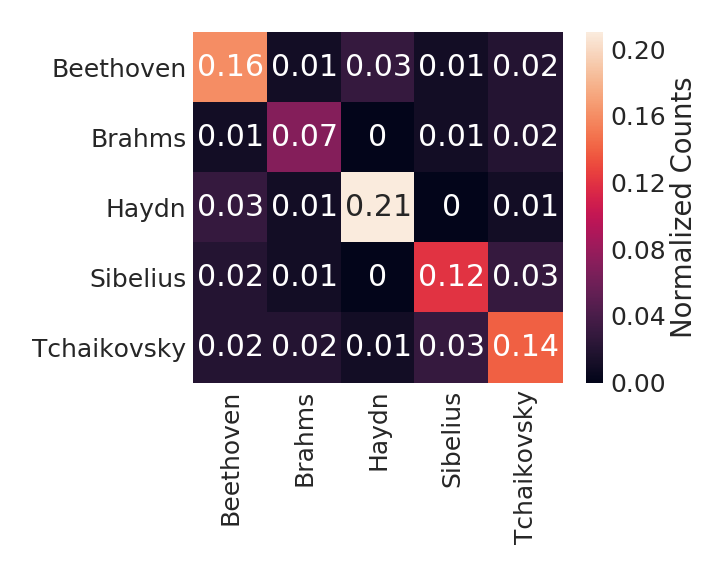}
\caption{Dilated}
\end{subfigure}

\begin{subfigure}[b]{.32\linewidth}
\includegraphics[width=\linewidth]{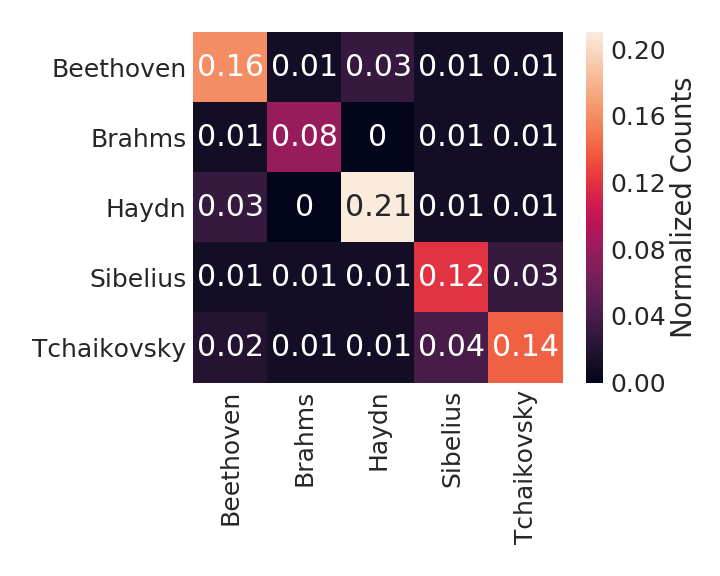}
\caption{1dF}
\end{subfigure}
\begin{subfigure}[b]{.32\linewidth}
\includegraphics[width=\linewidth]{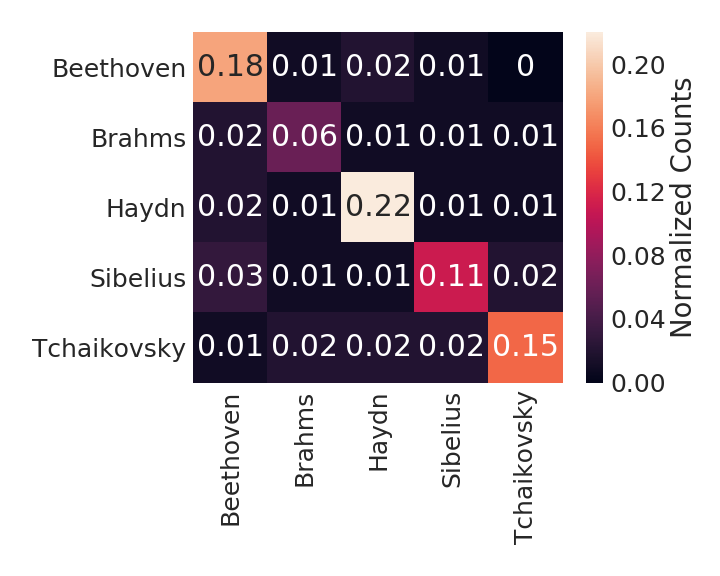}
\caption{1dT}
\end{subfigure}
\begin{subfigure}[b]{.32\linewidth}
\includegraphics[width=\linewidth]{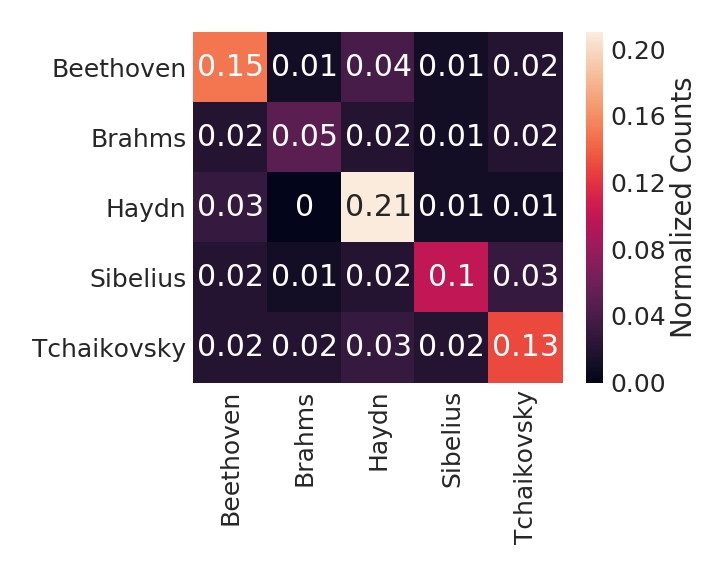}
\caption{Mean Wavelet (25)}
\end{subfigure}
\caption{Confusion matrices for all of the deep architectures and the mean wavelet coefficients (bandwidth = 25) with logistic regression for the Composer dataset on the test set.  The counts are divided by the overall number of test examples.  In general, Haydn and Beethoven are mistaken for each other the most, which makes sense from a musical style perspective.  The confusion matrices of each architecture appear similar.}
\label{fig:Composer-confusion}
\end{figure*}

Confusion matrices for the deep architectures overall for the Composer dataset are given in Figure~\ref{fig:Composer-confusion}. In general, Haydn and Beethoven are mistaken for each other the most, which makes sense from a musical style perspective as these composers were closest to each other in time and were both from the Classical era.  Confusion matrices for the Regular (Figure~\ref{fig:Beethoven-Regular-confusion}) and Deformable architectures (Figure~\ref{fig:Beethoven-Deformable-confusion}) for the Beethoven dataset are also shown.  LSO and Vienna tend to be confused with each other the most; this also makes sense, as these are the two most recent recordings considered and are both by European orchestras with similar performance styles.  

\section{Are the Learned Deep Features Similar to the Hand-Crafted Features?}\label{suppsec:experiment_2}

\subsection{Similarity by Architecture}

\begin{table*}[t]
\begin{center}\scalebox{0.8}{
\begin{tabular}{lccc} \hline
& NSynth (8) & Composer (5) & Beethoven (10)  \\\hline
\textit{Random Guessing} & \textit{26.50\%} &  \textit{25.20\%} & \textit{10.4\%} \\\hline
Deep Features \texttt{conv3} & $96.35\pm0.19$\% &  $72.41\pm1.21$\% &  $82.32\pm0.40$\% \\\hline
Deep + Mean Power & $96.36\pm0.20$\% &  $72.35\pm1.26$\% &  $82.35\pm0.34$\% \\
Deep + Time to -70 dB & $96.41\pm0.18$\% &  $72.85\pm1.19$\% &  $82.60\pm0.45$\% \\
Deep + Mean Wavelet (25) & $96.50\pm0.18$\% & $73.24\pm1.09$\% & $\bm{83.58\pm0.50}$\textbf{\%} \\
Deep + Wavelet Combined & $\bm{96.59\pm0.20}$\textbf{\%} &  $\bm{74.04\pm0.87}$\textbf{\%} &  $\bm{83.44\pm0.42}$\textbf{\%} \\
Deep + Top 4  Combined & $\bm{96.61\pm0.17}$\textbf{\%} & $\bm{74.71\pm0.78}$\textbf{\%} &  $\bm{84.23\pm0.31}$\textbf{\%} \\\hline
\end{tabular}}
\end{center}
\caption{Regular convolution deep features are concatenated to each top hand-crafted feature and classified with logistic regression on the main classification tasks for each dataset (number in parentheses for each dataset is the number of classes).  Hand-crafted features that are less similar to the Regular deep features (Figure~\ref{fig:sim-arch}), especially the Wavelet Combined and Top Combined features improve the classification accuracy when combined with the deep Regular features.}
\label{tab:sim-class-Regular}
\end{table*}%

\begin{figure}[htbp]
\begin{center}
\includegraphics[width=0.45\textwidth]{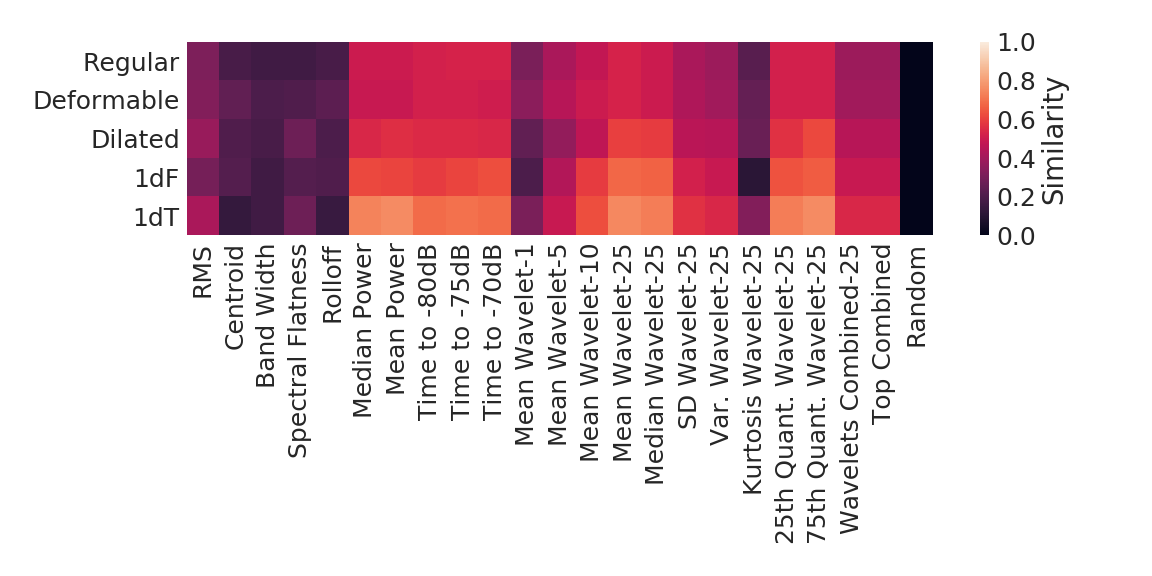}
\caption{Linear CKA similarity between the deep features from each architecture from the last convolutional layer (\texttt{conv3}) with all hand-crafted features for the NSynth dataset. We baseline with Random $\mathcal{N}(0,1)$ noise in the last column. Plots are the similarity value averaged across the deep features from each initialization. }
\label{fig:NSynth-sim-all}
\end{center}
\end{figure}

\begin{figure}[htbp]
\begin{center}
\includegraphics[width=0.45\textwidth]{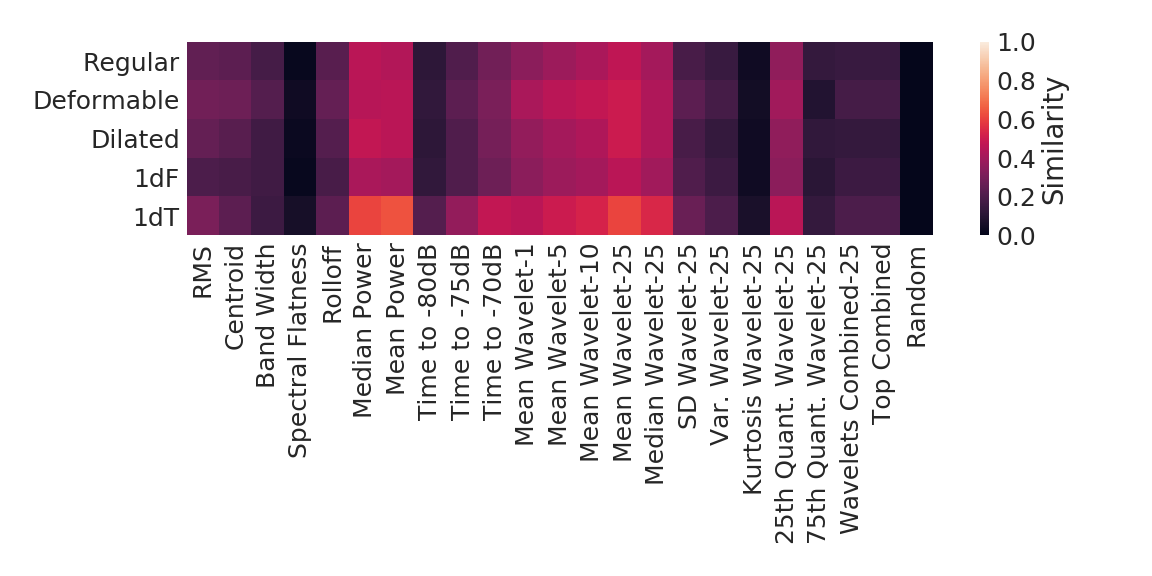}
\caption{Linear CKA similarity between the deep features from each architecture from the last convolutional layer (\texttt{conv3}) with all hand-crafted features for the Composer dataset.}
\label{fig:Composer-sim-all}
\end{center}
\end{figure}

\begin{figure}[htbp]
\begin{center}
\includegraphics[width=0.45\textwidth]{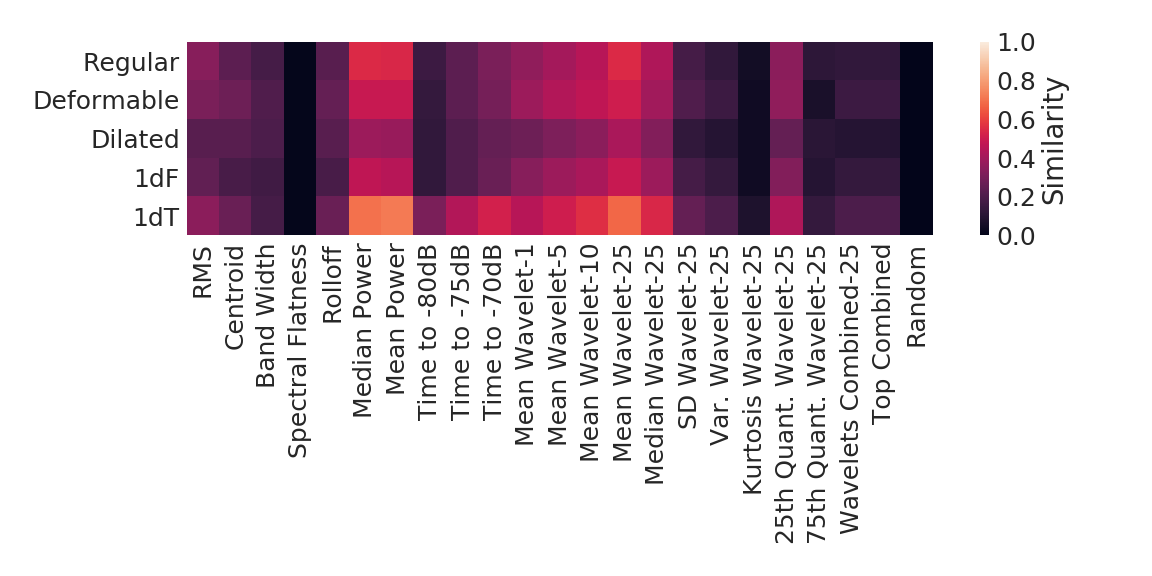}
\caption{Linear CKA similarity between the deep features from each architecture from the last convolutional layer (\texttt{conv3}) with all hand-crafted features for the Beethoven dataset.}
\label{fig:Beethoven-sim-all}
\end{center}
\end{figure}

We first compare the similarity between the deep features from the last convolutional layer (\texttt{conv3}) across architectures to all of the hand-crafted features for the NSynth (Figure~\ref{fig:NSynth-sim-all}), Composer (Figure~\ref{fig:Composer-sim-all}) and Beethoven (Figure~\ref{fig:Beethoven-sim-all}) datasets. We again find that the 1dT features are especially similar to the mean wavelet features, by design. We can confirm this similarity by concatenating the deep Deformable features to the top hand-crafted feature (Table~\ref{tab:sim-class}).  Compared to the Regular deep features, the Deformable deep features are more similar to the hand-crafted features, and indeed, we see less improvement when these hand-crafted features are concatenated to the Deformable deep features in Table~\ref{tab:sim-class} vs Table~\ref{tab:sim-class-Regular}.  However, the Deformable deep features are less similar to the Top 4 Combined features, and the last row of  Table~\ref{tab:sim-class} indicates that when combined with the Deformable deep features, these hand-crafted features can further improve the classification accuracy.

\begin{table*}[t]
\begin{center}\scalebox{0.8}{
\begin{tabular}{lccc} \hline
& NSynth (8) & Composer (5) & Beethoven (10)  \\\hline
\textit{Random Guessing} & \textit{26.50\%} &  \textit{25.20\%} & \textit{10.4\%} \\\hline
Deep Features \texttt{conv3} & $97.80\pm0.08$\% & $75.76\pm0.68$\% &  $84.11\pm0.70$\% \\\hline
Deep + Mean Power & $97.75\pm0.09$\% &  $75.92\pm0.85$\% & $84.17\pm0.79$\% \\
Deep + Time to -70 dB & $97.81\pm0.05$\% &  $76.23\pm0.65$\% &  $84.49\pm0.72$\% \\
Deep + Mean Wavelet (25) & $97.79\pm0.06$\% & $76.42\pm0.79$\% & $85.11\pm0.66$\% \\
Deep + Wavelet Combined & $97.84\pm0.12$\% &  $\bm{77.23\pm0.44}$\textbf{\%} &  $85.10\pm0.80$\% \\
Deep + Top 4  Combined & $97.87\pm0.09$\% &  $\bm{77.72\pm0.67}$\textbf{\%} &  $\bm{85.70\pm0.64}$\textbf{\%}\\\hline
\end{tabular}}
\end{center}
\caption{Deformable convolution deep features are concatenated to each top hand-crafted feature and classified with logistic regression on the main classification tasks for each dataset (number in parentheses for each dataset is number of classes).  Hand-crafted features that are less similar to the Deformable deep features, especially the Top Combined features improve the classification accuracy when combined with the deep Deformable features.}
\label{tab:sim-class}
\end{table*}%

\subsection{Similarity by Layer}

\begin{figure}[htbp]
\centering
\begin{subfigure}[b]{0.455\linewidth}
\centering
\includegraphics[width = \linewidth]{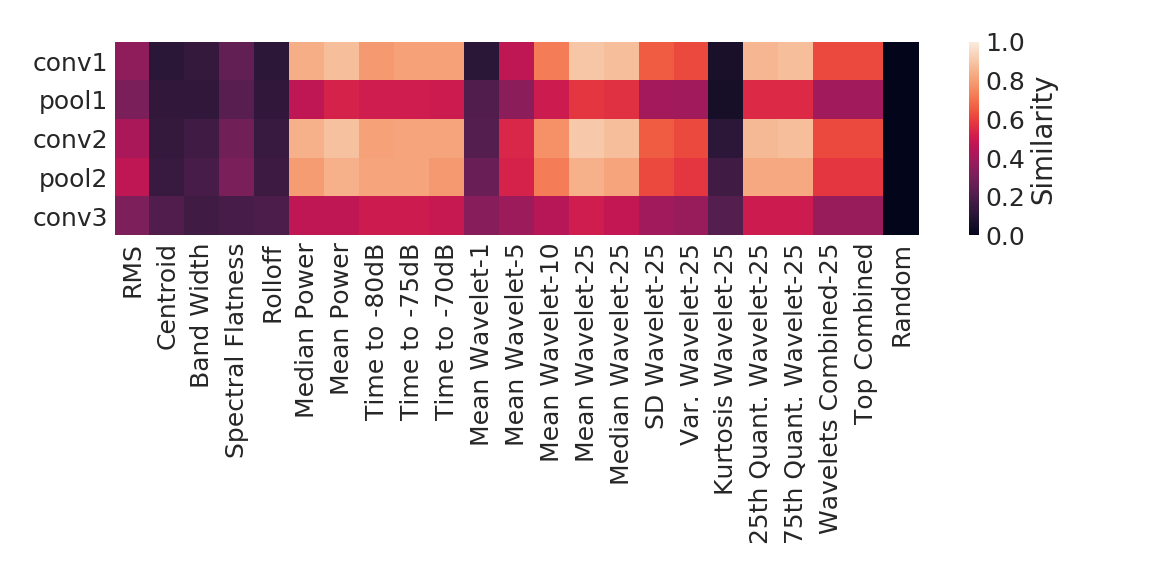}
\caption{Regular}\label{default}
\end{subfigure}
\begin{subfigure}[b]{0.455\linewidth}
\centering
\includegraphics[width =\linewidth]{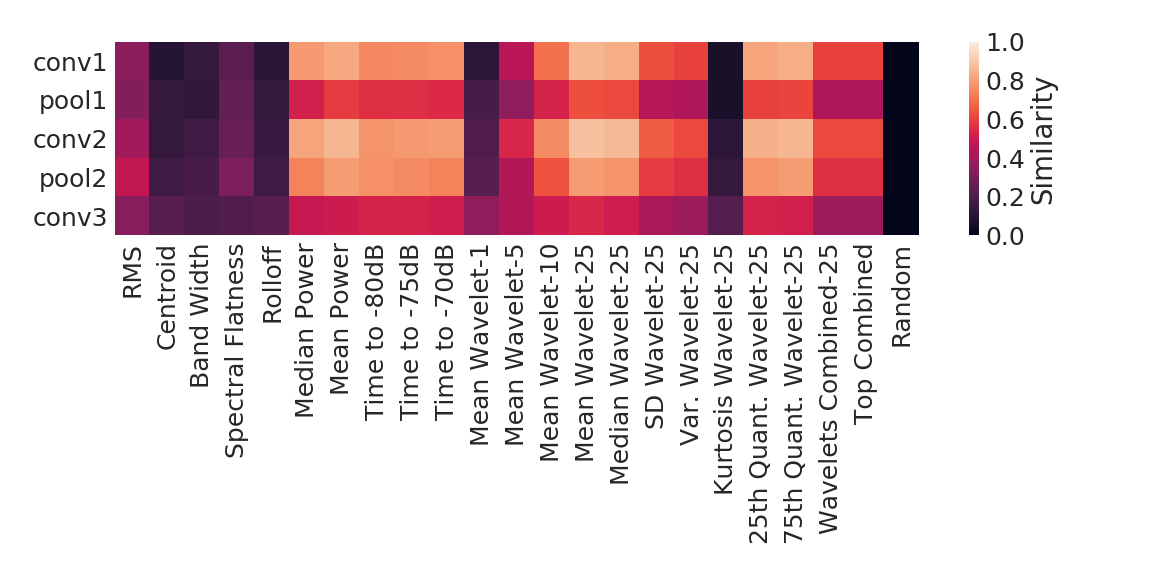}
\caption{Deformable}\label{default}
\end{subfigure}
\caption{Linear CKA similarity between the deep features from all layers for the NSynth dataset and all hand-crafted features for the (a) Regular and (b) Deformable deep features (trained to predict Instrument Family).}
\label{fig:NSynth-layer-IF-sim}
\end{figure}

\begin{figure}[htbp]
\centering
\begin{subfigure}[b]{0.455\linewidth}
\centering
\includegraphics[width=\linewidth]{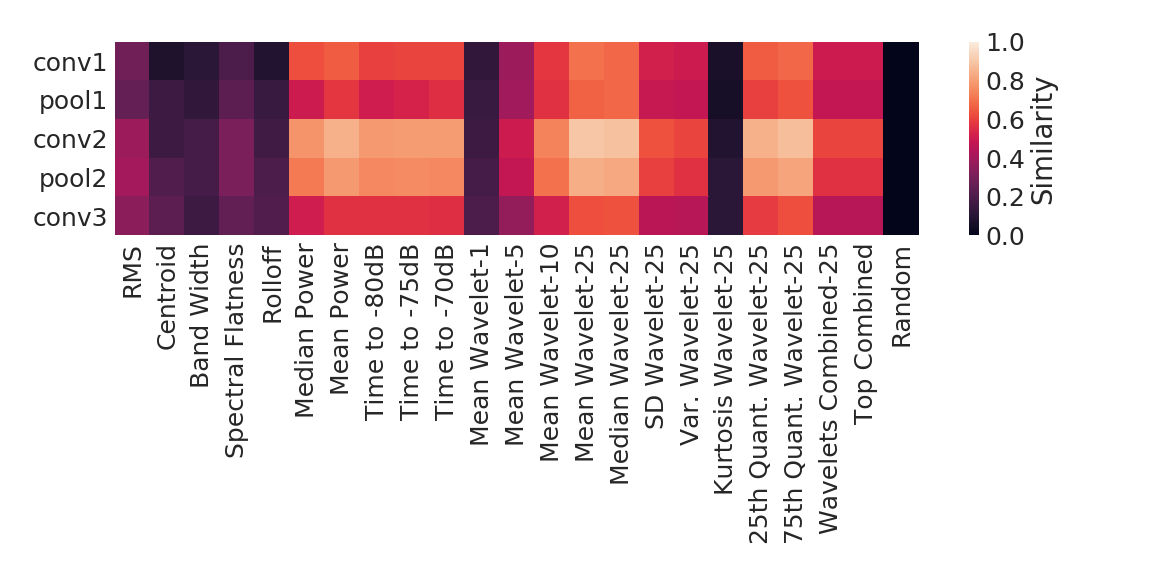}
\caption{Regular}\label{default}
\end{subfigure}
\begin{subfigure}[b]{0.455\linewidth}
\centering
\includegraphics[width=\linewidth]{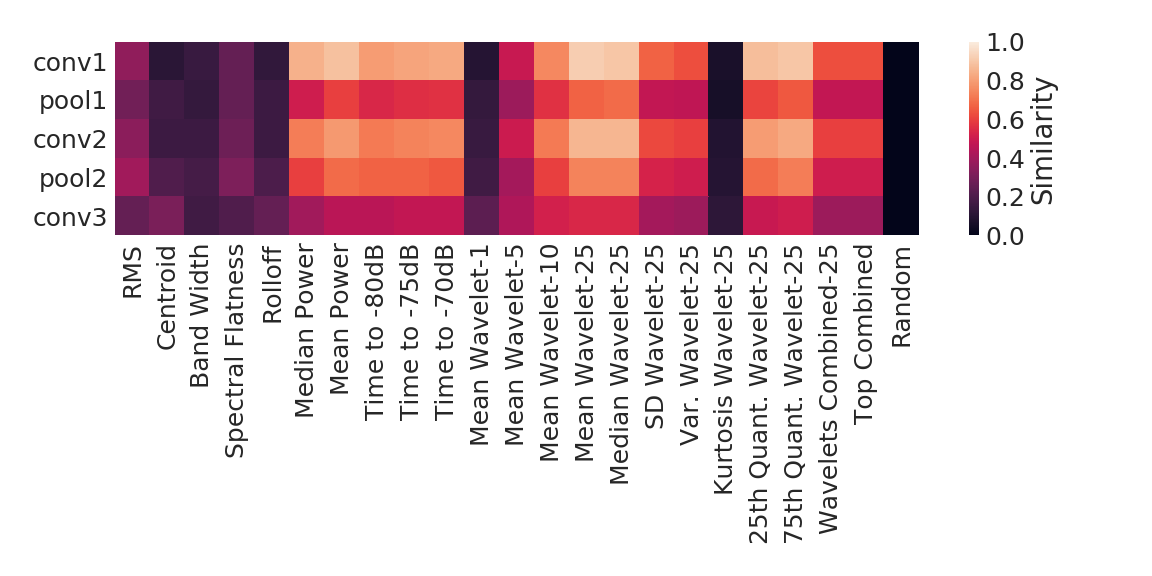}
\caption{Deformable}\label{default}
\end{subfigure}
\caption{Linear CKA similarity between the deep features from all layers for the NSynth dataset and all hand-crafted features for the (a) Regular and (b) Deformable deep features trained to predict Note Pitch.}
\label{fig:NSynth-layer-Pitch-sim}
\end{figure}

\begin{figure}[htbp]
\centering
\begin{subfigure}[b]{0.455\linewidth}
\centering
\includegraphics[width=\linewidth]{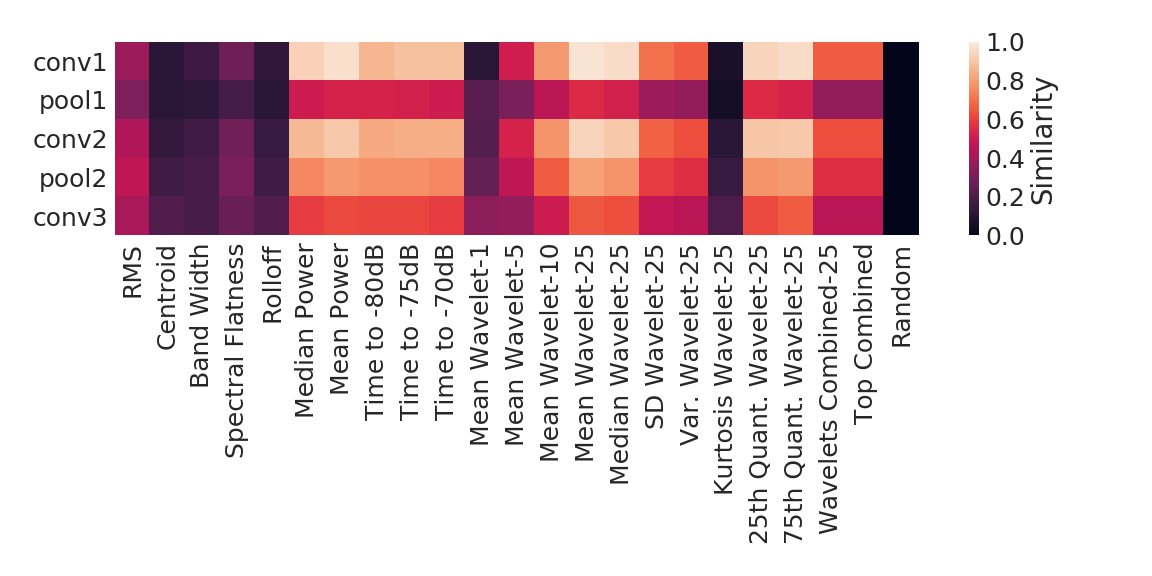}
\caption{Regular}\label{default}
\end{subfigure}
\begin{subfigure}[b]{0.455\linewidth}
\centering
\includegraphics[width=\linewidth]{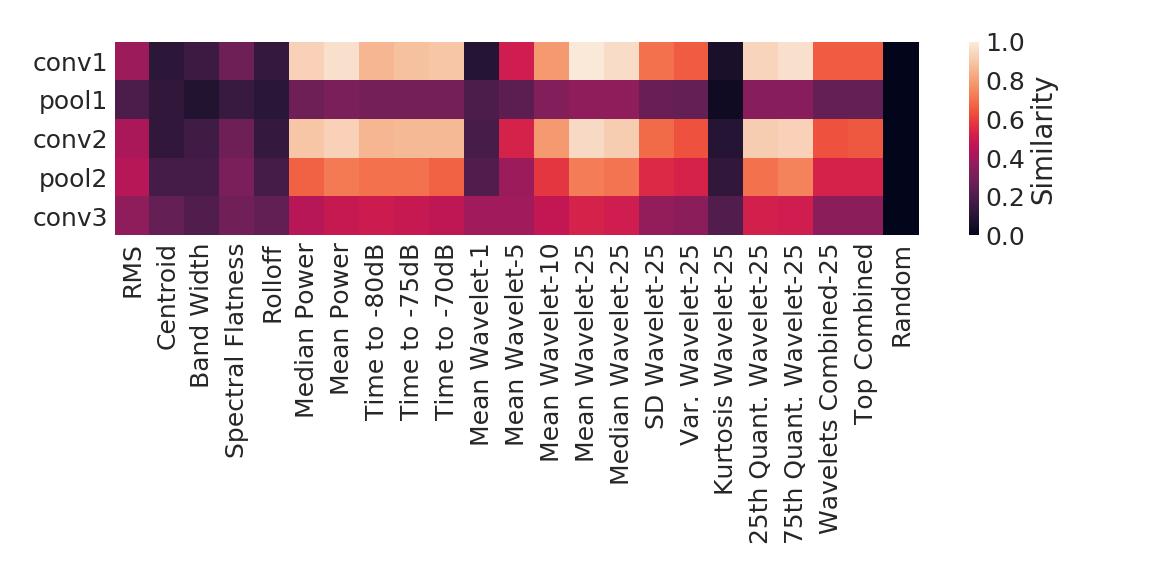}
\caption{Deformable}\label{default}
\end{subfigure}
\caption{Linear CKA similarity between the deep features from all layers for the NSynth dataset and all hand-crafted features for the (a) Regular and (b) Deformable deep features trained to predict Note Velocity. }
\label{fig:NSynth-layer-Velocity-sim}
\end{figure}

\begin{figure}[htbp]
\centering
\begin{subfigure}[b]{0.455\linewidth}
\centering
\includegraphics[width=\linewidth]{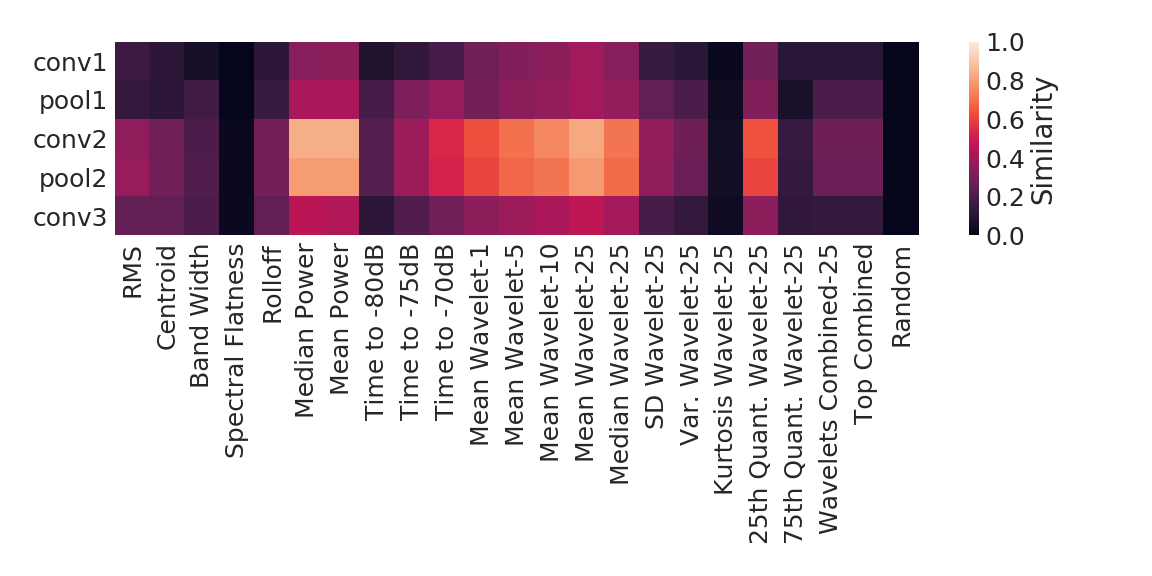}
\caption{Regular}\label{default}
\end{subfigure}
\begin{subfigure}[b]{0.455\linewidth}
\centering
\includegraphics[width=\linewidth]{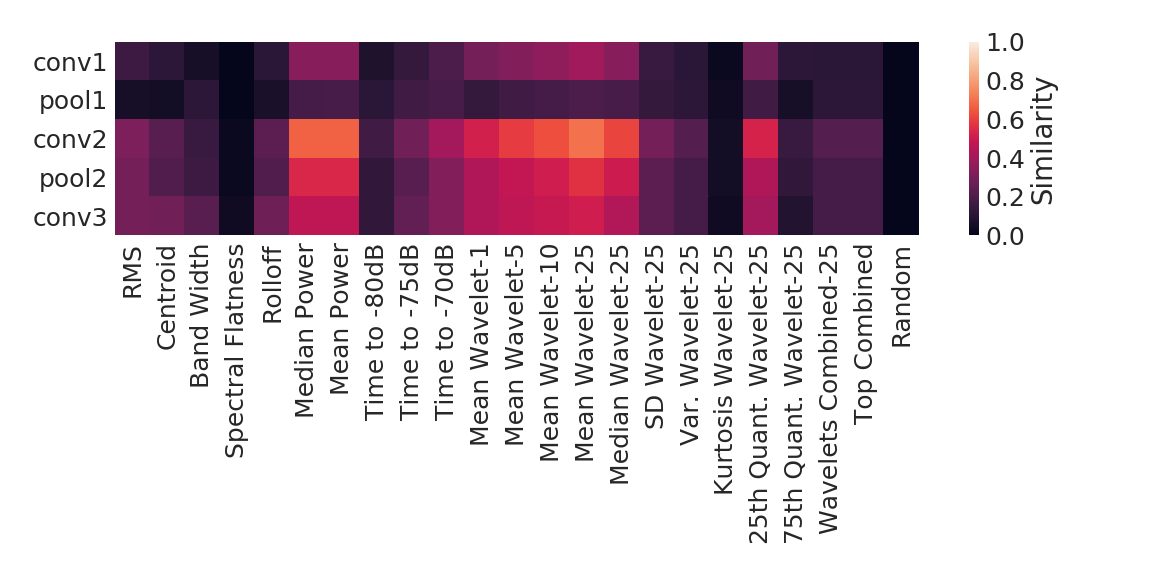}
\caption{Deformable}\label{default}
\end{subfigure}
\caption{Linear CKA similarity between the deep features from all layers for the Composer dataset and all hand-crafted features for the (a) Regular and (b) Deformable deep features.}
\label{fig:Composer-layer-sim}
\end{figure}

We next compare the similarity between the deep features from all layers to all of the hand-crafted features for the NSynth (Figure~\ref{fig:NSynth-layer-IF-sim}) and Composer (Figure~\ref{fig:Composer-layer-sim}) datasets for the Regular and Deformable convolutions.   The NSynth similarities trends hold whether the deep features are extracted from architectures trained to predict Instrument Family (Figure~\ref{fig:NSynth-layer-IF-sim}), Note Pitch (Figure~\ref{fig:NSynth-layer-Pitch-sim}) or Note Velocity (Figure~\ref{fig:NSynth-layer-Velocity-sim}), confirming the robustness of the deep features to multiple tasks.

\subsection{Additional Similarities}
\begin{figure}[h]
\centering
\begin{subfigure}[b]{0.325\linewidth}
\centering
\includegraphics[width=\linewidth]{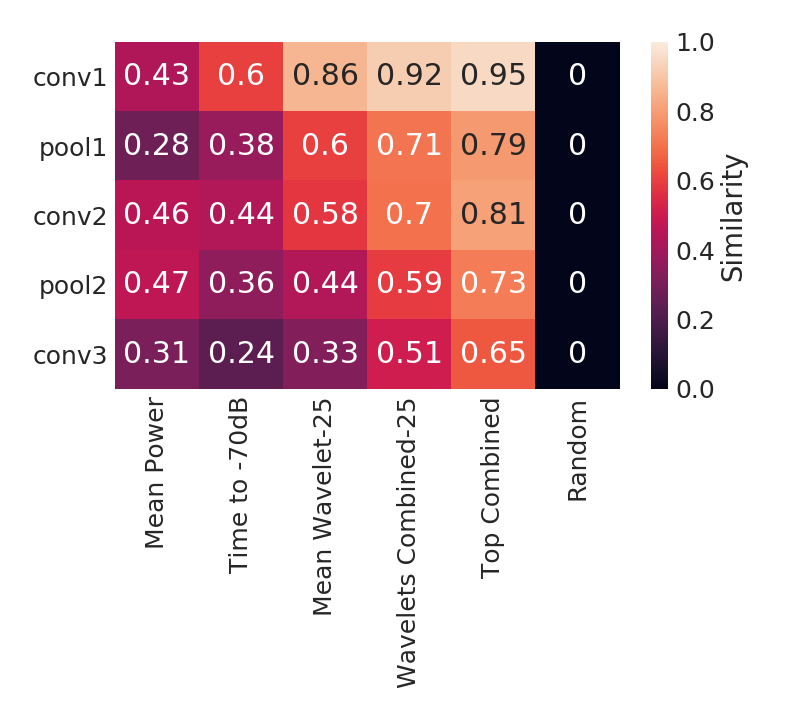}
\caption{Regular}
\end{subfigure}
\begin{subfigure}[b]{0.325\linewidth}
\centering
\includegraphics[width=\linewidth]{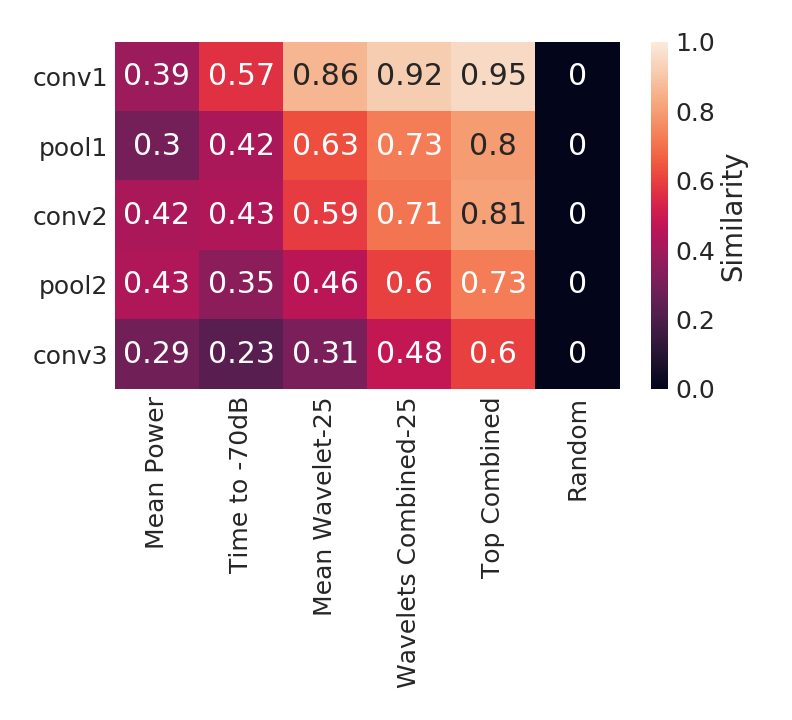}
\caption{Deformable}
\end{subfigure}
\caption{Linear Regression similarity between the deep features from each layer for the NSynth dataset for the (a) Regular and (b) Deformable architectures. The earlier layers exhibit very high similarity with the hand-crafted features. }
\label{fig:sim-NSynth}
\end{figure}

\begin{figure*}[h]
\centering
\begin{subfigure}[b]{.325\linewidth}
\includegraphics[width=\linewidth]{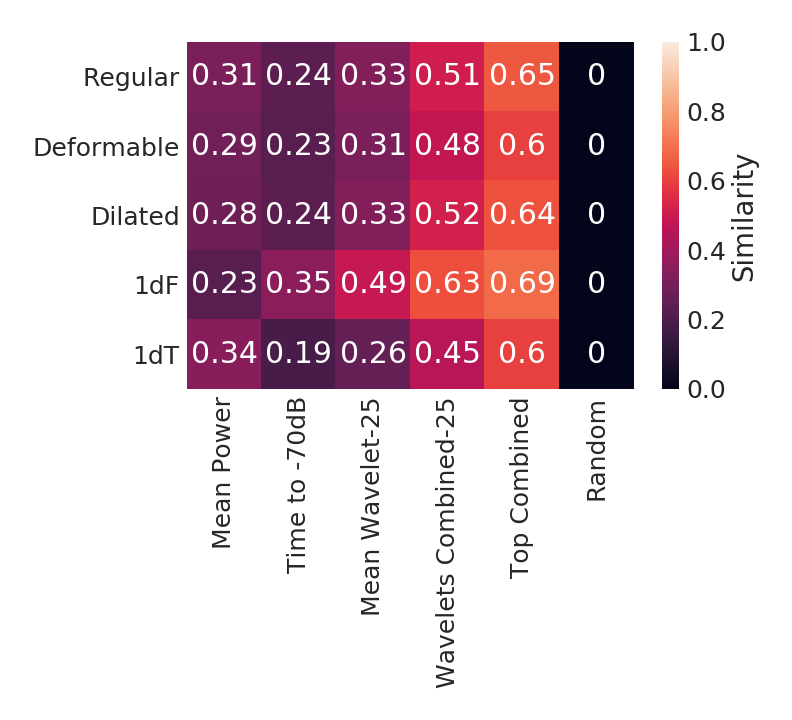}
\caption{NSynth}
\end{subfigure}
\begin{subfigure}[b]{.325\linewidth}
\includegraphics[width=\linewidth]{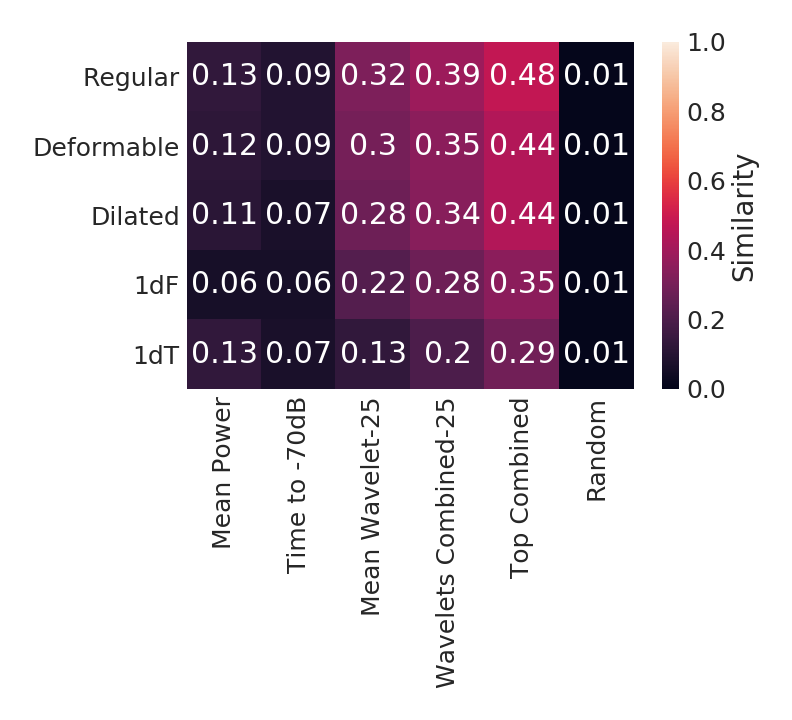}
\caption{Composer}
\end{subfigure}
\begin{subfigure}[b]{.325\linewidth}
\includegraphics[width=\linewidth]{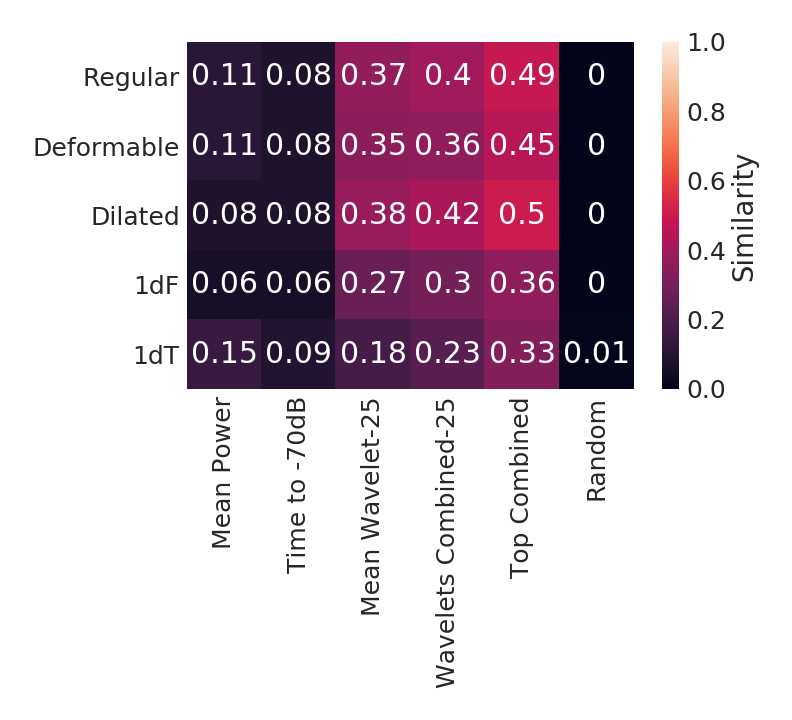}
\caption{Beethoven}
\end{subfigure}
\caption{Linear Regression similarity between the deep features from each architecture from the last convolutional layer (\texttt{conv3}) with the top hand-crafted features for the (a) NSynth, (b) Composer and (c) Beethoven datasets.}
\label{fig:sim-arch}
\end{figure*}

We also look at the Linear Regression similarity defined in \citet{pmlr-v97-kornblith19a} and Equation~\ref{eq:lr-sim} for all datasets.  The results across architectures again show high similarity between the deep features and the hand-crafted features, though there are differences to the plots with Linear CKA as the similarity (Figure~\ref{fig:sim-arch}), and similarly when comparing deep features across layers (Figure~\ref{fig:sim-NSynth}).  Linear CKA is found to be the best similarity measure by \citet{pmlr-v97-kornblith19a}, which is why it is the focus of the main paper, though the similarities are expected to change with the similarity measure used.

\begin{figure*}[t]
\centering
\begin{subfigure}[b]{.32\linewidth}
\includegraphics[width=\linewidth]{Plots/Orchestral/arch-spec-avg-top5}
\caption{Linear CKA}
\end{subfigure}
\begin{subfigure}[b]{.32\linewidth}
\includegraphics[width=\linewidth]{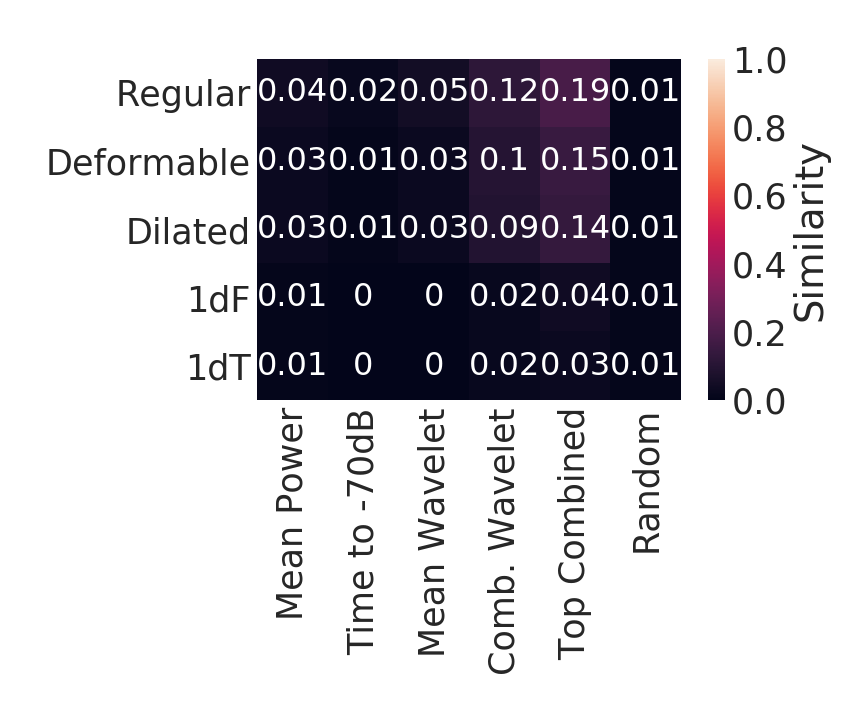}
\caption{$R^2_{CCA}$}
\end{subfigure}
\begin{subfigure}[b]{.32\linewidth}
\includegraphics[width=\linewidth]{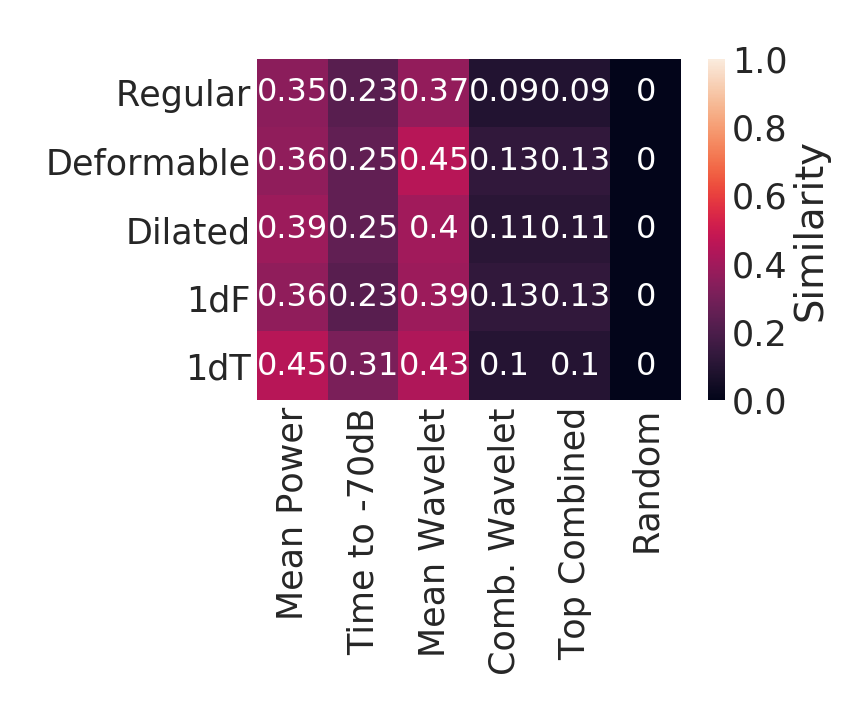}
\caption{$R^2_{SVCCA}$}
\end{subfigure}

\begin{subfigure}[b]{.32\linewidth}
\includegraphics[width=\linewidth]{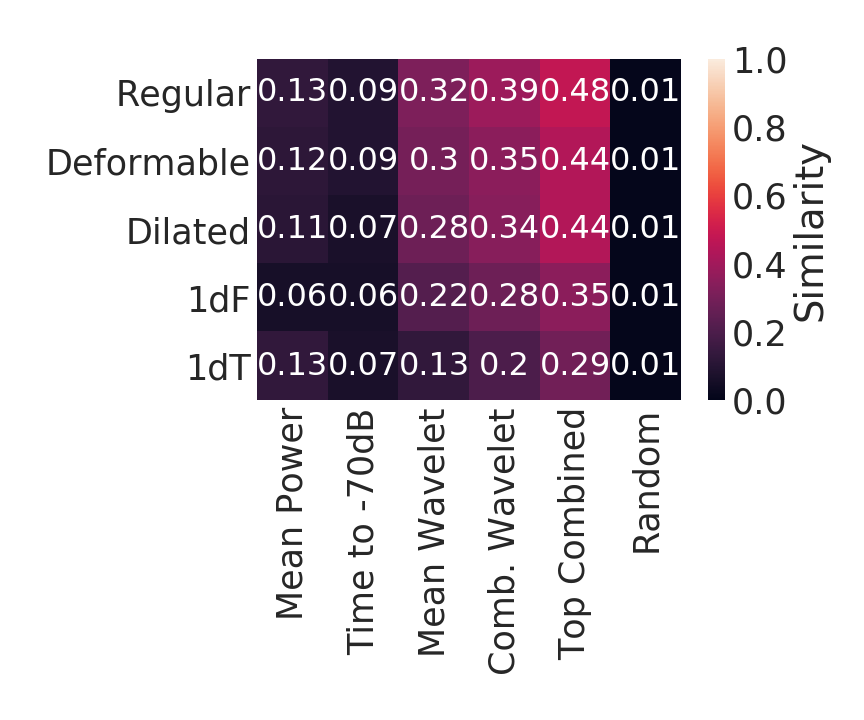}
\caption{Linear Regression, $R^2_{LR}$}
\end{subfigure}
\begin{subfigure}[b]{.32\linewidth}
\includegraphics[width=\linewidth]{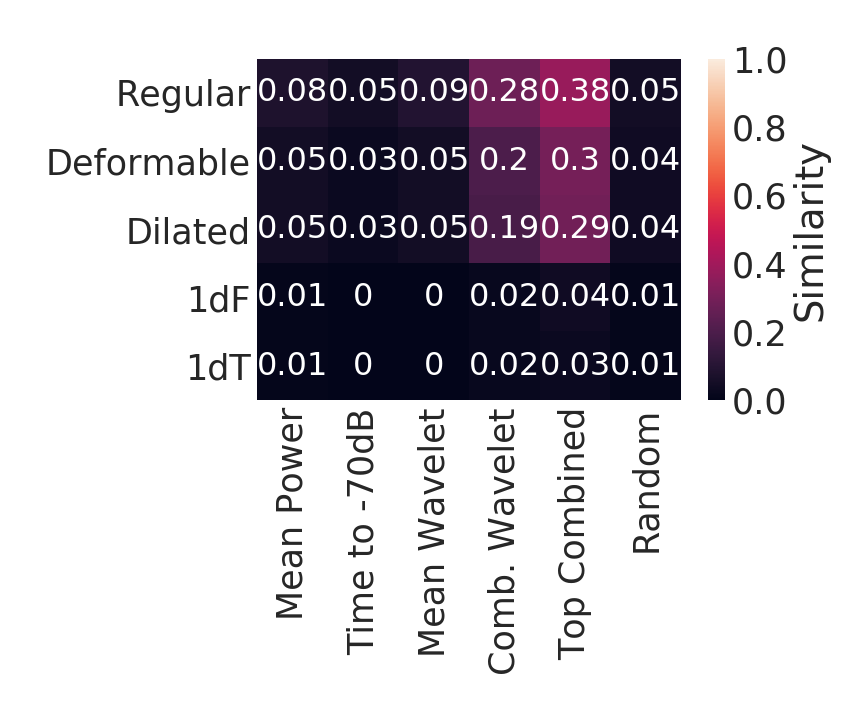}
\caption{$\bar{\rho}_{CCA}$}
\end{subfigure}
\begin{subfigure}[b]{.32\linewidth}
\includegraphics[width=\linewidth]{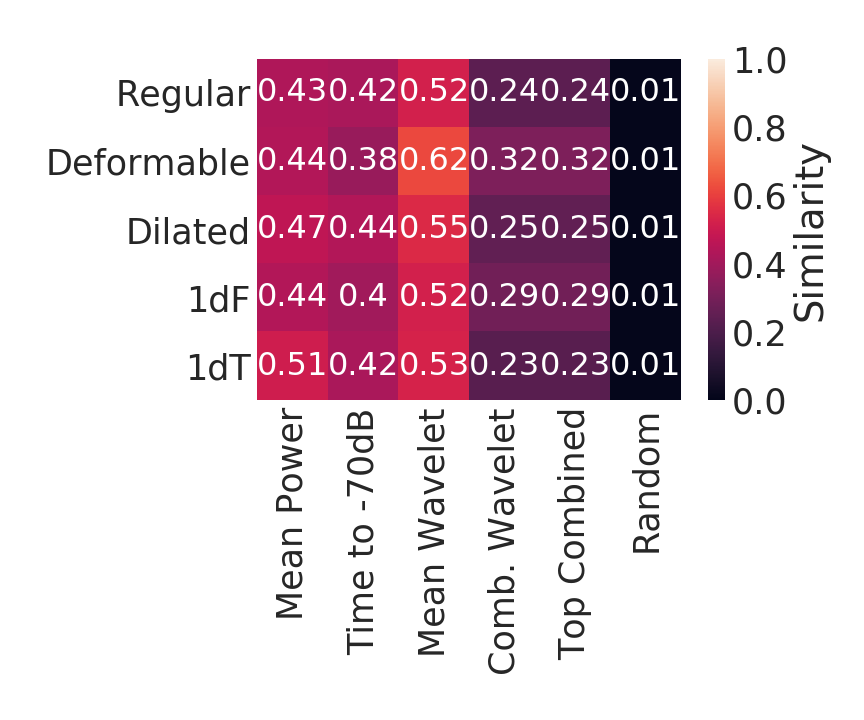}
\caption{$\bar{\rho}_{SVCCA}$}
\end{subfigure}
\caption{Similarity measures between the deep features from each architecture from the last convolutional layer (\texttt{conv3}) with the top hand-crafted features for the (a) Linear CKA, (b) $R^2_{CCA}$, (c) $R^2_{SVCCA}$, (d) Linear Regression, $R^2_{LR}$, (e) $\bar{\rho}_{CCA}$ and (f)  $\bar{\rho}_{SVCCA}$ similarities for the Composer dataset. We baseline with Random $\mathcal{N}(0,1)$ noise in the last column.  Plots are the similarity value averaged across the deep features from each initialization. }
\label{fig:sim-arch-Composer}
\end{figure*}

\begin{figure*}[t]
\centering
\begin{subfigure}[b]{.32\linewidth}
\includegraphics[width=\linewidth]{Plots/Orchestral/layer-spec-init-Regular-top5}
\caption{Linear CKA}
\end{subfigure}
\begin{subfigure}[b]{.32\linewidth}
\includegraphics[width=\linewidth]{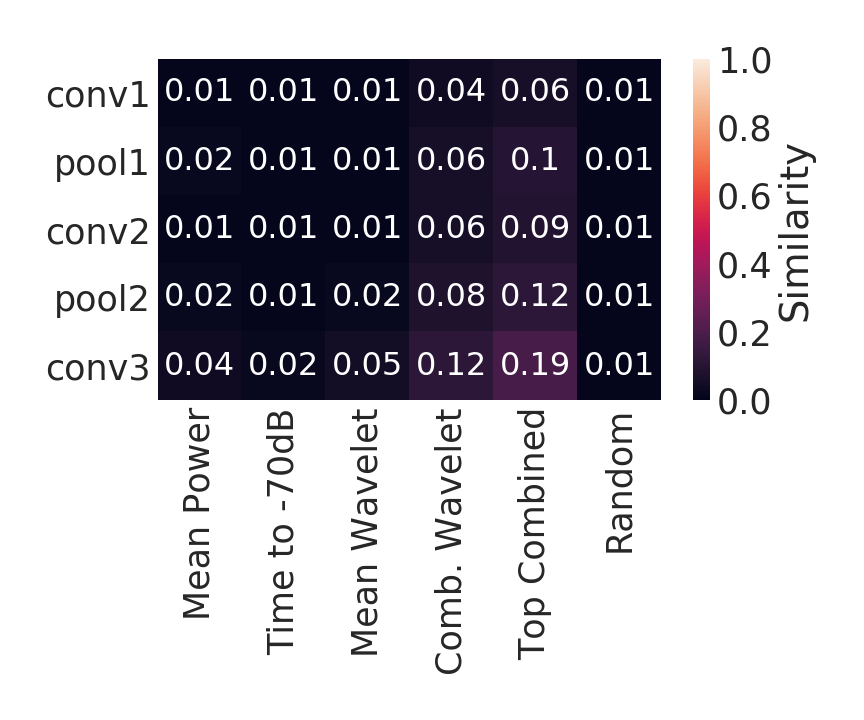}
\caption{$R^2_{CCA}$}
\end{subfigure}
\begin{subfigure}[b]{.32\linewidth}
\includegraphics[width=\linewidth]{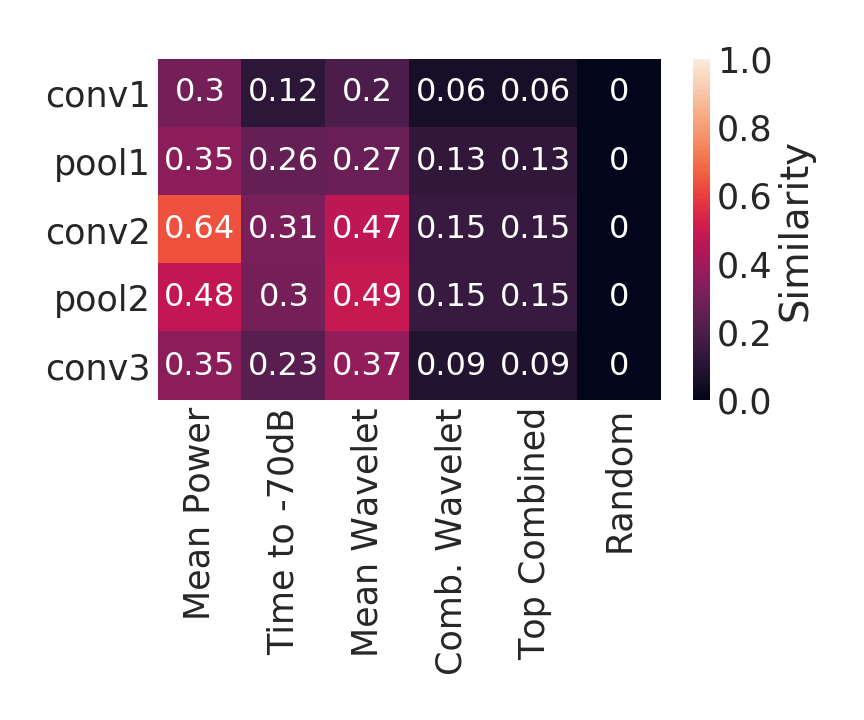}
\caption{$R^2_{SVCCA}$}
\end{subfigure}

\begin{subfigure}[b]{.32\linewidth}
\includegraphics[width=\linewidth]{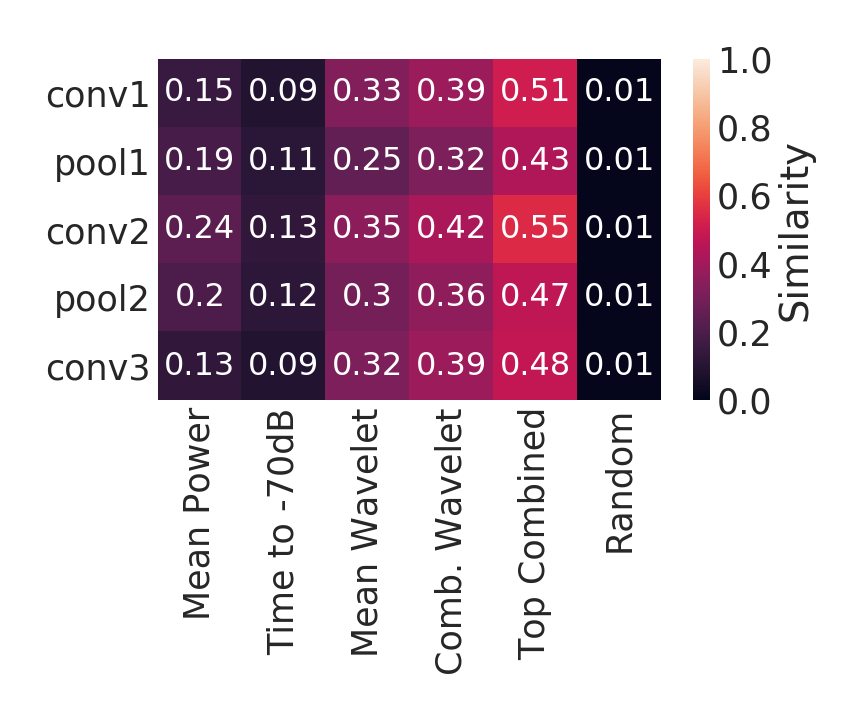}
\caption{Linear Regression, $R^2_{LR}$}
\end{subfigure}
\begin{subfigure}[b]{.32\linewidth}
\includegraphics[width=\linewidth]{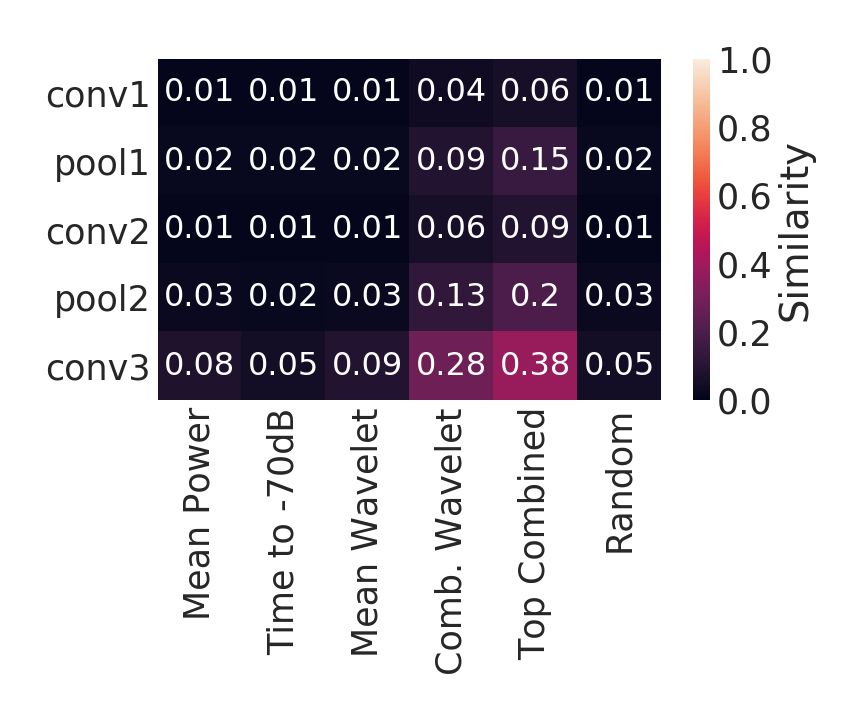}
\caption{$\bar{\rho}_{CCA}$}
\end{subfigure}
\begin{subfigure}[b]{.32\linewidth}
\includegraphics[width=\linewidth]{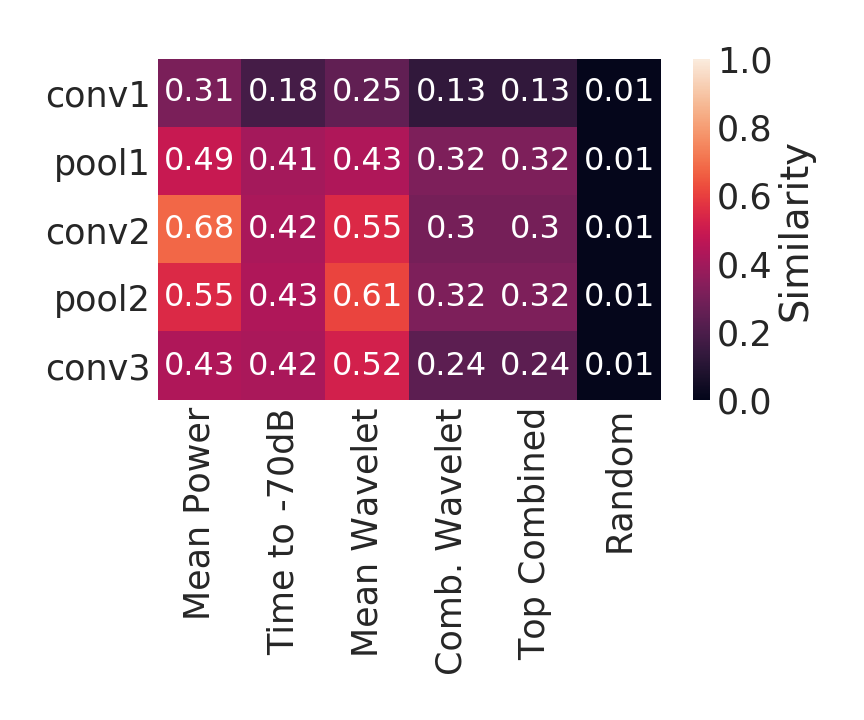}
\caption{$\bar{\rho}_{SVCCA}$}
\end{subfigure}
\caption{Similarity measures between the deep features from each layer for the Regular architecture with the top hand-crafted features for the (a) Linear CKA, (b) $R^2_{CCA}$, (c) $R^2_{SVCCA}$, (d) Linear Regression, $R^2_{LR}$, (e) $\bar{\rho}_{CCA}$ and (f)  $\bar{\rho}_{SVCCA}$ similarities for the Composer dataset. }
\label{fig:layer-spec-Composer-Regular}
\end{figure*}

\begin{figure*}[t]
\centering
\begin{subfigure}[b]{.32\linewidth}
\includegraphics[width=\linewidth]{Plots/Orchestral/layer-spec-init-Deformable-top5}
\caption{Linear CKA}
\end{subfigure}
\begin{subfigure}[b]{.32\linewidth}
\includegraphics[width=\linewidth]{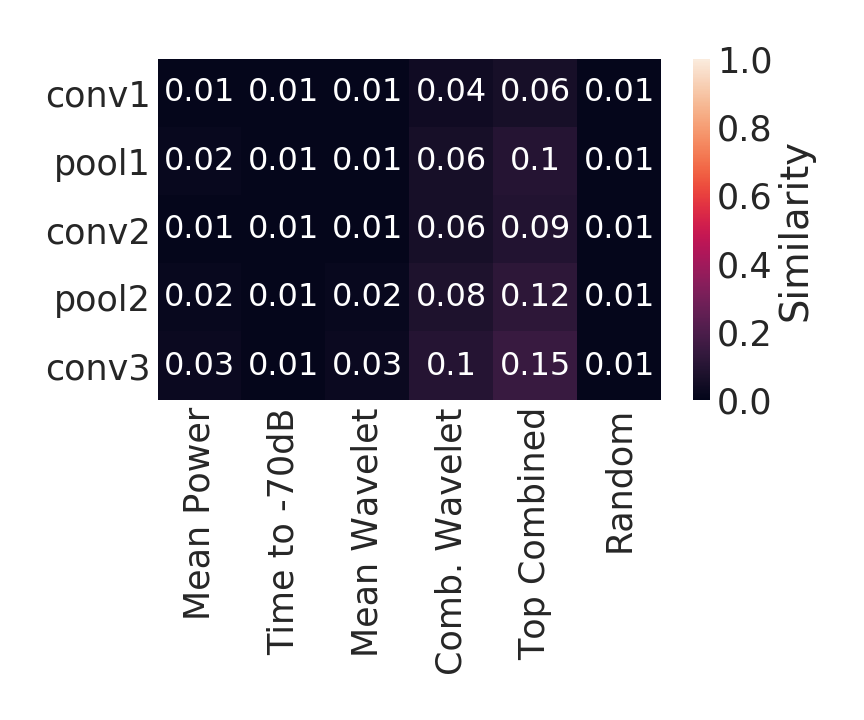}
\caption{$R^2_{CCA}$}
\end{subfigure}
\begin{subfigure}[b]{.32\linewidth}
\includegraphics[width=\linewidth]{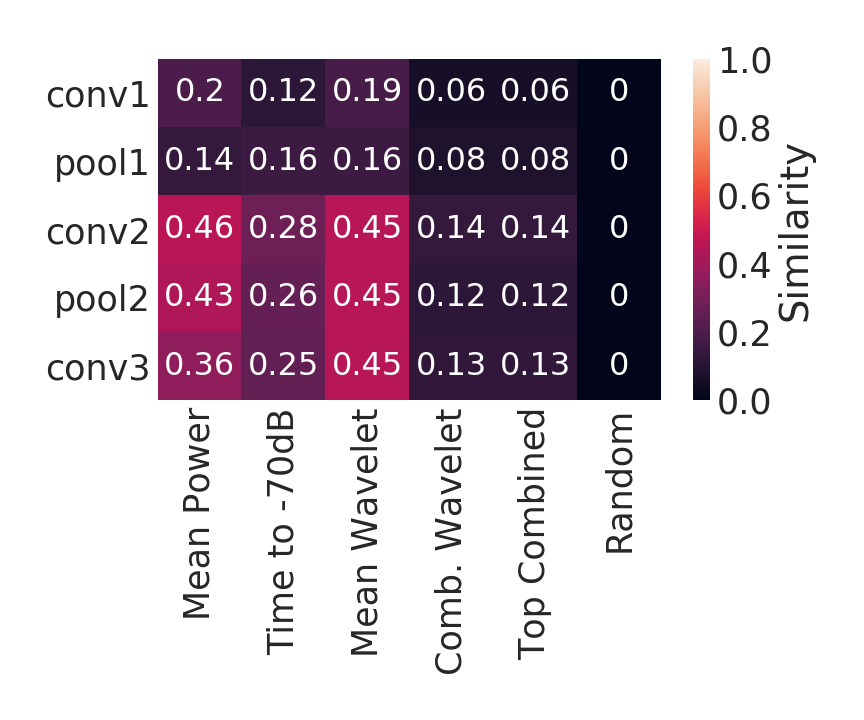}
\caption{$R^2_{SVCCA}$}
\end{subfigure}

\begin{subfigure}[b]{.32\linewidth}
\includegraphics[width=\linewidth]{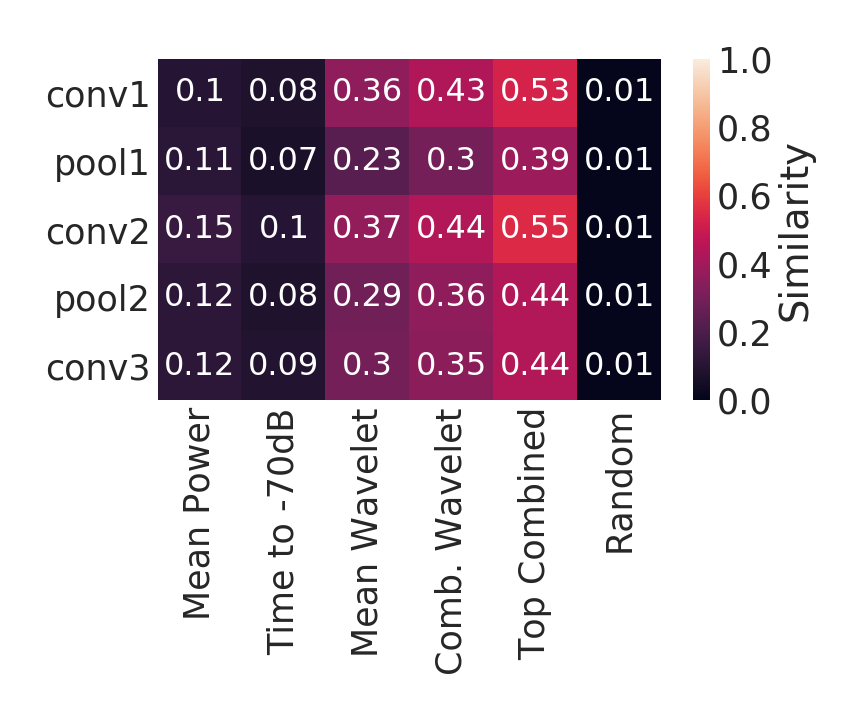}
\caption{Linear Regression, $R^2_{LR}$}
\end{subfigure}
\begin{subfigure}[b]{.32\linewidth}
\includegraphics[width=\linewidth]{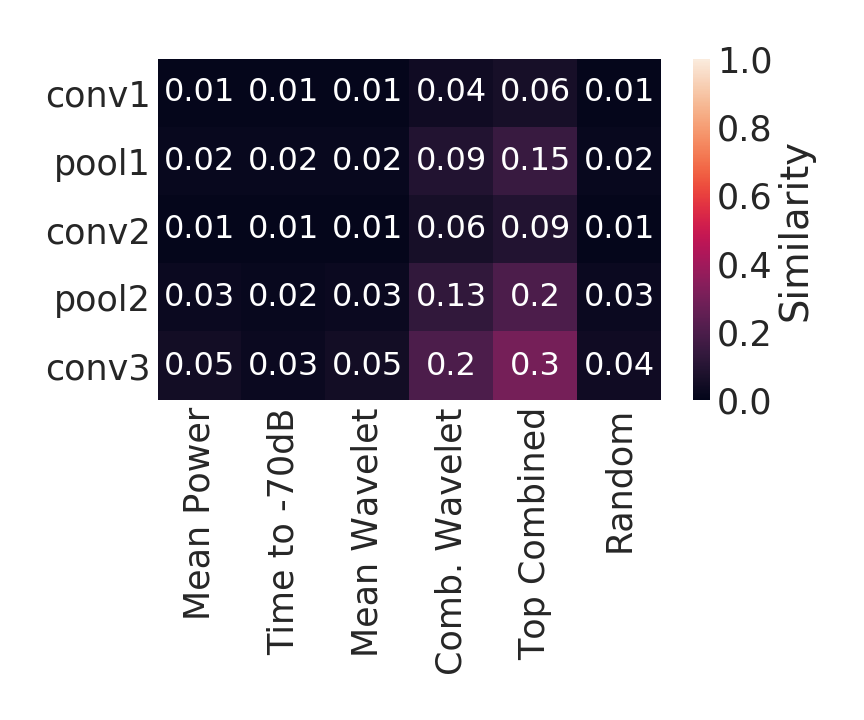}
\caption{$\bar{\rho}_{CCA}$}
\end{subfigure}
\begin{subfigure}[b]{.32\linewidth}
\includegraphics[width=\linewidth]{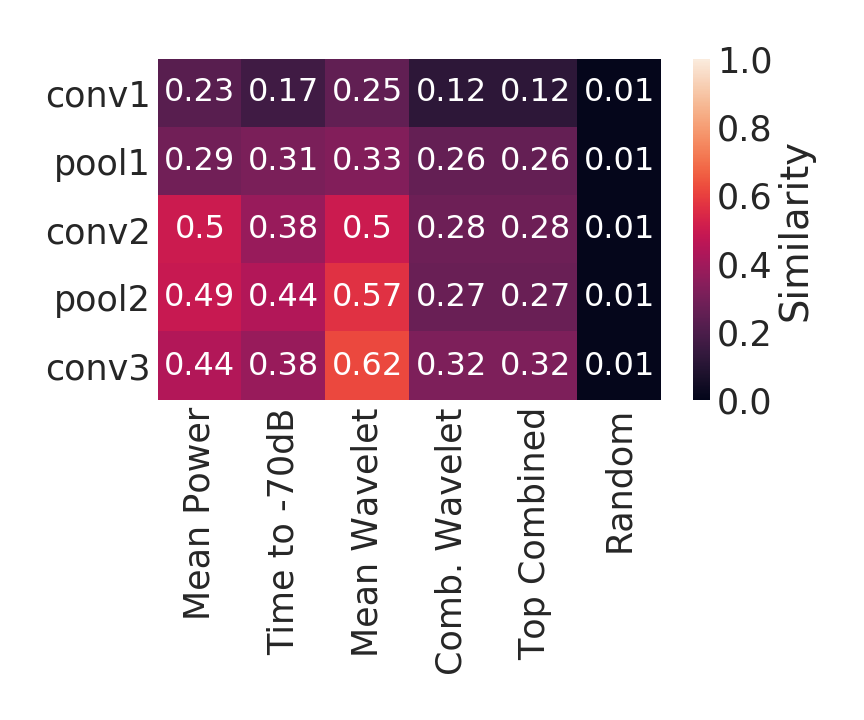}
\caption{$\bar{\rho}_{SVCCA}$}
\end{subfigure}
\caption{Similarity measures between the deep features from each layer for the Deformable architecture with the top hand-crafted features for the (a) Linear CKA, (b) $R^2_{CCA}$, (c) $R^2_{SVCCA}$, (d) Linear Regression, $R^2_{LR}$, (e) $\bar{\rho}_{CCA}$ and (f)  $\bar{\rho}_{SVCCA}$ similarities for the Composer dataset. }
\label{fig:layer-spec-Composer-Deformable}
\end{figure*}

Finally, for the Composer dataset, we compare the Linear CKA similarity used in the main paper to the linear regression, $R^2_{CCA}$, $\bar{\rho}_{CCA}$, $R^2_{SVCCA}$ and $\bar{\rho}_{SVCCA}$ similarities defined in Equations~\ref{eq:cca-sim} and \ref{eq:svcca-sim} above.  Again, these results confirm the findings in \citet{pmlr-v97-kornblith19a}: the linear regression and SVCCA similarity measures give similar results to the Linear CKA measure, while the CCA similarities appear less useful overall.  Similarities are calculated across architectures for \texttt{conv3} (Figure~\ref{fig:sim-arch-Composer}), and for all layers for the Regular (Figure~\ref{fig:layer-spec-Composer-Regular}) and Deformable (Figure~\ref{fig:layer-spec-Composer-Deformable}) architectures. 

\clearpage\newpage
\section{Do Untrained Convolutional Architectures Have a Useful Inductive Bias for Music Audio Discriminative Tasks?}\label{suppsec:experiment_3}

\begin{figure*}
\centering
\begin{subfigure}[b]{.325\linewidth}
\includegraphics[width=\linewidth]{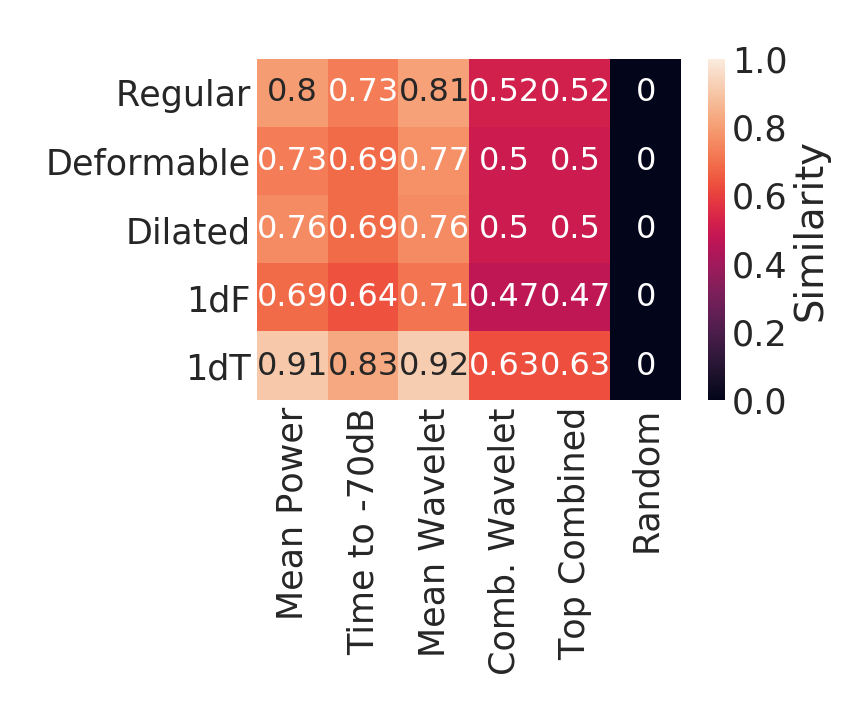}
\caption{NSynth}
\end{subfigure}
\begin{subfigure}[b]{.325\linewidth}
\includegraphics[width=\linewidth]{Plots/Orchestral/Untrained_arch-spec-init-all-top5}
\caption{Composer}
\end{subfigure}
\begin{subfigure}[b]{.325\linewidth}
\includegraphics[width=\linewidth]{Plots/Beethoven/Untrained_arch-spec-init-all-top5}
\caption{Beethoven}
\end{subfigure}
\caption{Linear CKA similarity between the \textbf{untrained} deep features from each architecture from the last convolutional layer (\texttt{conv3}) with the top hand-crafted features for the (a) NSynth, (b) Composer and (c) Beethoven datasets.}
\label{fig:untrained-arch-sim}
\end{figure*}

\begin{figure*}
\centering
\begin{subfigure}[b]{.325\linewidth}
\includegraphics[width=\linewidth]{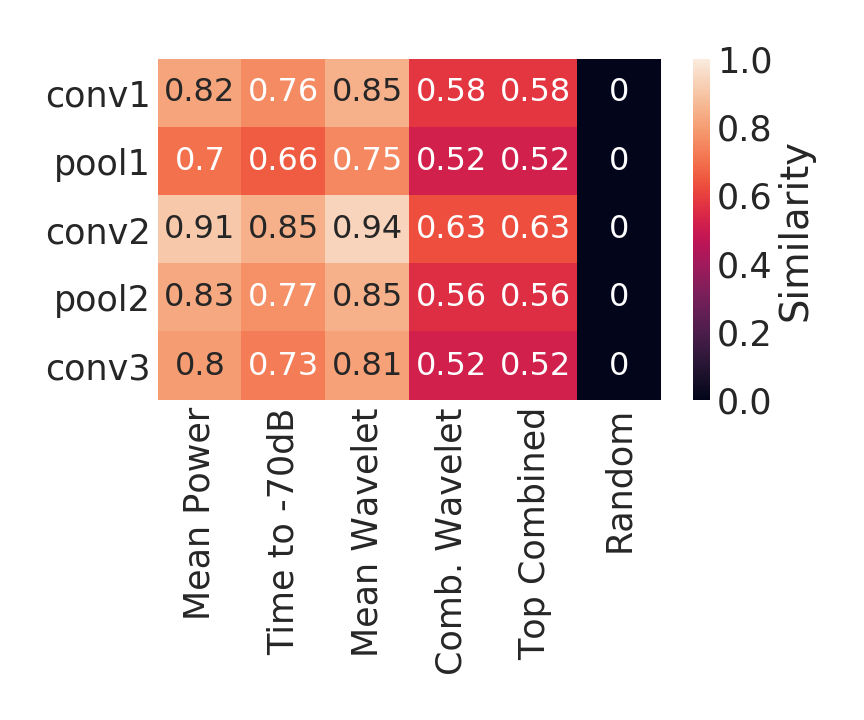}
\caption{Regular - NSynth}
\end{subfigure}
\begin{subfigure}[b]{.325\linewidth}
\includegraphics[width=\linewidth]{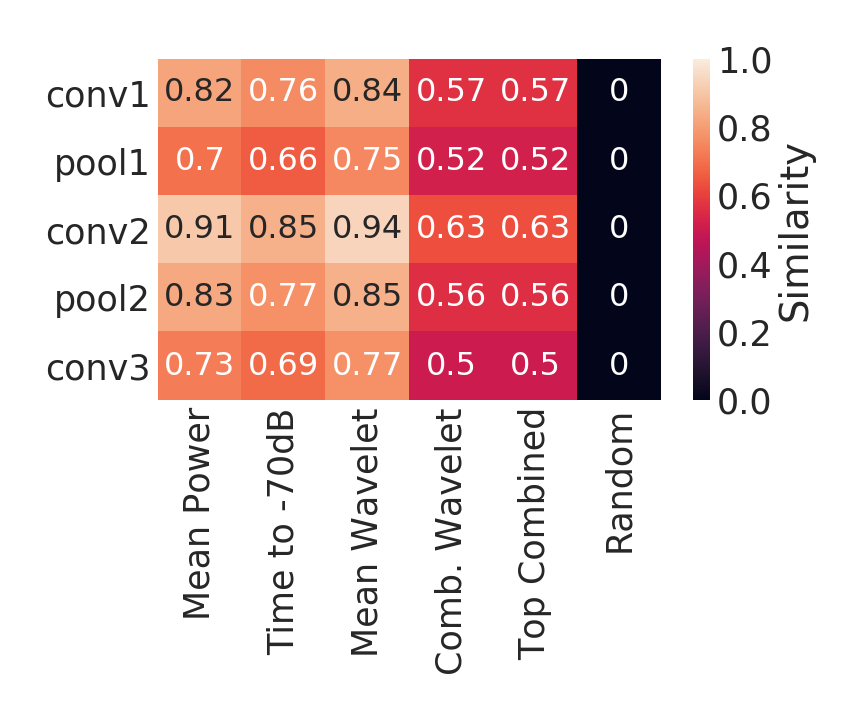}
\caption{Deformable - NSynth}
\end{subfigure}
\begin{subfigure}[b]{.325\linewidth}
\includegraphics[width=\linewidth]{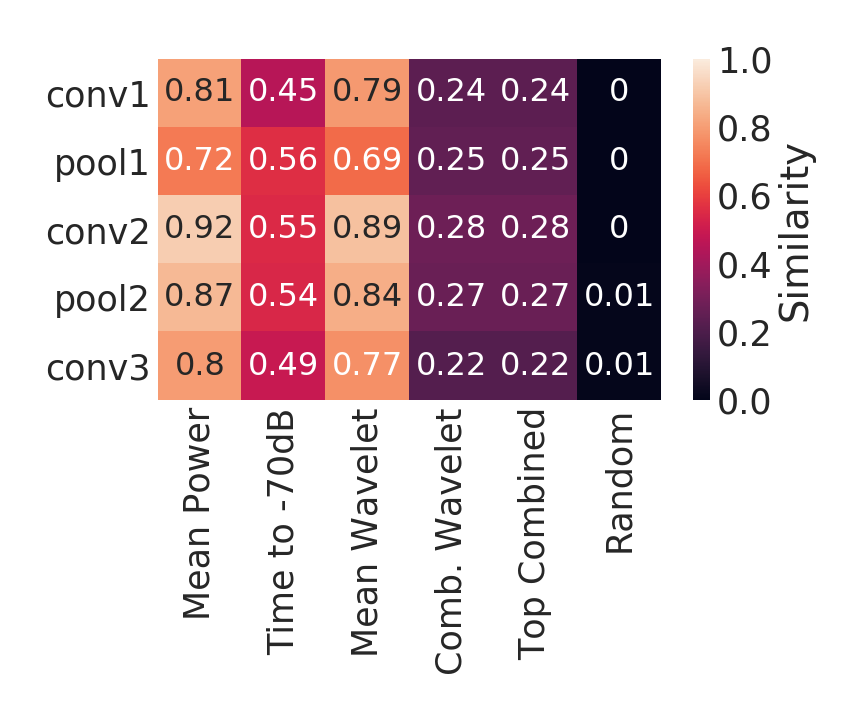}
\caption{Regular - Composer}\label{fig:untrained-Composer-sim}
\end{subfigure}
\caption{Linear CKA similarity between the \textbf{untrained} deep features from each layer for the NSynth dataset for the (a) Regular and (b) Deformable architectures, and for the Composer dataset for the (c) Regular architecture. The earlier layers exhibit very high similarity with the hand-crafted features. }
\label{fig:untrained-NSynth-sim}
\end{figure*}

For this experiment, we explore how well features from the \textbf{untrained} deep architectures can perform the main classification tasks and how similar the untrained features are to the hand-crafted features.  In Figure~\ref{fig:untrained-class} we look at how well each layer for the Regular and Deformable untrained architectures performs for the main classification tasks. We find that for both the NSynth and Composer datasets, all untrained layers are able to classify well above random guessing, and that only the later layers improve in accuracy after training.   This again confirms earlier results that the last convolutional layers are the most accurate and extract higher-level concepts than the earlier layers. Next, in Figure~\ref{fig:untrained-arch-sim} we compare the untrained deep features from the last convolutional layer to the hand-crafted features for all architectures.  Across all datasets, the untrained deep features are \textbf{more} similar to the hand-crafted features than the trained deep features in  Figure~\ref{fig:sim-arch}. The untrained deep features from the earlier convolutional layers for the NSynth (Figure~\ref{fig:untrained-NSynth-sim}) and Composer datasets (Figure~\ref{fig:untrained-Composer-sim}) are especially similar to the hand-crafted features, while the later layers become less similar to the hand-crafted features after training.



\begin{figure*}[h]
\centering
\begin{subfigure}[b]{.32\linewidth}
\includegraphics[width=\linewidth]{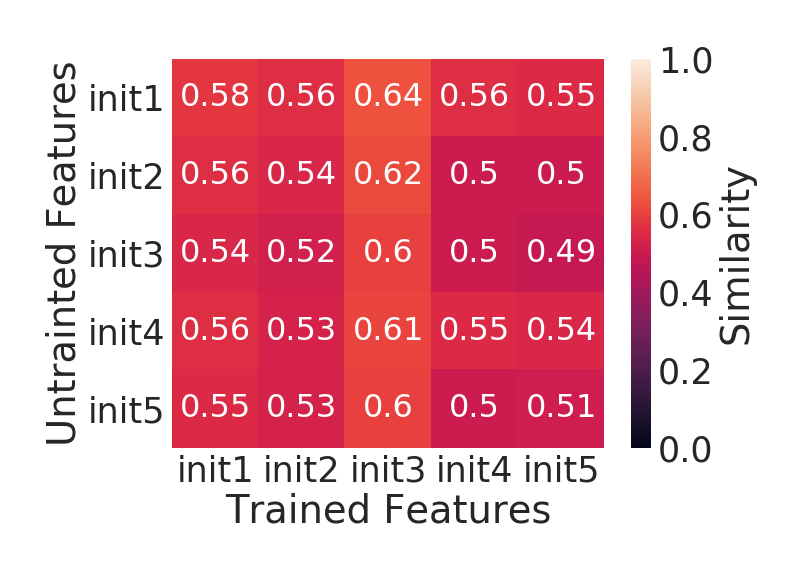}
\caption{NSynth}
\end{subfigure}
\begin{subfigure}[b]{.32\linewidth}
\includegraphics[width=\linewidth]{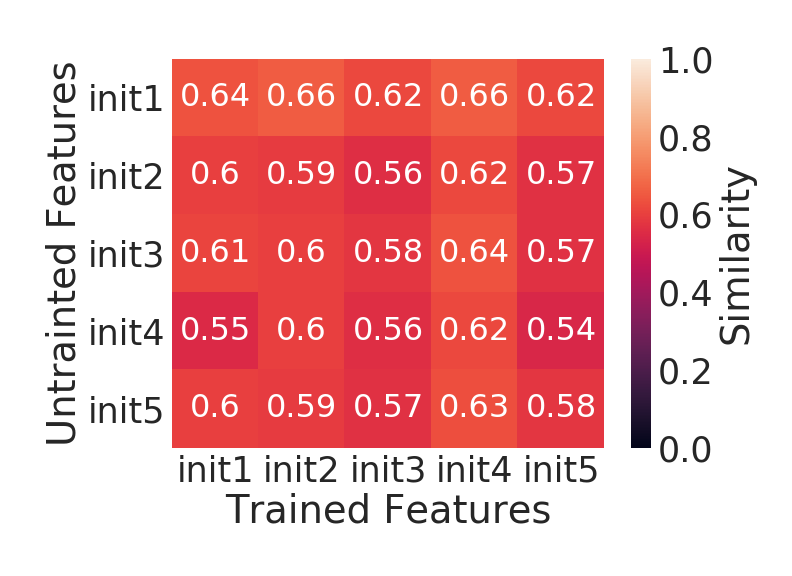}
\caption{Composer}
\end{subfigure}
\begin{subfigure}[b]{.32\linewidth}
\includegraphics[width=\linewidth]{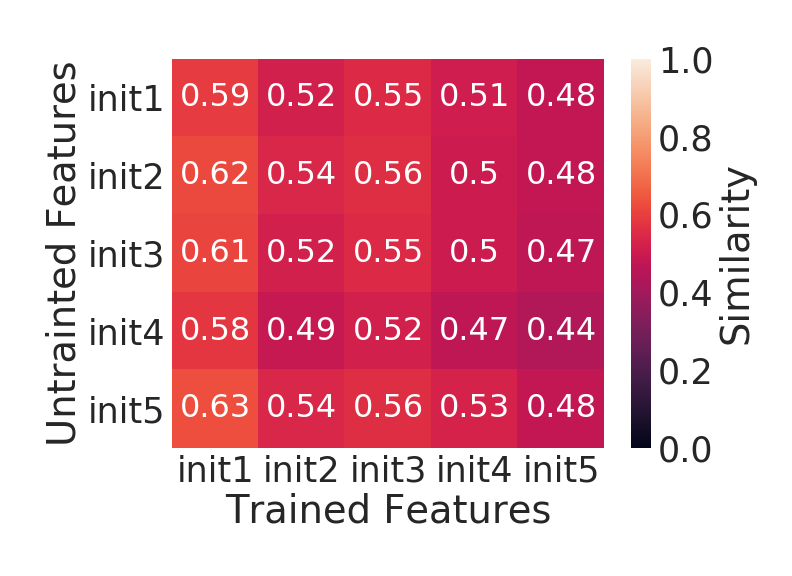}
\caption{Beethoven}
\end{subfigure}

\caption{Linear CKA similarity between untrained and trained features from the last convolutional layer (\texttt{conv3}) of the Regular architectures by initialization for the (a) NSynth, (b) Composer and (c) Beethoven datasets.  Across initializations, the untrained features are similar to the trained features, though not identical. }
\label{fig:untrain-train-Regular}
\end{figure*}

\begin{figure*}[h]
\centering
\begin{subfigure}[b]{.32\linewidth}
\includegraphics[width=\linewidth]{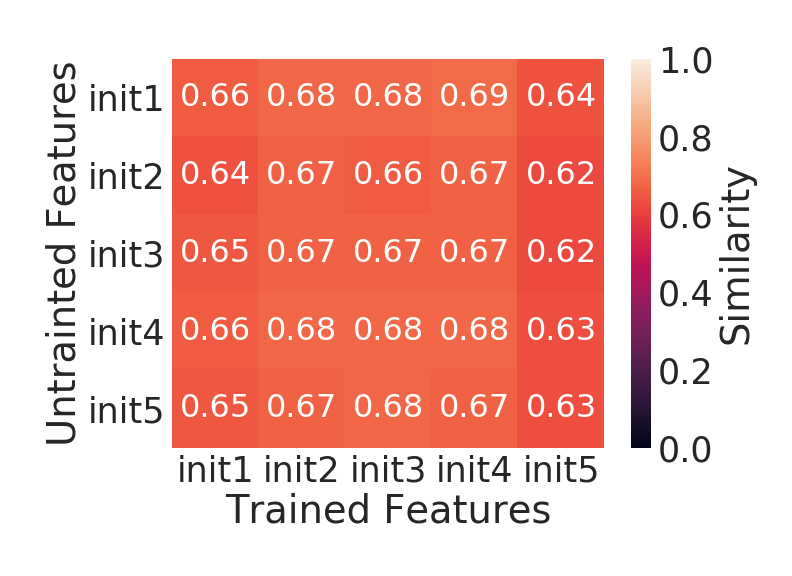}
\caption{NSynth}
\end{subfigure}
\begin{subfigure}[b]{.32\linewidth}
\includegraphics[width=\linewidth]{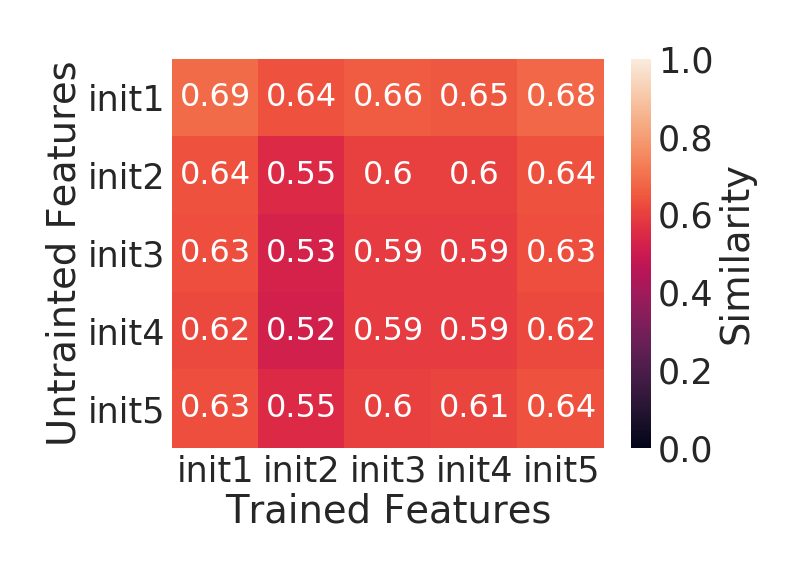}
\caption{Composer}
\end{subfigure}
\begin{subfigure}[b]{.32\linewidth}
\includegraphics[width=\linewidth]{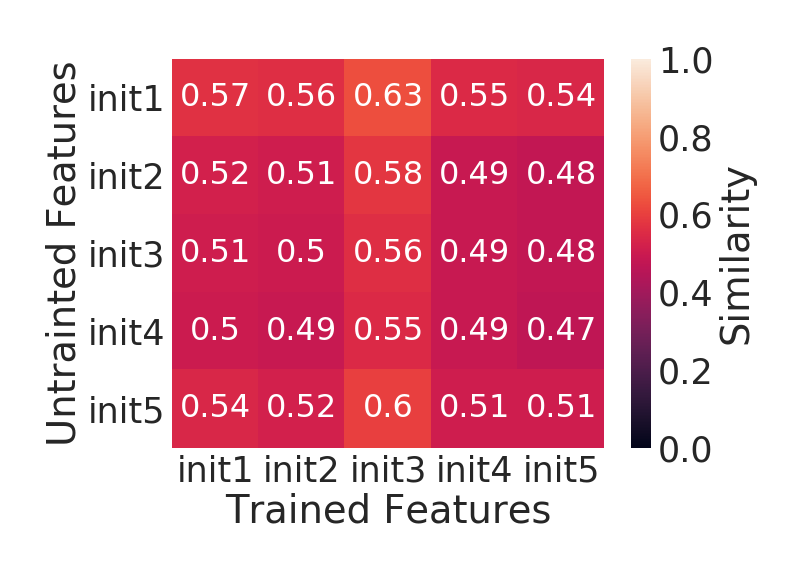}
\caption{Beethoven}
\end{subfigure}

\caption{Linear CKA similarity between untrained and trained features from the last convolutional layer (\texttt{conv3}) of the Deformable architectures by initialization for the (a) NSynth, (b) Composer and (c) Beethoven datasets.  Across initializations, the untrained features are similar to the trained features, though not identical. }
\label{fig:untrain-train-Deformable}
\end{figure*}

\begin{figure}[!htb]
\begin{center}
\includegraphics[width=0.45\textwidth]{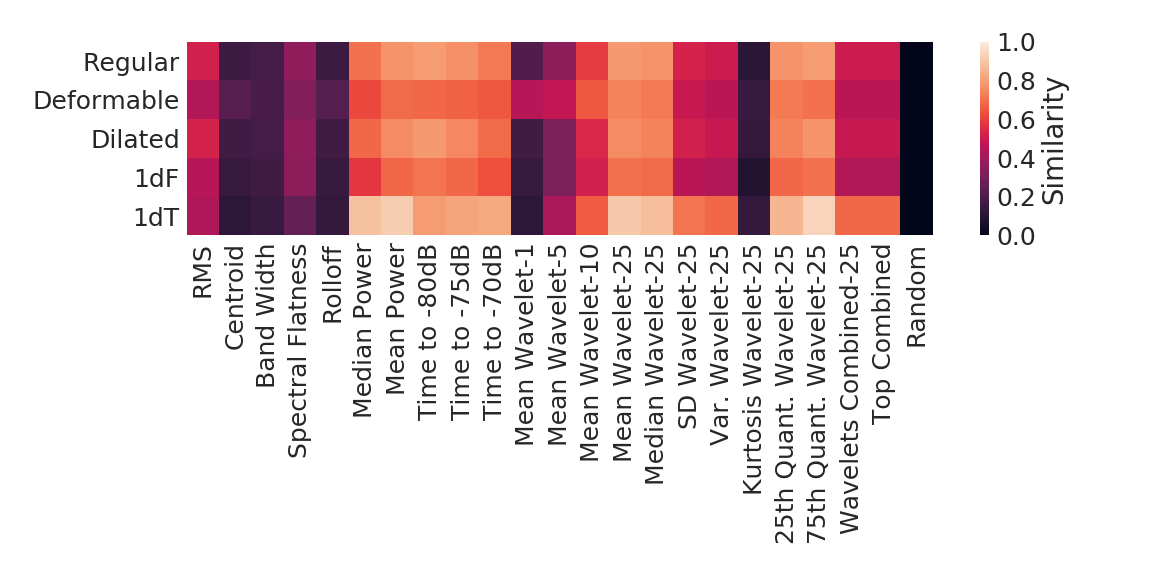}
\caption{Linear CKA similarity between the \textbf{untrained} deep features from each architecture from the last convolutional layer (\texttt{conv3}) with all hand-crafted features for the NSynth dataset.}
\label{fig:NSynth-sim-all-untrained}
\end{center}
\end{figure}

\begin{figure}[!htb]
\begin{center}
\includegraphics[width=0.45\textwidth]{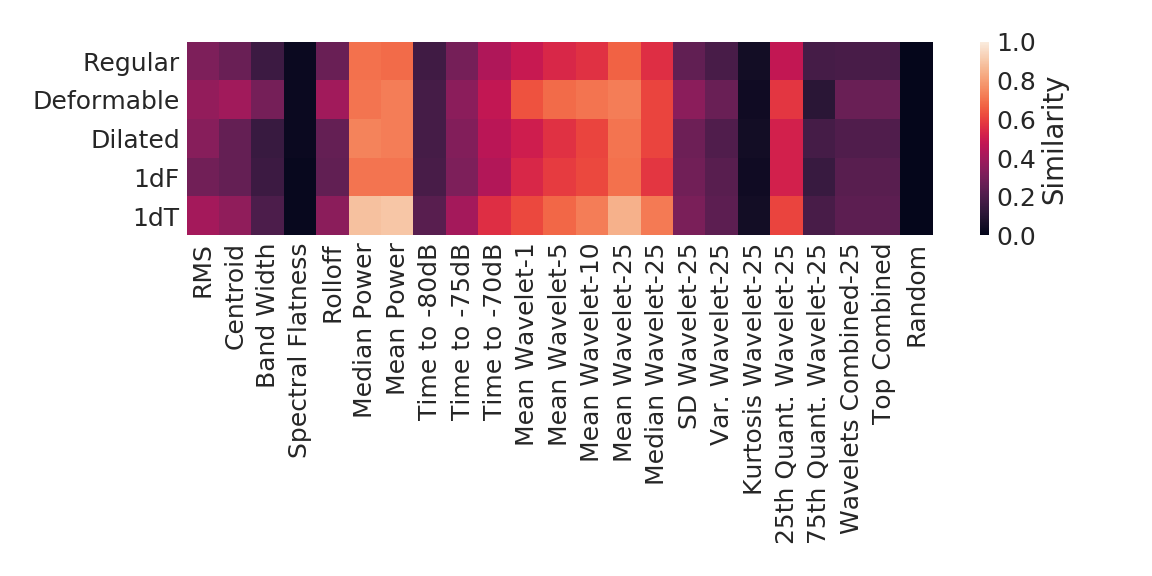}
\caption{Linear CKA similarity between the \textbf{untrained} deep features from each architecture from the last convolutional layer (\texttt{conv3}) with all hand-crafted features for the Composer dataset. }
\label{fig:Composer-sim-all-untrained}
\end{center}
\end{figure}

\begin{figure}[!htb]
\begin{center}
\includegraphics[width=0.45\textwidth]{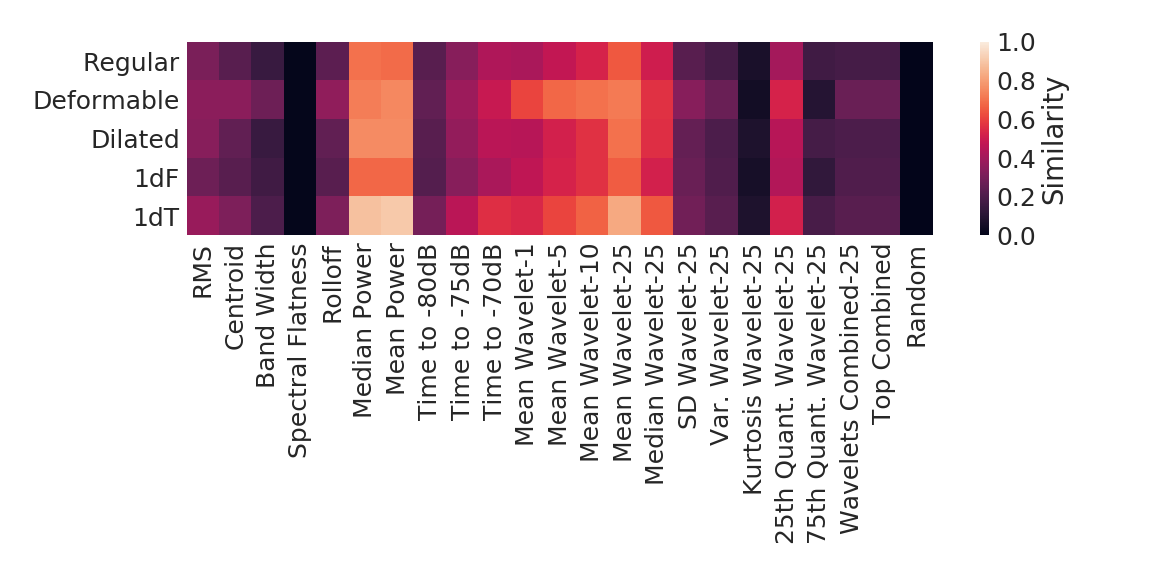}
\caption{Linear CKA similarity between the \textbf{untrained} deep features from each architecture from the last convolutional layer (\texttt{conv3}) with all hand-crafted features for the Beethoven dataset.}
\label{fig:Beethoven-sim-all-untrained}
\end{center}
\end{figure}

\begin{figure}[!htb]
\centering
\begin{subfigure}[b]{0.455\linewidth}
\includegraphics[width=\linewidth]{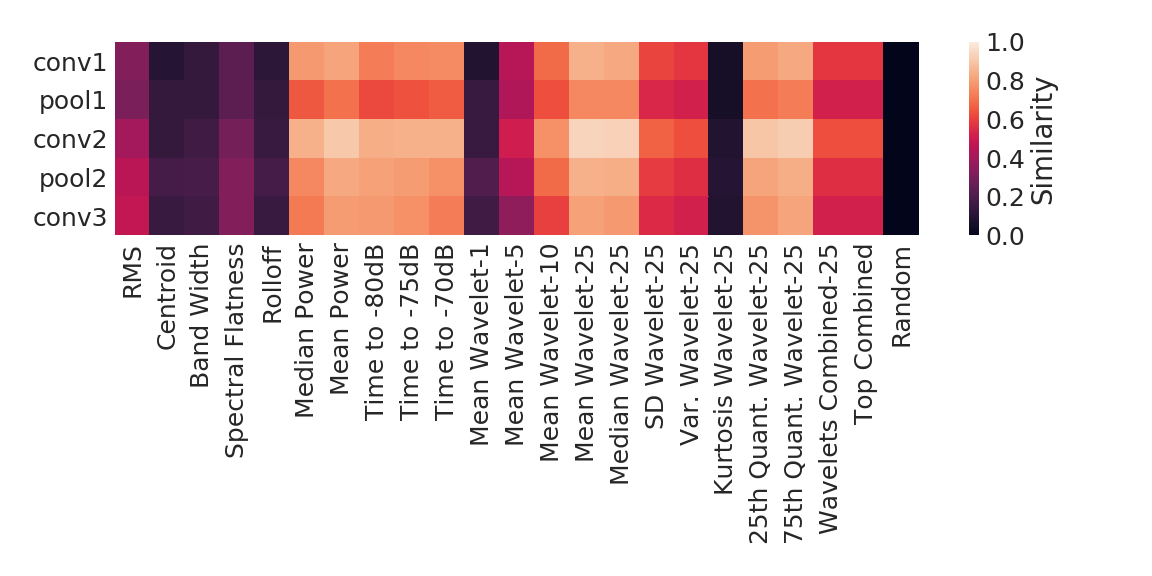}
\caption{Regular}\label{default}
\end{subfigure}
\begin{subfigure}[b]{0.455\linewidth}
\includegraphics[width=\linewidth]{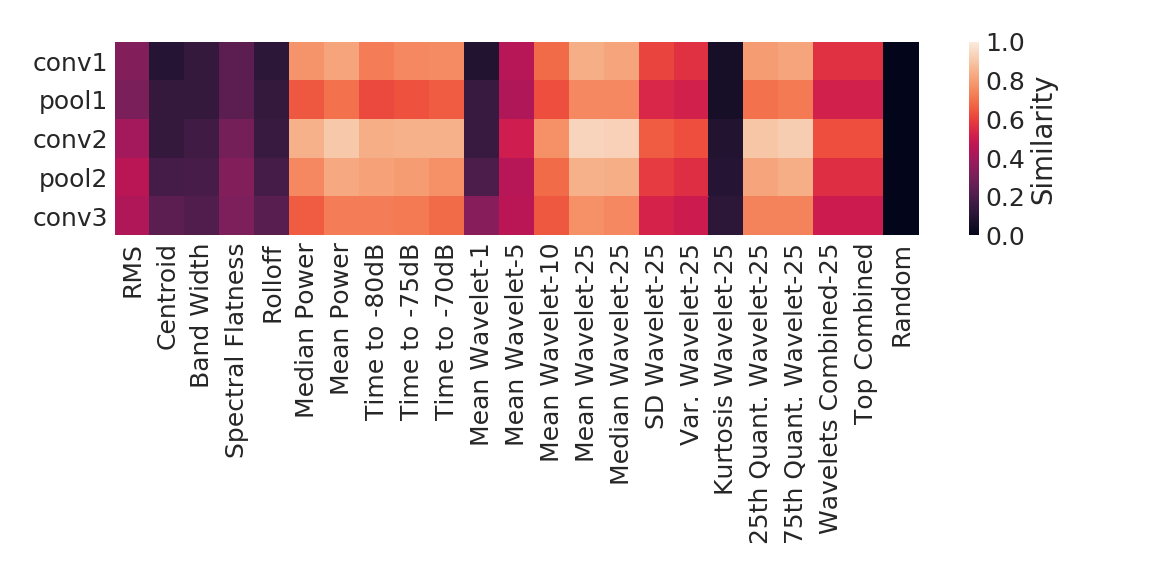}
\caption{Deformable}\label{default}
\end{subfigure}
\caption{Linear CKA similarity between the \textbf{untrained} deep features from all layers for the NSynth dataset and all hand-crafted features for the (a) Regular and (b) Deformable deep features.}
\label{fig:NSynth-layer-sim-untrained}
\end{figure}

\begin{figure}[!htb]
\centering
\begin{subfigure}[b]{0.455\linewidth}
\includegraphics[width=\linewidth]{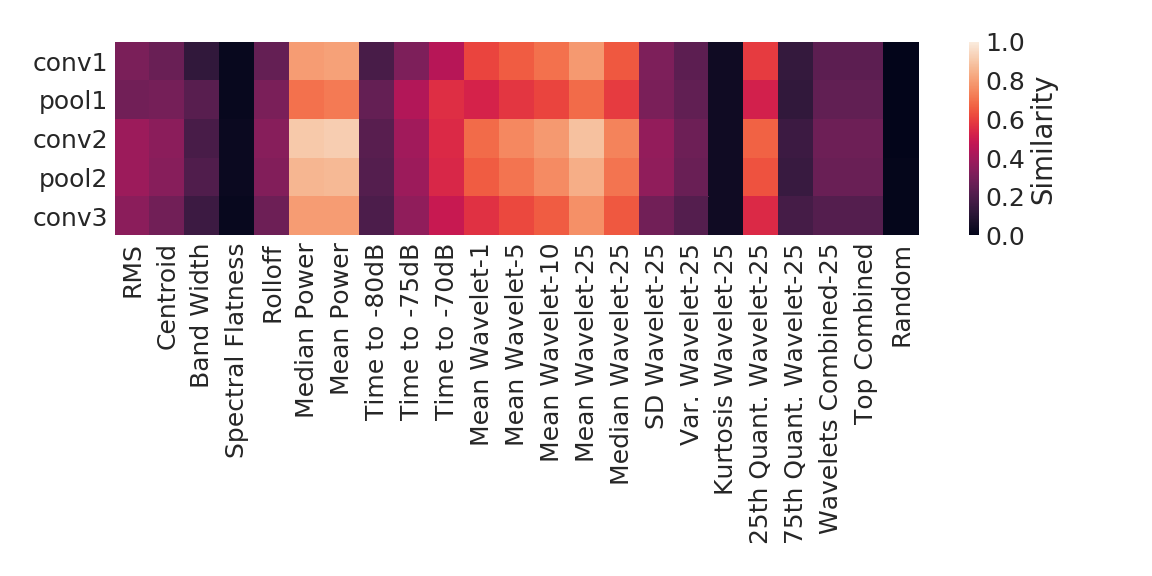}
\caption{Regular}\label{default}
\end{subfigure}
\begin{subfigure}[b]{0.455\linewidth}
\includegraphics[width=\linewidth]{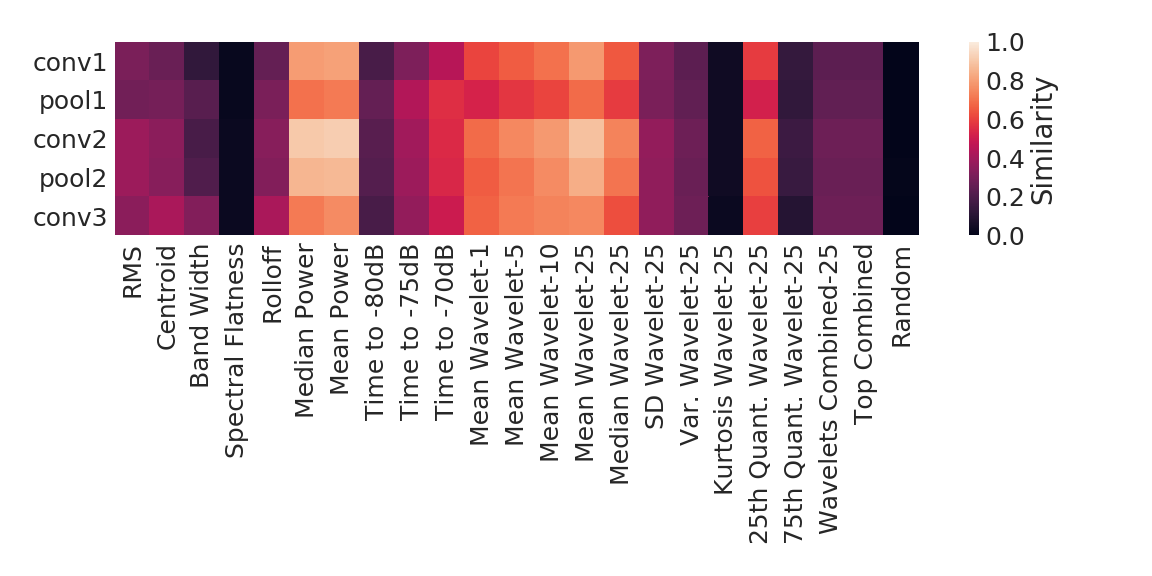}
\caption{Deformable}\label{default}
\end{subfigure}
\caption{Linear CKA similarity between the \textbf{untrained} deep features from all layers for the Composer dataset and all hand-crafted features for the (a) Regular and (b) Deformable deep features.}
\label{fig:Composer-layer-sim-untrained}
\end{figure}

We next compare the trained features from each initialization to the untrained features in terms of Linear CKA similarity for the Regular (Figure~\ref{fig:untrain-train-Regular}) and Deformable (Figure~\ref{fig:untrain-train-Deformable}) architectures across datasets.  We additionally compare the untrained deep features by architecture to all of the hand-crafted features for the NSynth (Figure~\ref{fig:NSynth-sim-all-untrained}), Composer (Figure~\ref{fig:Composer-sim-all-untrained}) and Beethoven (Figure~\ref{fig:Beethoven-sim-all-untrained}) datasets, again finding that the untrained features from the last convolutional layer for all architectures are highly similar to the hand-crafted features, especially the wavelet features. Additionally, across layers the untrained deep features are very similar to the hand-crafted wavelet features for the NSynth (Figure~\ref{fig:NSynth-layer-sim-untrained}) and Composer (Figure~\ref{fig:Composer-layer-sim-untrained}) datasets.

\begin{figure}[!htb]
\centering
\begin{subfigure}[b]{0.325\linewidth}
\centering
\includegraphics[width=\linewidth]{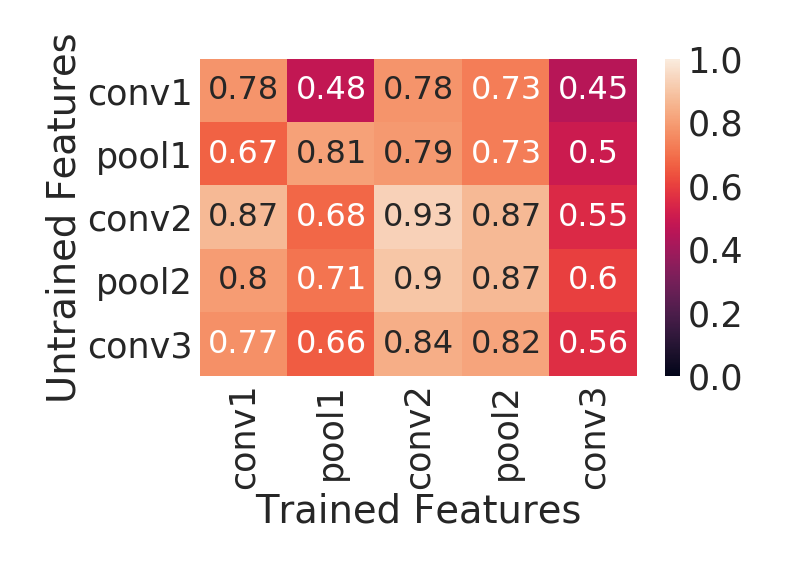}
\caption{Regular}
\end{subfigure}
\begin{subfigure}[b]{0.325\linewidth}
\centering
\includegraphics[width=\linewidth]{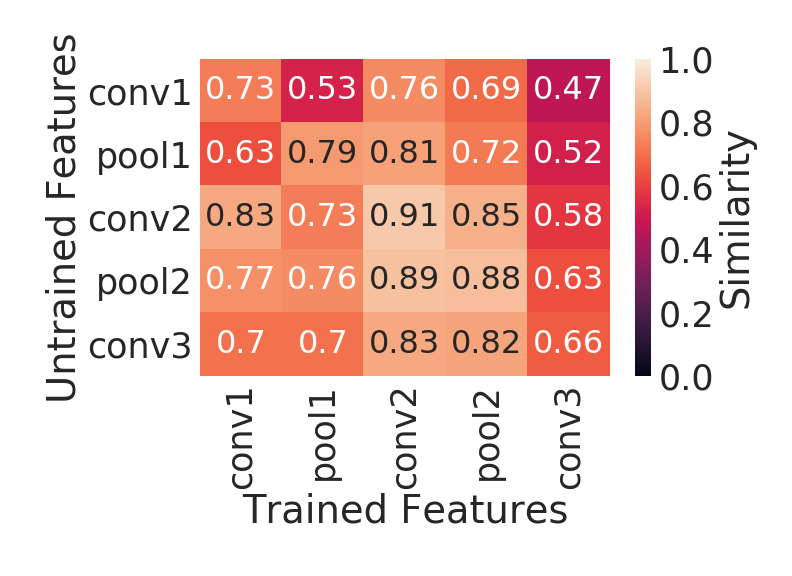}
\caption{Deformable}
\end{subfigure}
\caption{Linear CKA similarity between the \textbf{untrained} deep features from all layers and the trained deep features for the NSynth dataset and all hand-crafted features for the (a) Regular and (b) Deformable deep features.}
\label{fig:sim-NSynth-untrained-trained}
\end{figure}

Finally, we can compare the trained features by layer to the untrained features for the NSynth dataset for the Regular and Deformable architectures (Figure~\ref{fig:sim-NSynth-untrained-trained}).  The middle layers tend to be highly similar to each other, whether trained or not, while the last convolutional layer (\texttt{conv3}) is the least similar to the other layers and to itself before and after training.

\section{How do Different Architectures, Layers and Initializations Differ in their Learned Deep Features?}\label{suppsec:experiment_4}

In this section, we use the Linear CKA similarity measure to compare learned deep features across different architectures, layers, initializations and channels.

\subsection{Initializations}
\begin{figure*}[h]
\centering
\begin{subfigure}[b]{.32\linewidth}
\includegraphics[width=\linewidth]{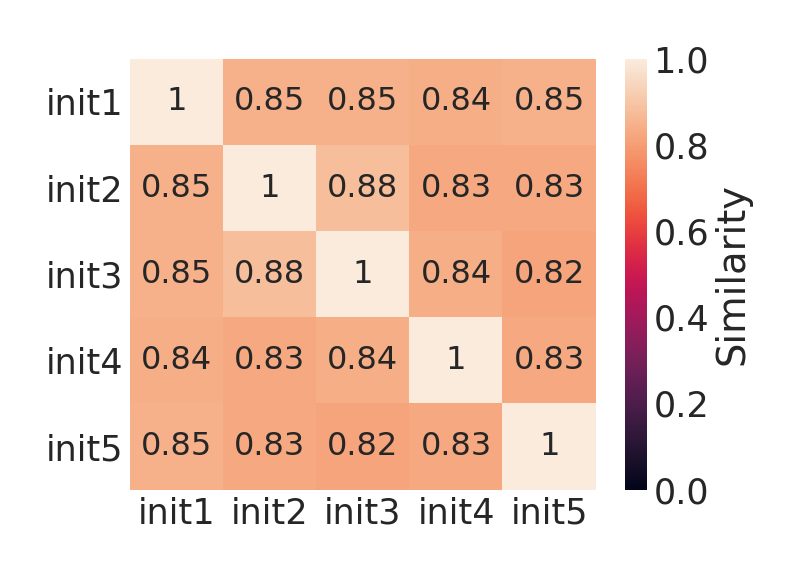}
\caption{NSynth}
\end{subfigure}
\begin{subfigure}[b]{.32\linewidth}
\includegraphics[width=\linewidth]{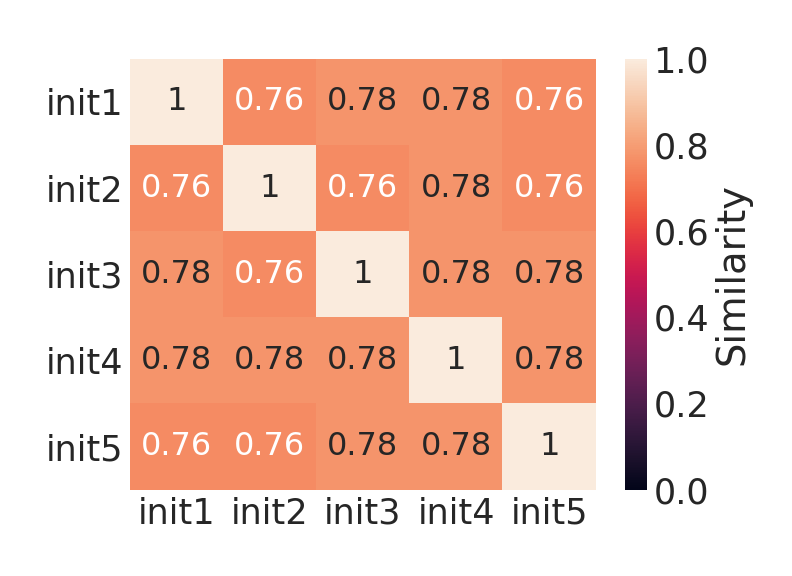}
\caption{Composer}
\end{subfigure}
\begin{subfigure}[b]{.32\linewidth}
\includegraphics[width=\linewidth]{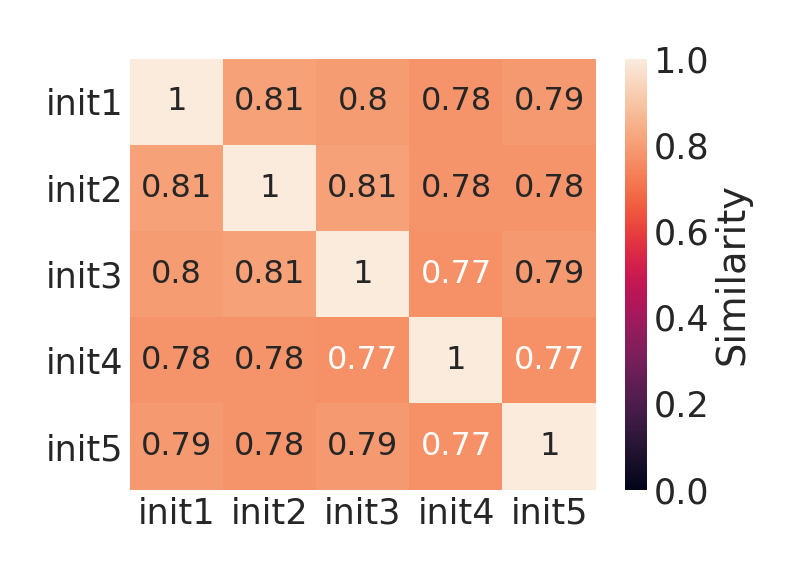}
\caption{Beethoven}
\end{subfigure}
\caption{Linear CKA similarity between initializations for Regular deep features from the last convolutional layer for the (a) NSynth, (b) Composer and (c) Beethoven datasets.  Deep features across initializations are highly similar, but not identical.}
\label{fig:init-init-sim-Regular}
\end{figure*}

\begin{figure*}[h]
\centering
\begin{subfigure}[b]{.32\linewidth}
\includegraphics[width=\linewidth]{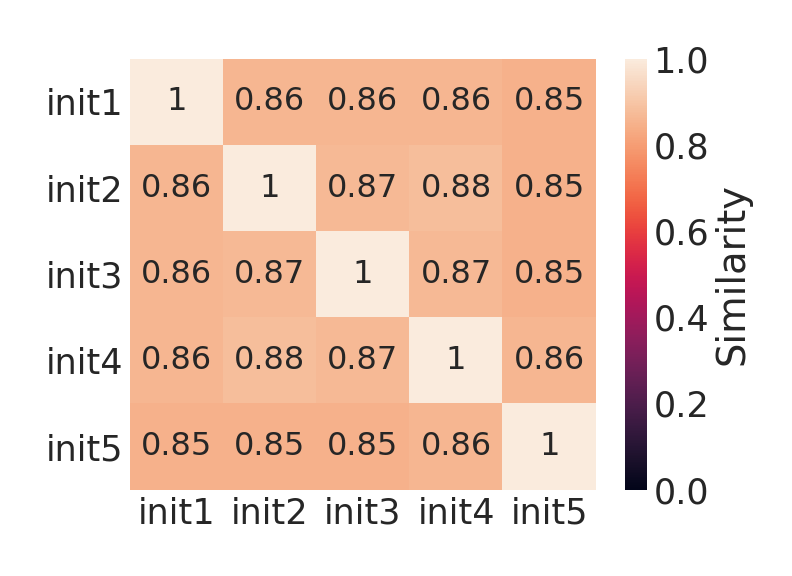}
\caption{NSynth}
\end{subfigure}
\begin{subfigure}[b]{.32\linewidth}
\includegraphics[width=\linewidth]{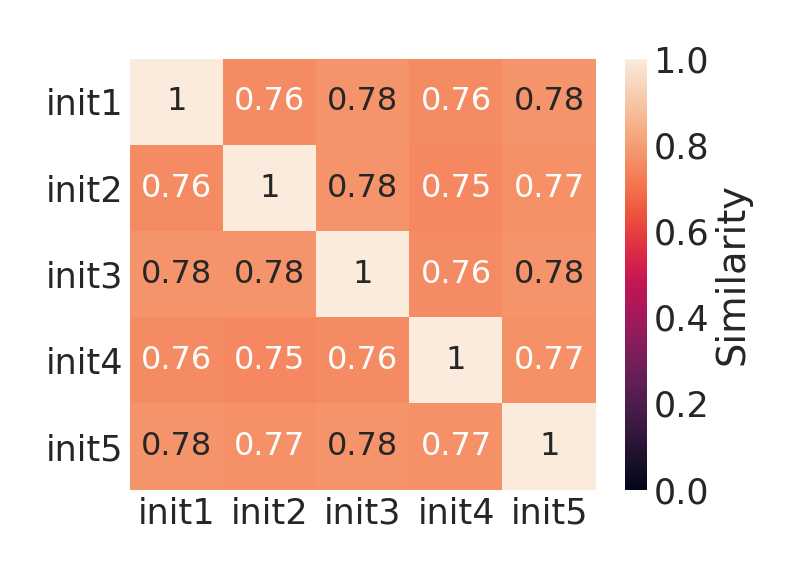}
\caption{Composer}
\end{subfigure}
\begin{subfigure}[b]{.32\linewidth}
\includegraphics[width=\linewidth]{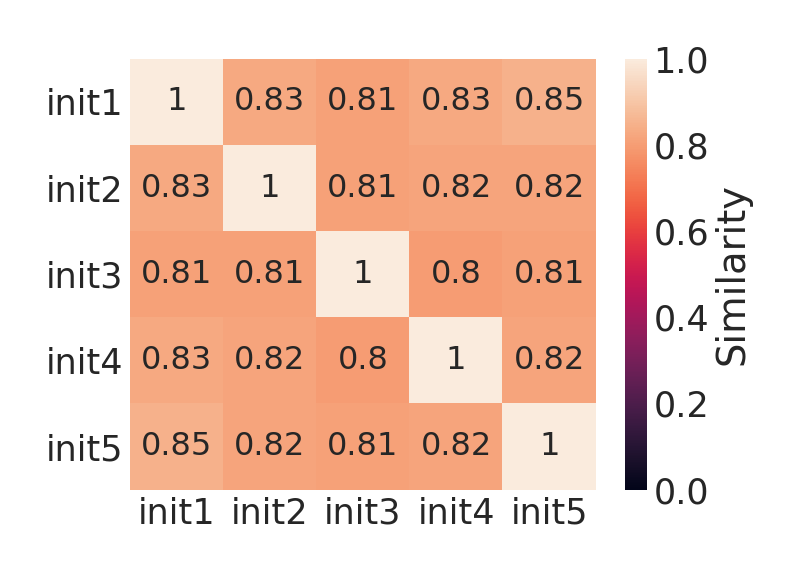}
\caption{Beethoven}
\end{subfigure}
\caption{Linear CKA similarity between initializations for Deformable deep features from the last convolutional layer for the (a) NSynth, (b) Composer and (c) Beethoven datasets.  Deep features across initializations are highly similar, but not identical.}
\label{fig:init-init-sim-Deformable}
\end{figure*}

We calculate the Linear CKA similarity between the learned features for the last convolutional layer (\texttt{conv3}) for different initializations for all datasets for the Regular (Figure~\ref{fig:init-init-sim-Regular}) and Deformable architectures (Figure~\ref{fig:init-init-sim-Deformable}).  Similar to the findings of \citet{DBLP:journals/corr/LiYCLH15, NEURIPS2018_5fc34ed3}, the deep features for different initializations are highly similar, but not identical, which reinforces the need to consider multiple initializations when exploring and using deep features. 

\begin{figure*}[h]
\centering
\begin{subfigure}[b]{.32\linewidth}
\includegraphics[width=\linewidth]{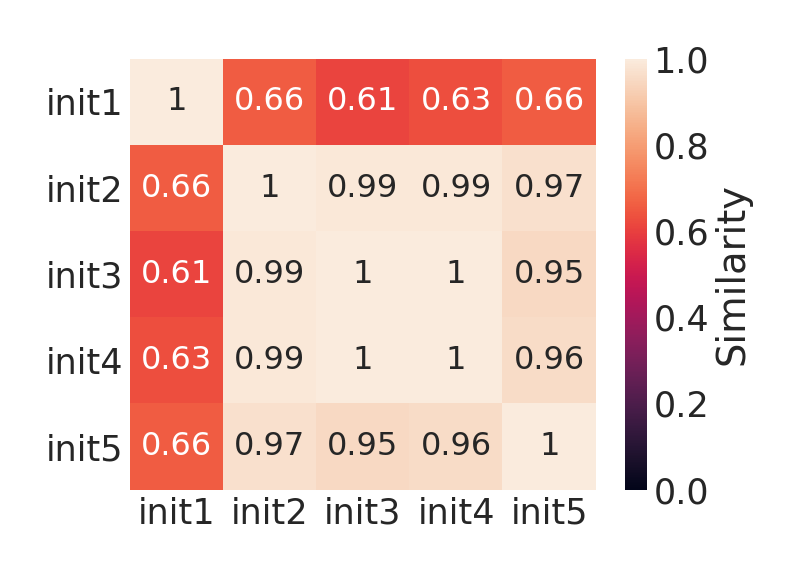}
\caption{\texttt{conv1}}
\end{subfigure}
\begin{subfigure}[b]{.32\linewidth}
\includegraphics[width=\linewidth]{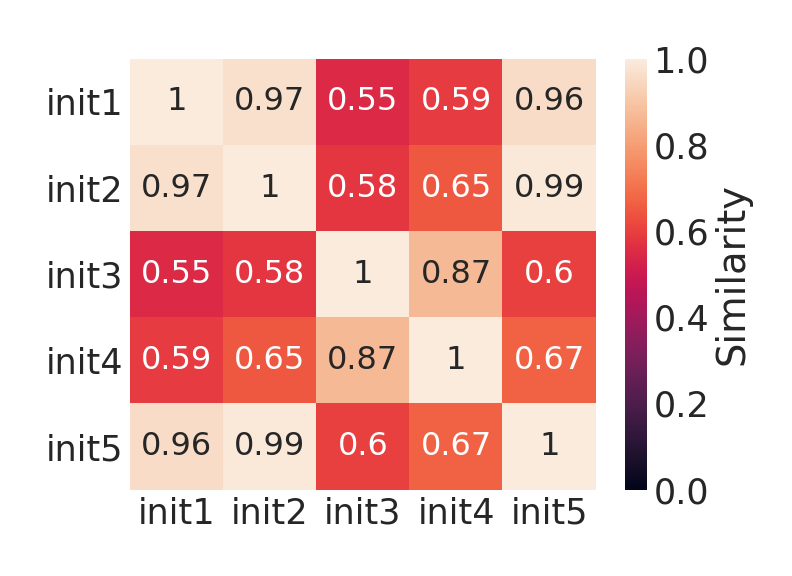}
\caption{\texttt{pool1}}
\end{subfigure}
\begin{subfigure}[b]{.32\linewidth}
\includegraphics[width=\linewidth]{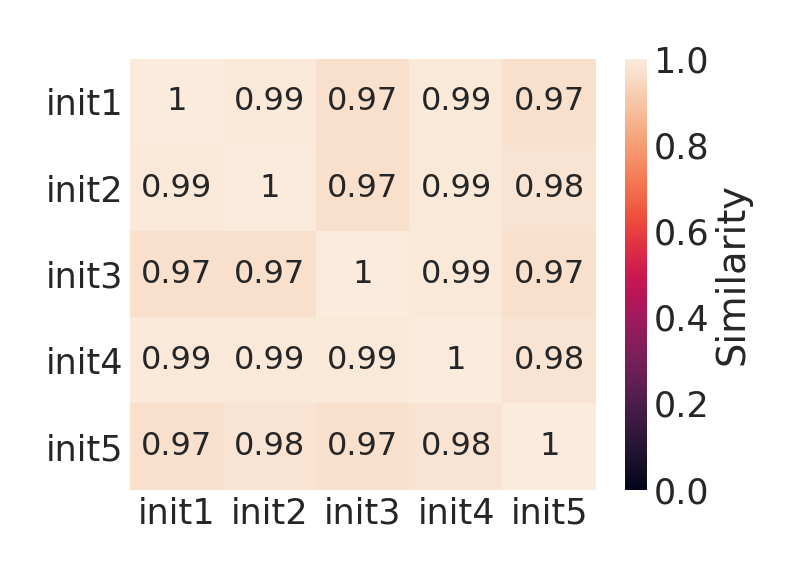}
\caption{\texttt{conv2}}
\end{subfigure}

\begin{subfigure}[b]{.32\linewidth}
\includegraphics[width=\linewidth]{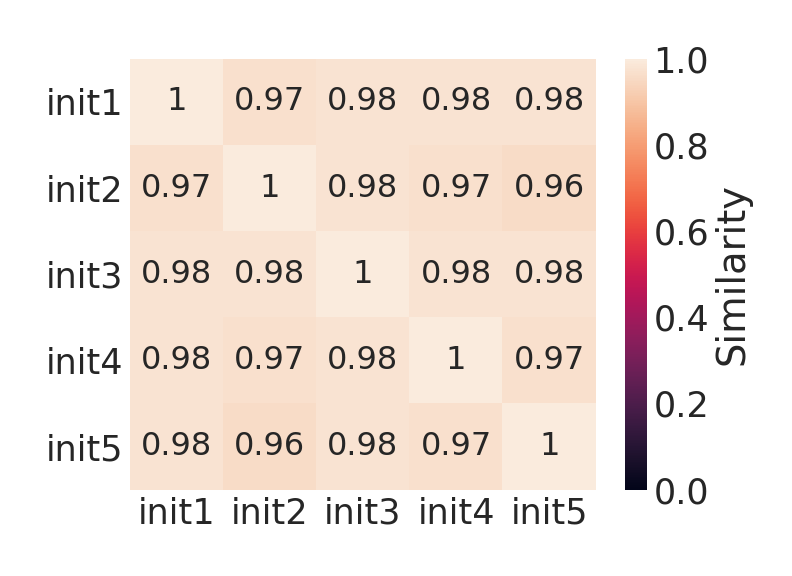}
\caption{\texttt{pool2}}
\end{subfigure}
\begin{subfigure}[b]{.32\linewidth}
\includegraphics[width=\linewidth]{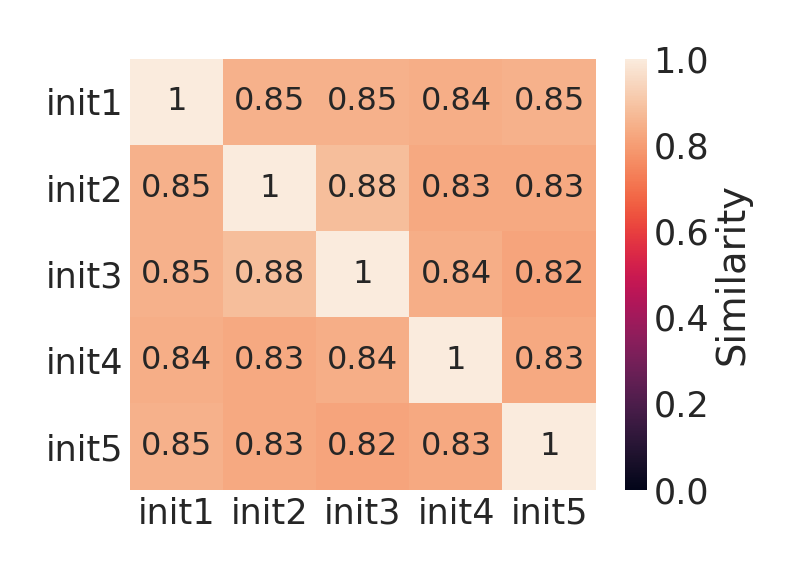}
\caption{\texttt{conv3}}
\end{subfigure}
\caption{Linear CKA similarity between initializations for the deep features from each layer for the Regular architectures on the NSynth dataset.}
\label{fig:init-layer-Regular}
\end{figure*}

\begin{figure*}[h]
\centering
\begin{subfigure}[b]{.32\linewidth}
\includegraphics[width=\linewidth]{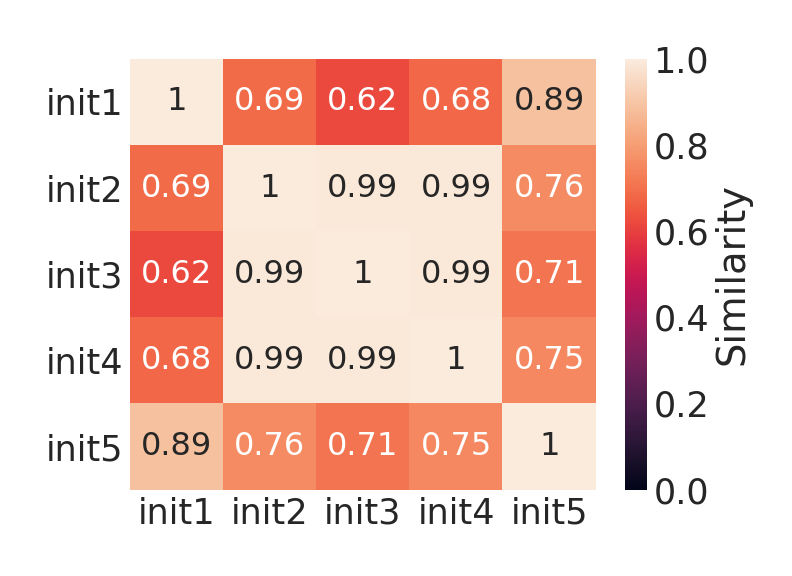}
\caption{\texttt{conv1}}
\end{subfigure}
\begin{subfigure}[b]{.32\linewidth}
\includegraphics[width=\linewidth]{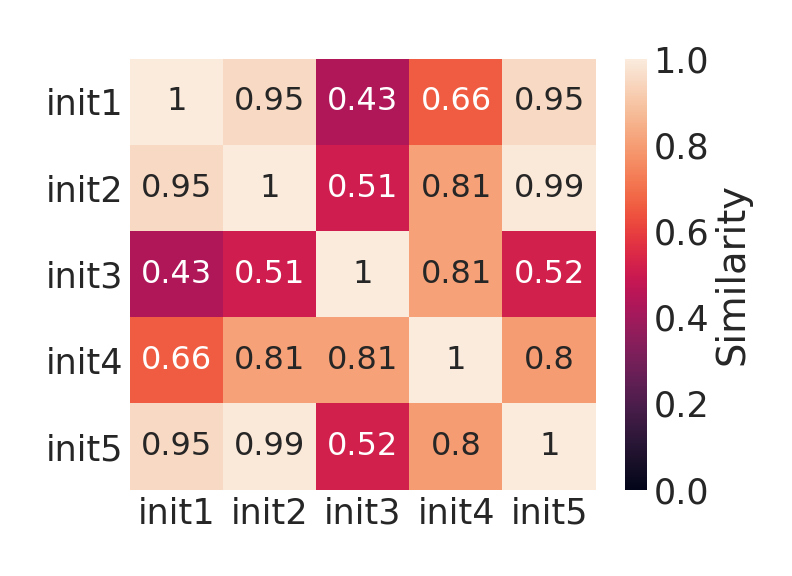}
\caption{\texttt{pool1}}
\end{subfigure}
\begin{subfigure}[b]{.32\linewidth}
\includegraphics[width=\linewidth]{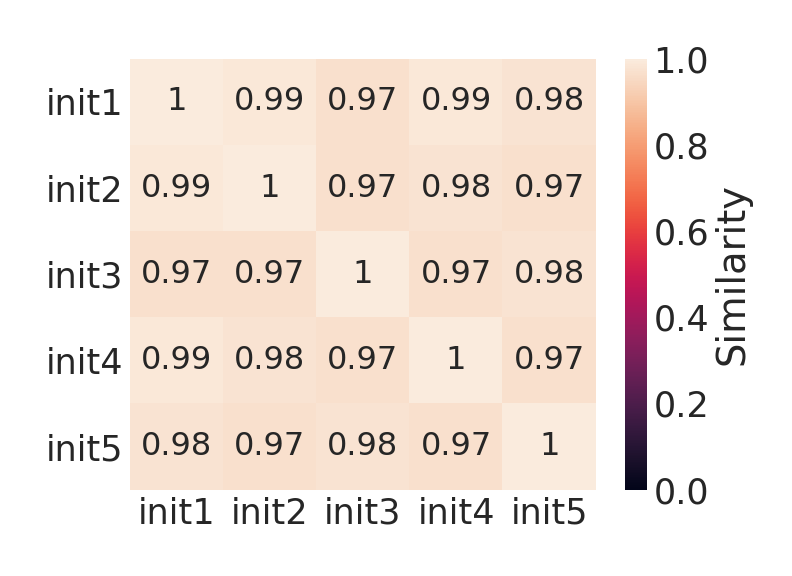}
\caption{\texttt{conv2}}
\end{subfigure}

\begin{subfigure}[b]{.32\linewidth}
\includegraphics[width=\linewidth]{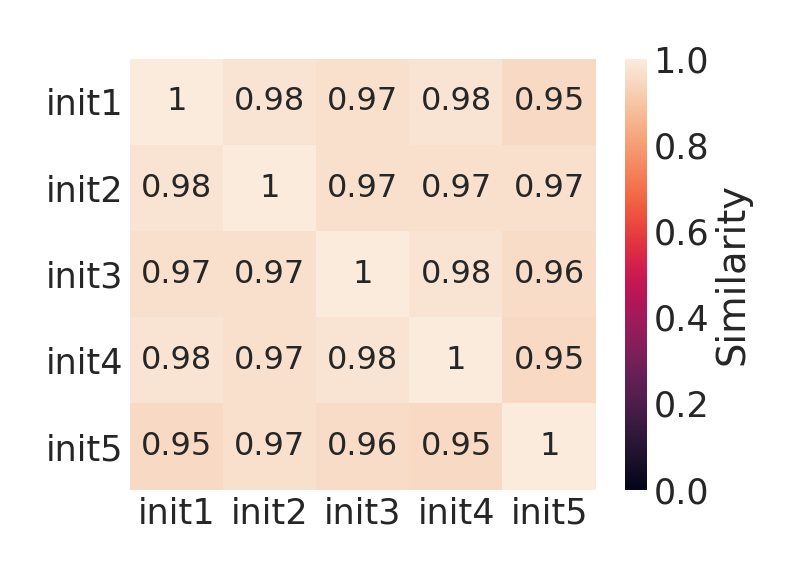}
\caption{\texttt{pool2}}
\end{subfigure}
\begin{subfigure}[b]{.32\linewidth}
\includegraphics[width=\linewidth]{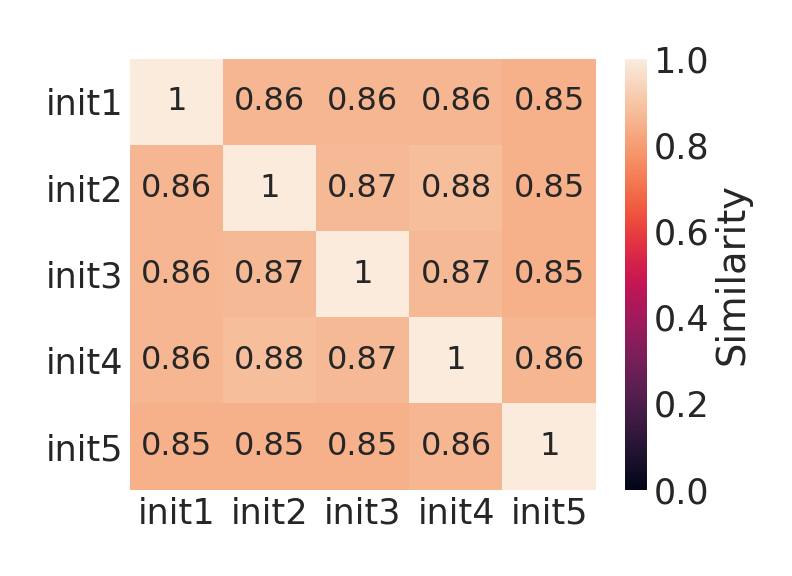}
\caption{\texttt{conv3}}
\end{subfigure}
\caption{Linear CKA similarity between initializations for the deep features from each layer for the Deformable architectures on the NSynth dataset.}
\label{fig:init-layer-Deformable}
\end{figure*}

We also explore the similarity between initializations for different layers for the NSynth dataset for the Regular (Figure~\ref{fig:init-layer-Regular}) and Deformable (Figure~\ref{fig:init-layer-Deformable}) architectures.  In general, the deep features for layers \texttt{conv2} and \texttt{pool2} are very similar to each other, while the learned features differ more for the earlier and later layers.

\subsection{Architectures}

\begin{figure}[h]
\centering
\begin{subfigure}[b]{0.325\linewidth}
\centering
\includegraphics[width=\linewidth]{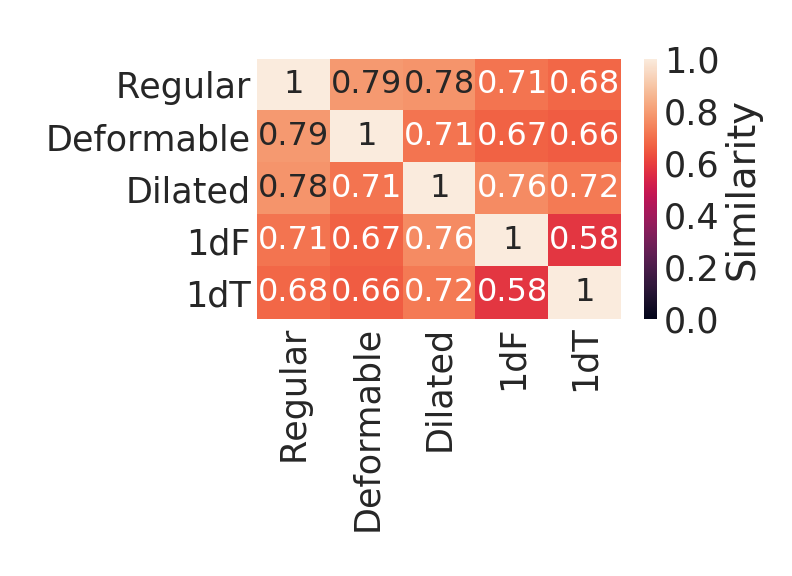}
\caption{NSynth}
\end{subfigure}
\begin{subfigure}[b]{0.325\linewidth}
\centering
\includegraphics[width=\linewidth]{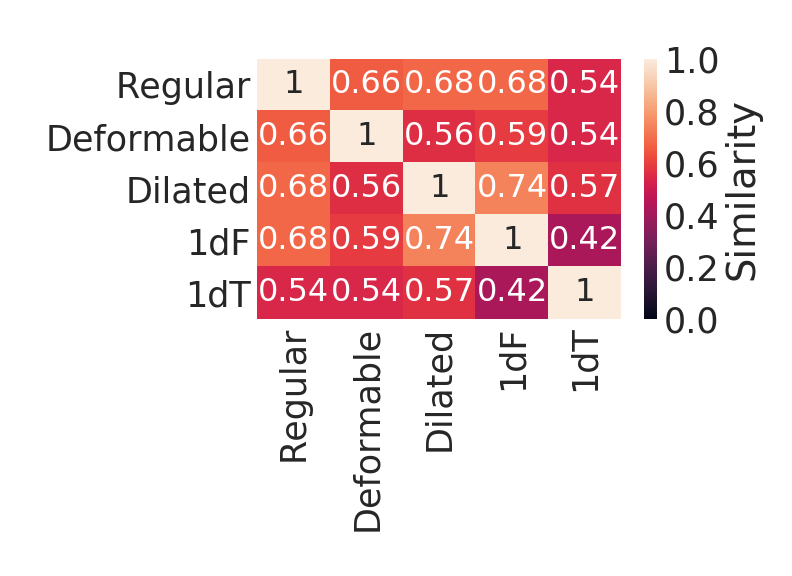}
\caption{Composer}
\end{subfigure}
\begin{subfigure}[b]{0.325\linewidth}
\centering
\includegraphics[width=\linewidth]{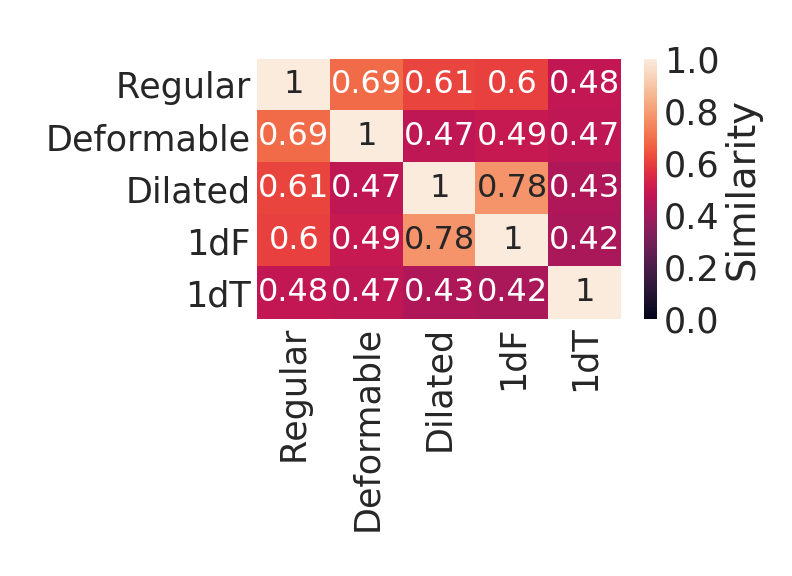}
\caption{Beethoven}
\end{subfigure}
\caption{Linear CKA similarity between architectures for deep features from the last convolutional layer for the (a) NSynth, (b) Composer and (c) Beethoven datasets.  The similarity is averaged over initializations. }
\label{fig:arch-arch-sim}
\end{figure}

We also compare the similarity of the deep features from the last convolutional layer  (\texttt{conv3}) between architectures, averaged over initializations, and find that the 1dT deep features tend to be the least similar to the features from the other architectures (Figure~\ref{fig:arch-arch-sim}).  The Regular features are similar to the Deformable and Dilated features, while the Deformable deep features are less similar to the Dilated and 1dF features than the Regular features are.  

\subsection{Channel Similarity}
\begin{figure}[!htb]
\centering
\begin{subfigure}[b]{0.325\linewidth}
\centering
\includegraphics[width=\linewidth]{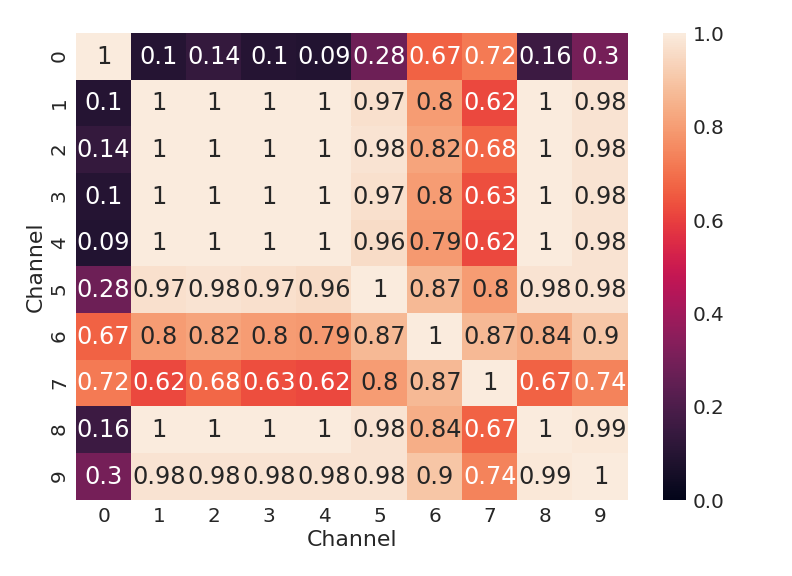}
\caption{\texttt{conv1}}
\end{subfigure}
\begin{subfigure}[b]{0.325\linewidth}
\centering
\includegraphics[width=\linewidth]{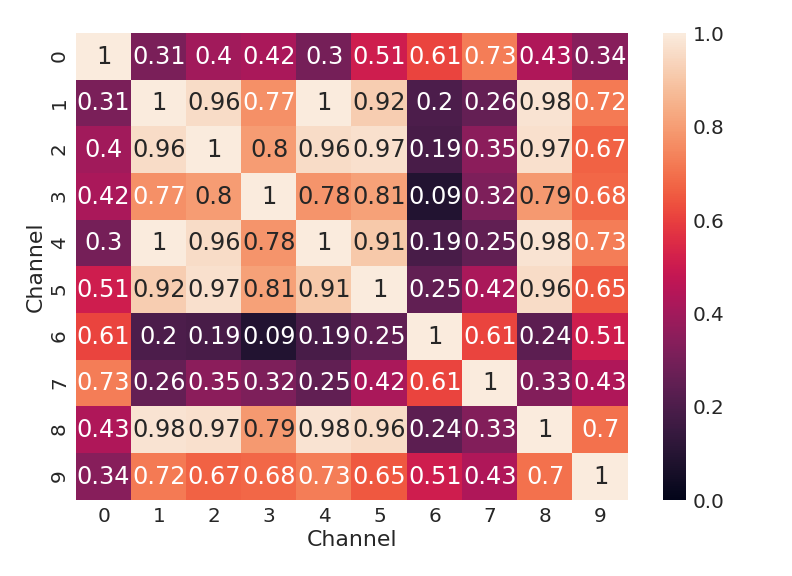}
\caption{\texttt{pool1}}
\end{subfigure}
\caption{Linear CKA similarity between channels for one initialization of the Regular deep features on the NSynth dataset for the (a) \texttt{conv1} and (b) \texttt{pool1} layers.  In general, the deep features are very similar across channels.  }
\label{fig:channel-channel-1}
\end{figure}

\begin{figure}[h]
\centering
\begin{subfigure}[b]{0.325\linewidth}
\includegraphics[width=\linewidth]{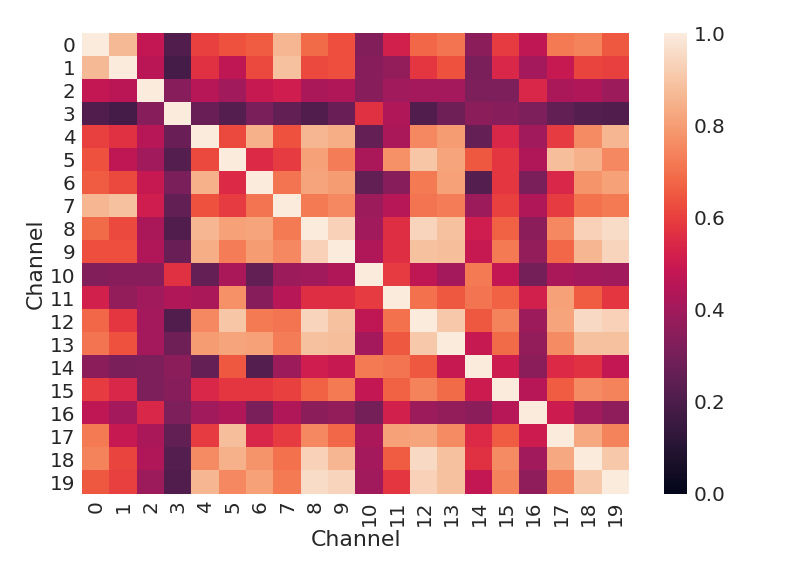}
\caption{\texttt{conv2}}
\end{subfigure}
\begin{subfigure}[b]{0.325\linewidth}
\includegraphics[width=\linewidth]{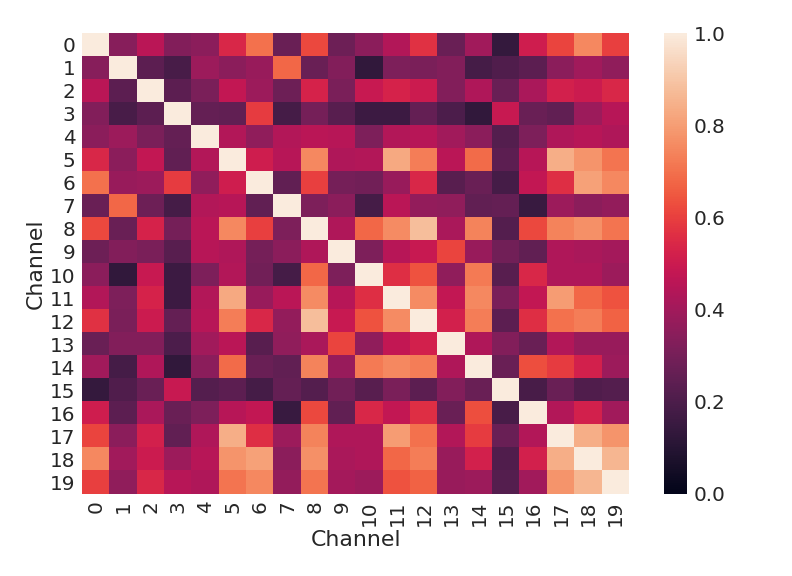}
\caption{\texttt{pool2}}
\end{subfigure}
\begin{subfigure}[b]{0.325\linewidth}
\includegraphics[width=\linewidth]{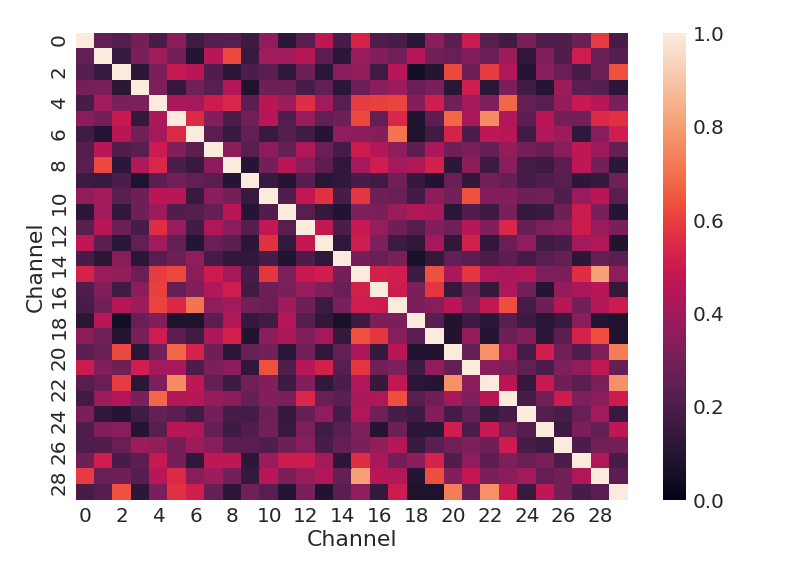}
\caption{\texttt{conv3}}
\end{subfigure}
\caption{Linear CKA similarity between channels for one initialization of the Regular deep features on the NSynth dataset for the (a) \texttt{conv2},  (b) \texttt{pool2} and (c) \texttt{conv3} layers. }
\label{fig:channel-channel-2}
\end{figure}

For computational reasons, for the main layer experiments, we average the features over channels for layers \texttt{conv1} and \texttt{pool1}.  This reduces the dimensionality of the learned features at these layers by a factor of 10 and allows for the matrix multiplications necessary for the Linear CKA calculations and for the logistic regression decoding experiments to occur in a reasonable time.  The dimensions of the learned features are each layer for the NSynth dataset are: \texttt{conv1}-(10, 124, 124), \texttt{pool1} - (10, 62, 62), \texttt{conv2} - (20, 22, 22), \texttt{pool2} - (20, 11, 11) and \texttt{conv3} - (30, 4, 4); averaging across channels reduces the dimensionality of the \texttt{conv1} features from $n\times 153760$ to $n\times 15376$.  This is reasonable, as except for one or two channels, the features across channels exhibit very high similarity for layers \texttt{conv1} and \texttt{pool1} (Figure~\ref{fig:channel-channel-1}). The later layers, however, are of smaller dimension and exhibit less similarity between channels (Figure~\ref{fig:channel-channel-2}), so we flatten over channels for the experiments using deep features from the \texttt{conv2}, \texttt{pool2} and \texttt{conv3} layers.  

\subsection{Layer Similarity}

\begin{figure}[!htb]
\centering
\begin{subfigure}[b]{0.325\linewidth}
\centering
\includegraphics[width=\linewidth]{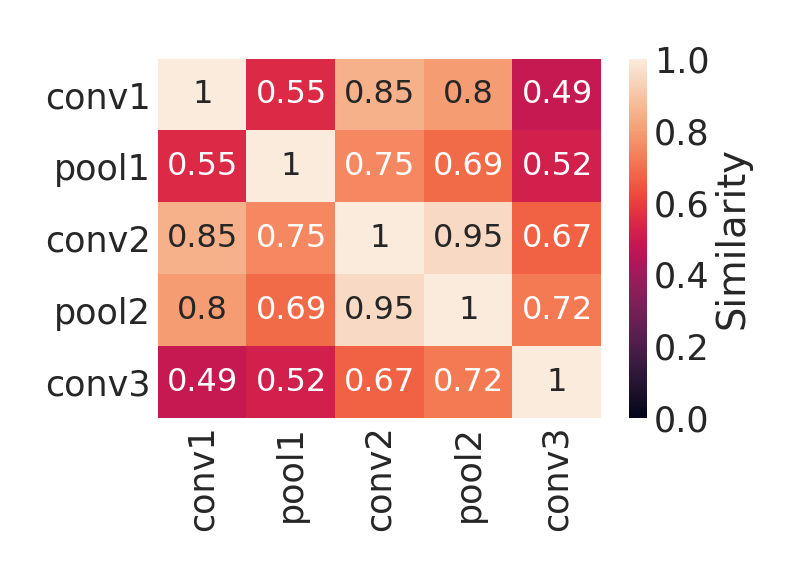}
\caption{Regular}
\end{subfigure}
\begin{subfigure}[b]{0.325\linewidth}
\centering
\includegraphics[width=\linewidth]{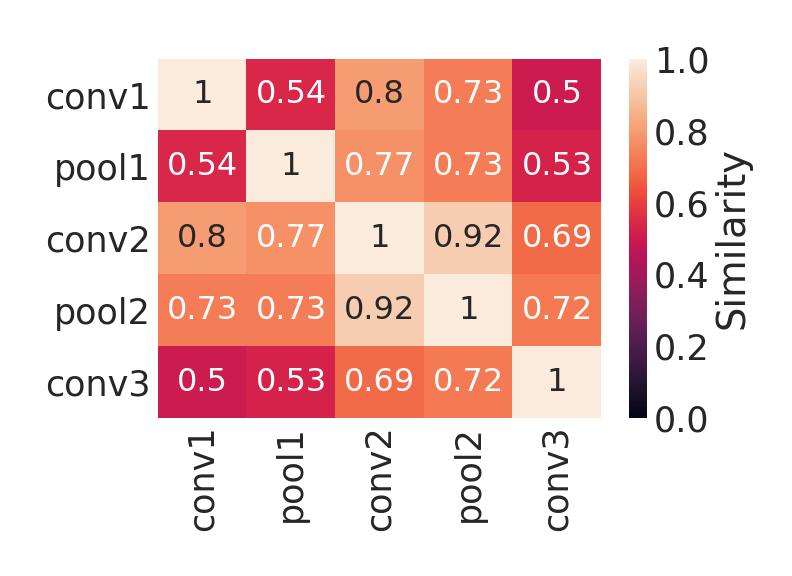}
\caption{Deformable}
\end{subfigure}
\caption{Linear CKA similarity between layers for the (a) Regular and (b) Deformable architectures trained on the NSynth dataset.  The similarity is averaged over initializations. }
\label{fig:layer-layer-sim-NSynth}
\end{figure}

\begin{figure}[!htb]
\begin{center}
\includegraphics[width=0.325\textwidth]{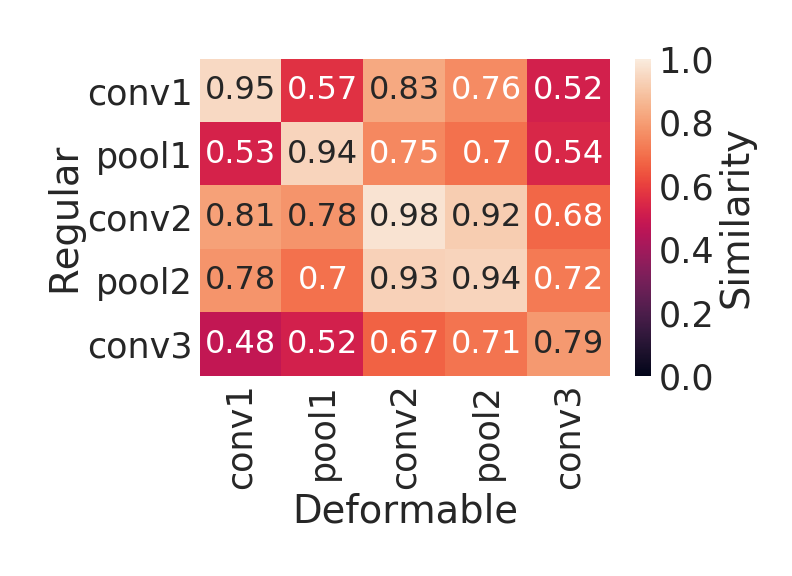}
\caption{Linear CKA similarity between layers from the Regular architecture with layers from the Deformable architecture on the NSynth dataset.  The similarity is averaged over initializations. }
\label{fig:layerReg-layerDeform-init}
\end{center}
\end{figure}

Finally, we compare the Linear CKA similarity of the deep features by layer for the Regular and Deformable architectures for the NSynth dataset to explore how learned features compare across layers within the same architecture (Figure~\ref{fig:layer-layer-sim-NSynth}) .  The middle layers, \texttt{conv2} and \texttt{pool2}, tend to be most similar to all other layers, while the earlier and later layers exhibit less similarity to each other.  We also compare the deep features by layer for the Regular architecture to the deep features by layer for the Deformable architecture for the NSynth data (Figure~\ref{fig:layerReg-layerDeform-init}).  As expected, the actual Deformable convolution layer (\texttt{conv3}) tends to have the least similarity with the same layer in the Regular architecture, as well as the earlier layers in the Regular architecture.  In general, all convolutional layers tend to differ more from each other between the Regular and Deformable architectures than the pooling layers do.

\section{Conclusion}\label{suppsec:discussion}

In the context of the music audio application for this work, we explore why deep convolutional architectures perform so well for music audio discriminative tasks, with a focus on understanding deep features in terms of feature classification and similarity to interpretable features. We find that deep features from Regular and Deformable convolutions in particular are useful for several discriminative tasks, whether or not originally trained on a specific task. We also find that deep features are very similar to hand-crafted wavelet features, whether the deep features are extracted from a trained or untrained architecture.  Earlier convolutional layers do not improve in the accuracy of their deep features after training, while the later layers do.  Indeed, the later layers become less similar to the hand-crafted features after training, while the deep features from earlier layers remain highly similar to the hand-crafted features. This suggests that deep convolutions may perform so well for discriminative audio tasks, even though mel-spectrograms are not natural images, because they recover hand-crafted features from classical signal processing that are known (and shown) to be useful for discriminative tasks.  

The similarity between the untrained deep features and the hand-crafted features, in particular, suggests future avenues of research in refining deep priors for music audio data, relating to and expanding on \citet{dap}, for example. Additionally, incorporating knowledge of the deep features into future deep models could enable reduced size architectures for related audio tasks, which currently tend to be massive in terms of parameters and required data, e.g., \citet{oord_wavenet:_2016}.


\end{document}